\documentclass[12pt]{article}

\usepackage[T1]{fontenc}
\usepackage{geometry}
\usepackage{alphabeta} 
\usepackage{physics}
\usepackage{charter,graphicx} 
\usepackage{epsf}
\usepackage{wrapfig}
\usepackage{graphicx}
\usepackage{graphicx}
\usepackage{amsmath, amsthm, amsfonts, amssymb}
\usepackage{slashed}
\usepackage{here}
\usepackage{bbm}
\usepackage[toc,page]{appendix}
\usepackage{mathrsfs} 
\usepackage[hidelinks]{hyperref}
\usepackage{cleveref}
\usepackage{slashed}
\usepackage{dsfont}
\usepackage{minibox}
\usepackage{graphics}
\usepackage{epsfig}
\usepackage{xcolor}
\usepackage{array}
\usepackage{tikz}
\usepackage{tikz-cd}
\usetikzlibrary{arrows,calc}
\usepackage{relsize}
\usetikzlibrary{decorations.pathmorphing,calc}
\usepackage{authblk}
\usepackage[labelfont=bf]{caption}
\usepackage{caption}
\usepackage{subcaption}
\usepackage{placeins}
\usepackage{braket}
\usepackage[compat=1.1.0]{tikz-feynman}

\newcommand{\be}{\begin{equation}}
\newcommand{\ee}{\end{equation}}
\newcommand{\bea}{\begin{eqnarray}}
\newcommand{\eea}{\end{eqnarray}}
\newcommand{\bal}{\begin{aligned}}
\newcommand{\eal}{\end{aligned}}

\def\a{\alpha}
\def\b{\beta}
\def\g{\gamma}
\def\G{\Gamma}
\def\d{\delta}
\def\D{\Delta}
\def\ve{\varepsilon}
\def\m{\mu}
\def\n{\nu}
\def\l{\lambda}

\def\r{\rho}
\def\s{\sigma}

\def\thintablerule{\hrule height0.4pt}

\allowdisplaybreaks

\begin{document}

\centerline
{\Large \bf Energy-Momentum tensor correlators in $\phi^4$ theory I:}
\vskip .5cm
\centerline {\Large \bf The spin-zero sector}

\vskip 2 cm
\centerline{\large Nikos Irges and Leonidas Karageorgos}
\vskip1ex
\vskip.5cm
\centerline{\it Department of Physics}
\centerline{\it National Technical University of Athens}
\centerline{\it Zografou Campus, GR-15780 Athens, Greece}
\vskip .4cm
\begin{center}
{\it e-mail: irges@mail.ntua.gr, leokarageorgos@mail.ntua.gr}
\end{center}
\vskip 1.5 true cm
\centerline {\bf Abstract}
\vskip 3.0ex
\thintablerule
\vskip 2.0ex

We revisit the construction of the renormalized trace $\Theta$ of the Energy-Momentum tensor in the four-dimensional $\lambda\phi^4$ theory,
using dimensional regularization in $d=4-\ve$ dimensions.
We first construct several basic correlators such as $\braket{\phi^2 \phi\phi}$, $\braket{\phi^4 \phi \phi}$ to order $\lambda^2$ and from these the correlators $\braket{K_I \phi \phi}$ 
and $\braket{K_I K_J}$ with $K_I$ the basis of dimension $d$ operators.
We then match the limit of their expressions on the Wilson-Fisher fixed point 
to the corresponding expressions obtained in Conformal Field Theory.
Then, using the 3-point function $\braket{\Theta\phi\phi}$, we construct the operator $\Theta$ as a certain linear combination of the basis operators, 
using the requirements that $\Theta$ should vanish on the fixed point and that it should have zero anomalous dimension.
Finally, we compute the 2-point function $\braket{\Theta\Theta}$ and we show that it obeys an 
eigenvalue equation that gives additional information about the internal structure of the Energy-Momentum tensor operator
to what is already contained in its Callan-Symanzik equation.

\vskip 2.0ex
\thintablerule

\vskip-0.2cm
\newpage

\tableofcontents
\numberwithin{equation}{section}
\newpage
\section{Introduction}

In a couple of older papers \cite{Brown,Collins_Brown} Brown and Collins constructed the renormalized 
Energy-Momentum Tensor (EMT) of the four-dimensional $\lambda\phi^4$ theory.
We will be interested in the massless limit, in which case the bare Lagrangean is
\be \label{flat space action}
\mathcal{L}_0 = \frac{1}{2} \partial_\mu \phi_0\partial^\mu \phi_0 - \frac{\l_0}{4!} \phi_0^4 \, .
\ee
In particular, \cite{Brown,Collins_Brown} considering a deformed by a term $\eta_0 R\phi_0^2$ curved space 
version of the $\l\phi^4$ theory, showed that in the massless, flat space limit the  trace 
$\Theta$ of the renormalized EMT (in $d=4$) is of the form
\be\label{CollinsTheta}
\Theta = \beta_\lambda \phi^4 + d_\eta \Box \phi^2\, ,
\ee
where $\beta_\l$ is the beta function of the coupling $\l$ and $d_\eta$ is a coefficient that 
contains renormalization factors and an arbitrary renormalized constant $\eta$. 
This $\eta$ is the renormalized version of the bare $\eta_0$, which denotes the deviation of the coupling constant 
of the $R \phi^2$ term from the standard coefficient $\xi_0 =\frac{1}{2}\frac{d-2}{4(d-1)} $. 
Then, calling it "a convenient choice", they set $d_\eta=0$ in order to define the renormalized $\eta$, thus defining
at the same time a $\Theta$ that does not contain any dimension $4$ operator other than $\b_\l \phi^4$.   
Although this definition of the trace is minimal, in the sense that the trace contains only one operator, it is not unique.
According to Brown \cite{Brown} defining a renormalized energy-momentum tensor with non-minimal trace, is allowed. 
The non-minimal definitions differ from the minimal one by a finite constant multiple of the renormalized $\Box \phi^2$  operator. 

The anomalous dimensions of composite spin operators in the $\l\phi^4$ theory have been 
computed to a rather high order already some time ago, see for example \cite{Manashov_Traceless}. 
The renormalization of 3-point functions involving composite spin-0 operators 
has been extensively studied in \cite{Skenderis 3K}, even though in the context of a CFT in momentum space. 
Renormalization techniques closely related to those presented here
can be found in \cite{Claudio}, where a generalized conformal symmetry that imposes the constraints of 
conformal symmetry on the correlation functions are exploited.

One of the goals of this paper is to rederive $\Theta$ using a number of Quantum Field Theory (QFT) operator correlators rather than just renormalization counterterms.
The computation of the correlators involved in the process is by itself interesting and for this reason we present it in great detail.
Another novel aspect of the present approach is that we give the simple rules that connect the computed form of the QFT 
correlators projected on the interacting IR 
(or Wilson-Fisher (WF)) fixed point with their corresponding expressions that arise in the context of Conformal Field Theory (CFT).

Last but not least, we show that the self 2-point function of $\Theta$ satisfies an eigenvalue-like equation
\footnote{Eigenvalue-like because $e_\Theta$ is a function of $\l$. 
We will continue using the term eigenvalue for simplicity.} of the form
\be\label{eigTheta}
\m \frac{\partial} {\partial\m} \braket{\Theta\Theta} =- e_\Theta \braket{\Theta\Theta}
\ee
and we fix the leading order form of the eigenvalue $e_\Theta$ within perturbation theory, with Dimensional Regularization (DR) in 
$d=4-\ve$ dimensions, with regularization  scale $\m$.
To see the meaning of \eqref{eigTheta} consider the 2-point correlator of an operator $\mathcal{O}$ of 
anomalous dimension $\Gamma_{\mathcal{O}}$ with itself and its Callan-Symanzik (CS) equation
\be
\left(\m \frac{\partial} {\partial\m} + \b_\l \frac{\partial}{\partial\l} + 2 \Gamma_{\mathcal{O}} \right) \braket{\mathcal{O}\mathcal{O}} = 0\, .
\ee
It is easy to see that in a  perturbative series in powers of $\l$, at a given order, the middle term analogous to $\b_\l$ is higher order with respect to the other two terms  ,
provided that $\Gamma_{\mathcal{O}}\ne 0$. Then, the CS equation can be also read as the eigenvalue equation 
\be\label{eOO}
\m \frac{\partial} {\partial\m}  \braket{\mathcal{O} \mathcal{O}} = - 2 \Gamma_{\mathcal{O}}  \braket{\mathcal{O} \mathcal{O}}, 
\ee
a rather trivial statement, as it contains no more information than the CS equation from which it originates. 
Things become non-trivial when the operator 
has zero anomalous dimension. Such an operator is the EMT whose trace satisfies 
\be\label{CSTheta}
\left(\m \frac{\partial} {\partial\m} + \b_\l \frac{\partial}{\partial\l} \right) \braket{\Theta\Theta} = 0\, 
\ee
and it is not clear if it can be read as an eigenvalue equation. By adding and subtracting 
$e_\Theta$ as in \cite{Ising} a possible eigenvalue equation emerges,
if the two parentheses in the bracket vanish separately:
 \be\label{eTheta}
 \left[ \left(\m \frac{\partial} {\partial\m} + e_\Theta\right) + \left(\b_\l \frac{\partial}{\partial\l} - e_\Theta\right) \right] \braket{\Theta\Theta} = 0\, .
 \ee
 Any operator $\mathcal O$ constructed from the field and its derivatives can be inserted as a term in the quantum effective action,
with its dimensionality adjusted appropriately by some dimensionful parameter.
Effective field theory instructs us that the insertion must be of the form $g {\mathcal O}$ where $g$ is the coupling associated with the operator.
This ties the RG flow of the coupling to the RG flow of the operator because the renormalization process requires the
effective action itself to be free of divergencies. The $g {\mathcal O}$ form of the insertion gives also a meaning to the operator through
physical processes like scattering amplitudes that are ultimately expressed in terms of the couplings.
It is therefore fortunate that standard algorithmic computational schemes like perturbation theory can be connected to experimental data
and it is even more fortunate that certain quantities, not in the perturbative regime in general, are also possible to probe via (semi) perturbative schemes.
Such are for example the critical exponents of interacting fixed points in some field theories 
that can be reproduced by the $\epsilon$-expansion.
It is in this sense that perturbative schemes are special for renormalizable theories.
Nothing of the above seems to hold for the trace operator $\Theta$. Since however it is believed to be an operator with a physical meaning,
the question that arises is what is its associated coupling. We argue that the eigenvalue $e_\Theta$ is precisely this coupling.
The peculiarity of $\Theta$ is that classically it is identically zero in the massless theory and the standard chain of renormalization steps
classical $\to$ bare $\to$ renormalized operator does not exist.
As a result, we do not know if the perturbative running that $e_\Theta$ defines is in any sense the preferred running that drives the system to its interacting fixed point.
In fact the generic process can be inverted: first, a consistent definition of a renormalized $\Theta$ can be given out of which, 
the eigenvalue $e_\Theta$ can be extracted. Then, by solving the two coupled differential equations inside
\eqref{eTheta}, the correlator $\braket{\Theta\Theta}$ may be extracted.
Clearly, if such a chain of reasoning exists, it gives new information about the internal quantum structure of the EM tensor.
The general solution to the coupled system in \eqref{eTheta},
\bea\label{eigsystem}
\left(\m \frac{\partial} {\partial\m} + e_\Theta\right)\braket{\Theta\Theta}&=&0 \nonumber\\
\left(\b_\l \frac{\partial}{\partial\l} - e_\Theta\right)\braket{\Theta\Theta}&=&0
\eea
is quite complicated. 
We can construct however straightforwardly a simple perturbative solution with $e_\Theta=2\Gamma_{\phi^4}$, in an expansion in $\l$.
We are looking for a solution $\braket{\Theta\Theta}(\l, \frac{-p^2}{\m^2})$ with an overall scaling $p^4$ that respects 
the vanishing of the trace operator at a fixed point, that is $\Theta\sim \b_\l$.
Inputs are the leading order values $\b_\l=\frac{3\l^2}{(4\pi)^2}$ and $\Gamma_{\phi^4}=\frac{6\l}{(4\pi)^2}$.
The solution is then 
\be\label{ThetaThetapert1}
\braket{\Theta\Theta} \sim c p^4 \b_\l^2 
\left(1 + \frac{6\l}{(4\pi)^2}\ln \left(\frac{-p^2}{\m^2}\right)+ O(\l^2)\right) \equiv p^4 c_\Theta\, .
\ee
A central result of this paper is the verification of this expression via a leading order diagrammatic calculation.
Other solutions, with different eigenvalues, can be constructed by the assumption that near fixed points
the eigenvalue represented by some critical exponent does not vary much \cite{Ising}. 
In this case the solution to \eqref{eigsystem} is separable in the variables $\l$ and $\m$.
Choosing for $e_\Theta$ the value of the critical exponent $\eta$ (twice the value of 
the wave function renormalization $2\g_\phi$ evaluated on the WF fixed point)
the solution takes the form
\be
\braket{\Theta \Theta} = c p^4 \b_\l^2 e^{\int_1^\l dy \frac{\eta-2 \frac{\partial \b_\l(y)}{\partial y}}{\b_\l(y)}}
\left(\frac{-p^2}{\m^2}\right)^{\eta/2}\, .
\ee
Expanding in small $\eta$ this can be also written as
\be
\braket{\Theta \Theta} = c\, p^4 \b_\l^2 e^{-2 \int_1^\l dy \frac{\partial \ln\b_\l(y)}{\partial y}}
\left(1 + \frac{\eta}{2} \ln \left(\frac{-p^2}{\m^2} \right)+\eta \int_1^\l \frac{dy}{\b_\l(y)} + O(\eta^2)\right)\, .
\ee
The Fourier transform of both \eqref{ThetaThetapert1} and of this solution is of the form 
\be
\braket{\Theta(x) \Theta(0)} = \b_\l^2 \left(\frac{-1}{|x|^2\m^2}\right)^{e_\Theta/2} \frac{{\tilde c}(d,\l)}{|x|^{2d}}\, .
\ee
In $\tilde c$ we have collected all $x$-independent quantities  and
$e_\Theta$ has in each case its corresponding value.

While QFT correlation functions are less constrained than those in a CFT, there are several 
techniques that are similar in the two cases. 
The most common technical issue in this type of calculations is the evaluation of higher loop integrals. 
We give all the necessary for completeness loop computations that arise in our analysis in detail 
in an Appendix, despite the fact that some of them
are either trivial, like those associated with the 1 and 2-loop renormalization process and some others, such as the 3 and 4-loop integrals,
are similar to those performed in recent papers. 
Similar to the loop integrals computed here have been presented in \cite{Claudio, Giombi int, Rychkov_Trace}. 
More concretely, recall that as long as the system does not sit on the fixed point, correlation functions such as $\braket{\phi^4 \phi^2}$
do not vanish. This latter correlator has been studied in detail in \cite{Rychkov_Trace} who showed that
at next to leading order it involves certain 4-loop integrals. All of these can be reduced to easily computable lower order integrals
except from one that proves to be irreducible, which then they compute via two different methods. 
Here we are interested in correlators of dimension $d$ derivative operators and eventually in correlators of $\Theta$,
apparently a very different context.
We will see nevertheless that the renormalization process does involve the above irreducible integral (along with the reducible ones).
This is in fact the case despite the fact that
the model under consideration in \cite{Rychkov_Trace} is a non-local theory which generally has a non-standard kinetic term.
It is however easy to see that their model (or at least the loop integrals involved) has a local limit that can be reached by setting $d$ 
in their work equal to $4-\epsilon$, where the theory (the loop integrals) reduces to the standard $\l\phi^4$ theory examined here. 
As a result their irreducible 4-loop integral, for $d=4-\epsilon$, coincides with our irreducible 4-loop integral.
We show this in detail in the Appendix and consequently use their result without further discussion.

 \section{Composite operators}
 
 The insertion of composite operators in a correlator results in divergences due to multiple fields being defined at the same point.
 These divergences are beyond those that arise during the renormalization of couplings (and also wave function renormalization). This is the reason why we must choose from the beginning 
 a strategy for the renormalization of composite operator correlators that takes care of all of the divergences.
 The most straightforward way would be to renormalize all couplings, field $\phi$ and composite operators simultaneously, however this mixes in a rather complicated way
 coupling, $\phi$ and operator counter-terms. For this reason we will follow a different strategy. First we renormalize the bare coupling $\l_0$ in DR by taking the limit $\ve\to 0$
thus obtaining at the end a four-dimensional finite renormalized coupling $\l$. This is the coupling that enters the Feynman rules that follow in the next step,
where we renormalize the correlator, by opening again the dimension $d=4-\ve$ and performing the integrals associated with the new divergences in a second stage of DR. 
There is a price to be paid for this shortcut which is that in an expression that contains, say the coupling beta function $\b_\l=\frac{3\l^2}{16\pi^2}+\cdots$, we will have to adjust to
$\b_\l\to {\hat\b_\l} = -\ve\l + \frac{3\l^2}{16\pi^2}+\cdots$ if we need its $d$-dimensional version.
The first stage of coupling (and wave function) renormalization is by now trivial textbook material, nevertheless we review it (to order $\l^2$) in some detail in an Appendix
since it serves as a good introduction to the methods and some of the Feynman integrals involved. 
 
Since we are especially interested in the vicinity of the WF fixed point, we should also remind of a few subtle points.
The first is that since the coupling at the WF point is large, strictly speaking performing expansions in powers of $\l$ and neglecting higher powers with respect to lower ones
is not really correct. The second is that the $\ve$-expansion requires the DR operation to take $\ve\to 0$ everywhere on the phase diagram (rendering the theory four-dimensional everywhere)
except exactly on the WF fixed point where one is instructed to take $\ve=1$, turning the theory at that point abruptly into a three-dimensional CFT.
Quantities therefore that are computed in a series in powers of $\l$ have in addition, issues of convergence.
These two issues are related of course and the spirit of the $\ve$-expansion instructs to ignore both, as long as it generates numbers that are checked to be correct, mainly on the lattice.
Such are the anomalous dimensions of operators that are reproduced quite well in this context, provided that they are computed at a rather high order.
However for other quantities like the eigenvalue $e_\Theta$ that we are about to compute, we simply do not know. It would be interesting to check its
non-perturbative value on the lattice.
Apart from these disclaimers, the method of computing correlators of operators that we will present is completely straightforward. 

We start from the simplest class of objects, the 'primary' bare operators $\phi_0^n$.
We define a bare composite operator (from now on simply operator) $\mathcal{O}_{n,0}(x)$ as:
\be\label{On0}
\mathcal{O}_{n,0}(x) = \lim_{x_{n-1} \rightarrow x}  \cdots \lim_{x_2\rightarrow x_3}  \lim_{x_1 \rightarrow x_2} 
\phi_0(x_1) \phi_0(x_2) \cdots \phi_0(x_{n-1})\phi_0(x) \equiv \phi_0^n(x)
\ee
We will also encounter derivative or 'descendant' operators. Their construction is similar and we give just a simple example, more complicated ones being easy to construct:
\be
 \Box \phi^2_0(x) \equiv \Box^{(x)}\lim_{y\to x}  \phi_0(x) \phi_0(y)
\ee
and so forth. With descendants there is an additional complication, when they appear inside correlators. We have that
\be
\partial_{\n}^{(x)} \partial_{\r}^{(x_1)} \left< \phi_{0}(x) \phi_0(x_1) \cdots \phi_0(x_n) \right> \neq \left< \partial_{\n}^{(x)} \partial_{\r}^{(x_1)} \phi_{0}(x) \phi_0(x_1) \cdots \phi_0(x_n) \right>
\ee
as the correlation functions are given by time-ordered products and the time derivatives also act on the Heaviside step functions in them, generating extra terms. 
These are the so called 'contact terms' that are multiplied by $\delta$-functions since the derivative of the Heaviside is a delta function.
For every correlator that contains a descendant we have to therefore take care of its contact terms. We will be doing this case by case.

In this work we start from the basic statement that the renormalized $4$-dimensional operator $\Theta$ 
can be defined as a linear combination of operators of the same dimension. 
In DR it will be a linear combination of dimension $d$ operators. 
This is a non-trivial statement as the classical operator $\Theta_{\rm cl}$ in the massless theory is identically zero.
This also suggests that perhaps there is no unique way to define a renormalized $\Theta$.
We will elaborate more on this later. 

There are four bare operators of dimension $d$ in our case. The three of them are 
\be
\bal
K_{1,0}(x) &= \partial_\n \phi_0(x)\partial^{\n} \phi_0(x)\\
K_{2,0}(x) &= \Box \phi_0^2(x) = \Box \mathcal{O}_{2,0}(x)\\
K_{3,0}(x) &= \phi_0(x) \Box \phi_0(x)
\eal
\ee
Clearly they are not independent, by differentiation they are related to each other through the identity :
\be \label{F-identity}
F_0(x)\equiv K_{2,0}(x) - 2 K_{1,0}(x) - 2 K_{3,0}(x) =0\, .
\ee
The fourth bare operator is 
\be\label{K4 def}
K_{4,0}(x) = \l_0 \phi_0^4(x) \equiv \l_0 \mathcal{O}_{4,0}(x)\, .
\ee
This is not an independent object either. Multiplying the equation of motion 
\be
\Box \phi_0 = - \frac{\l_0}{6}\phi_0^3
\ee
by $\phi_0$, we can construct the new vanishing quantity
\be
\bal
E_0(x)\equiv \phi_0(x) \left[\Box \phi_0(x) + \frac{\l_0}{6}\phi_0^3(x) \right]=K_{3,0}(x) + \frac{1}{6} K_{4,0} (x) = 0 \, .
\eal
\ee
The somewhat not immediately obvious fact is that the relations $F=0$ and $E=0$ hold up to the quantum level, as operator identities.
It is important to prove this because if this is the case, the basis of independent operators reduces from four to two.
The way we construct an answer is to insert (or project) the bare quantities into a correlator that contains along with $F_0$ and $E_0$, a number of fundamental fields $\phi_0$, equal
to the number of fields $\phi_0$ contained in the $K_{I,0}, I=1,2,3,4 $ basis. That is, we compute 
$\braket{F_0 \phi_0 \phi_0}$ and $\braket{E_0 \phi_0 \phi_0}$ and we show diagrammatically up to a given order that they vanish. 
The standard renormalization process ensures that to that order, $\braket{F \phi \phi}=0$ and $\braket{E \phi \phi}=0$ and then by removing
the correlator and the auxiliary fields $\phi$ we can promote $F=E=0$ to renormalized operator identities.

\subsection{Renormalization of correlators}
 
The relevant to this work correlators are of the form
\be
\braket{{\mathcal O}_{n,0} \phi \phi}
\ee
for $n=2,4$ (from which the $\braket{K_I \phi\phi}$ will be extracted) and
\be
\braket{K_I K_J}\, .
\ee
Since the latter can be obtained from the former by applying the operator identities, 
inserting derivatives and taking limits, we only need to describe the
construction of the 3-point functions in detail. We will take the following steps:
 \begin{enumerate}
\item We start from the bare $(n+2)$-point function of fundamental fields:
\bea
&&\braket{\phi_{0}(x_1) \cdots \phi_{0}(x_{n-1}) \phi_{0}(x) \phi_{0}(y) \phi_{0}(z)} = \nonumber\\ 
&&\frac{\bra{0} T \left\{\phi_0(x_1) \cdots \phi_{0}(x_{n-1}) \phi_{0}(x) \phi_0 (y) \phi_0 (z) e^{iS_{int}^{(0)}[\phi_0 ; \l_0]}\right\}  \ket{0}}{\bra{0} T  e^{iS_{int}^{(0)}[\phi_0 ; \l_0]} \ket{0}}
\eea
with  $\ket{0}$ the vacuum of the free theory. This yields a number of diagrams at each order in perturbation theory, which can be computed.
The final expression will be a power series in the bare coupling $\lambda_0$. 
\item In order to form the correlator with the operator $\mathcal{O}_{n,0}$ and two fields $\phi_0$, we use \eqref{On0},
to arrive at an expression for
\be
\braket{ \mathcal{O}_{n,0}(x)  \phi_0(y) \phi_0(z) } \, .
\ee
\item We apply a Fourier transformation in order to obtain the expression of the bare 3-point function in momentum space:
\be
\braket{ \mathcal{O}_{n,0}(p_1)  \phi_0(p_2) \phi_0(p_3) } = 
\int \mathrm{d}^d x \mathrm{d}^d y \mathrm{d}^d z \braket{ \mathcal{O}_{n,0}(x)  \phi_0(y) \phi_0(z) } e^{ip_1 x} e^{ip_2 y} e^{ip_3 z}
\ee
The general form of the bare 3-point function will be:
\be\label{Onbaremom}
\braket{ \mathcal{O}_{n,0}(p_1)  \phi_0(p_2) \phi_0(p_3) } = p_1^{m_1} p_2^{m_2} p_3^{m_3} \times \sum_{n=n_{\min}}^{n_{\max}}  \sigma_n \lambda_0^n \left[ \text{Loop integral} \right]_n (2\pi)^d \d(p_1 +p_2 +p_3)
\ee
where $n_{\min}$ is the power of $\lambda_0$ in the leading order contribution to the correlation function, 
and $n_{\max}$ is the order of perturbation theory at which we truncate the process. 
The numerical factor $\sigma_n$ represents the symmetry factor of the loop diagram.
The powers of the external momenta, $p_1^{m_1}, p_2^{m_2}, p_3^{m_3}$ are determined by the first non-vanishing diagram, they must be consistent with the mass dimensions of the correlation function
and the momentum conservation rule must be obeyed. In other words for $\braket{\mathcal{O}_A \mathcal{O}_B \mathcal{O}_C}$:
\be\label{sumruledims}
m_1 + m_2 +m_3 = \left[ \mathcal{O}_A \right]+ \left[ \mathcal{O}_B \right] +\left[ \mathcal{O}_C \right] -2d\, .
\ee
\item We express the bare coupling $\l_0$ in \eqref{Onbaremom} in terms of the renormalized coupling $\l$, employing the results of the renormalization of the Lagrangean,
which has yielded the counterterms $\d_\l$ and $\d_\phi$:
\be\label{lambdarenorm}
\l_0 = Z_\l Z_\phi^{-2} \l\, .
\ee 
All (unless otherwise specified) renormalization factors contain the corresponding counterterm 
according to the usual convention, for example  $Z_\l = 1 + \d_\l$ etc. 
As we already know  $Z_\l$ and $Z_\phi$ are  expressed in power series of the renormalized $\l$. So we should keep only terms up to order $n_{max}$.
The 3-point function takes now the form
\be
\left< \mathcal{O}_{n,0}(p_1)  \phi_0(p_2) \phi_0(p_3) \right> = 
p_1^{m_1} p_2^{m_2} p_3^{m_3} \times \sum_{n=n_{\min}}^{n_{\max}}  \sigma_n \rho_n(\m) \lambda^n \left[ \text{Loop integral} \right]_n\, ,
\ee
where $\rho_n(\m)$  contains the information from the counterterms $\d_\phi$ and $d_\l$ due to \eqref{lambdarenorm}.
\item The renormalization of the operator is implemented by the standard definition (see for example the textbook \cite{Peskin}):
\be
\mathcal{O}_{n,0} = Z_{\mathcal{O}_n} \mathcal{O}_n\, .
\ee
Using the renormalization of the field $\phi_0 = Z_\phi^{1/2} \phi$ on the left hand side, we have
\be \label{renorm eq for O}
Z_{\mathcal{O}_n}  Z_\phi \braket {\mathcal{O}_n(p_1) \phi (p_2) \phi(p_3) } =\braket{ \mathcal{O}_{n,0}(p_1)  \phi_0(p_2) \phi_0(p_3) }\, .
\ee
In the case where the operator $\mathcal{O}_n$ mixes under renormalization with other operators, the $Z_{\mathcal{O}_n}$ is promoted to a  matrix. 
\item We impose a renormalization condition at a certain  energy scale $\m$, defined by the conditions $p^2=-\m^2$ for the 2-point functions and 
$p_1^2 =p_2^2 =p_3^2 = - \m^2 $ and $p_i\cdot p_j = \frac{1}{2} \m^2$ for $i\ne j$ for the 3-point functions. 
This latter choice is called the 'Symmetric Point (S.P.)' and will be used throughout this work for the 3-point functions.
The renormalization condition will therefore have the following general form:
\be
\left< \mathcal{O}_n(p_1) \phi (p_2) \phi(p_3) \right> = p_1^{m_1} p_2^{m_2} p_3^{m_3} \times\s_{n_{min}} \l^{n_{min}} \; \text{at the $S.P.$}
\ee
The specific form depends on the correlator that is computed and will be given when necessary.
Using this renormalization condition we solve \eqref{renorm eq for O} for $Z_{\mathcal{O}_n}$.
\item Having obtained the expression for  $Z_{\mathcal{O}_n}(\m^2)$ we can evaluate the form of the 
renormalized 3-point function, by expanding
\be
\braket{\mathcal{O}_n(p_1) \phi (p_2) \phi(p_3)} = \frac{\braket {\mathcal{O}_{n,0}(p_1)  
\phi_0(p_2) \phi_0(p_3) }}{  Z_{\mathcal{O}_n}  Z_\phi}\, .
\ee
in powers of $\l$. One should arrive at an expression free of UV divergences of the form $1/\epsilon^p$.
\item Finally we apply the Callan-Symanzik equation in order to obtain the anomalous dimension of the operator $\mathcal{O}_n
$\footnote{In the case that operators are mixed under renormalization we have to think of the anomalous dimension as a matrix $\G_{IJ}$ in the Callan-Symanzik equation.}.
\be
\left[ \m \frac{ \partial}{\partial \m} + \b_\l \frac{\partial}{\partial \l} + 2\g_\phi + \G_{\mathcal{O}_n} \right] \braket{\mathcal{O}_n(p_1) \phi (p_2) \phi(p_3) } =0\, .
\ee
We compare our results for the anomalous dimensions obtained from the CS equation with standard expressions obtained 
directly from the counter-terms, for example those summarized in \cite{Henriksson_2023}.
\end{enumerate}

\subsection{Conformal limit of correlators}
\label{conlimsect}

After renormalization, we take the extra step of determining a way to connect the computed from QFT form of the correlator to the corresponding expression in conformal field theory.
We will use a slightly different method for the 2-point functions and the 3-point functions because the simplicity of the momentum space form of the conformal 2-point 
function allows us to take a shortcut. From now on by $*$ we will be denoting the WF point, for example $\l^* = \frac{16\pi^2}{3}\epsilon$ etc.

The general form of a conformal 2-point function  in momentum space is:
\be \label{2pt CFT form}
\braket{\mathcal{O}_A(p) \mathcal{O}_B(-p)}^* = c\, \d_{A B} p^{2\D_{\mathcal{O}_A} -d}\, ,
\ee
where $\D_{\mathcal{O}_A}$ denotes  the total scaling dimension of the operator $\mathcal{O}_A$, defined as
\be
 \D_{\mathcal{O}_A}=\left[ \mathcal{O}_A \right] + \G_{\mathcal{O}_A}^* \, .
\ee
By $\left[ \mathcal{O}_A \right]$ we define the engineering dimension of the operator $\mathcal{O}_A$. 
$\G_{\mathcal{O}_A}^*$ is the anomalous dimension of the operator on the WF  fixed point. Using the perturbative results 
for the anomalous dimension functions $\G_{\mathcal{O}_A}$, we can obtain the value of $\G_{\mathcal{O}_A}^*$ as  power series of $\epsilon$, by setting $\l \to \l^*$. 
\be
\G_{\mathcal{O}_A}^* = \g_1^* \epsilon  + \g_2^* \epsilon^2 + \cdots
\ee
Expanding \eqref{2pt CFT form} in powers of $\epsilon$ we obtain the following 
"QFT-like" form for the 2-point function 
\be\label{2ptcft}
\braket{\mathcal{O}_A (p) \mathcal{O}_A (-p)}^* = c p^{2 \left[\mathcal{O}_ A\right] -d } \left[1 + \g_1^* \epsilon \ln p^2 + \cdots \right]
\ee
This form is valid for generic operators and it is of course not valid for $\Theta$ that has a vanishing
anomalous dimension and is proportional to $\b_\l$. 
For $\Theta$ instead of \eqref{2ptcft} we expect an expression like \eqref{ThetaThetapert1}, as discussed in the Introduction. It is the $\b_\l^2$ sitting in front of the expression that invalidates the (otherwise similar) above form.
Returning to generic operators, by inspecting the above expression we can deduce that this form can be obtained by considering the following substitution in the QFT correlation function
(apart from the obvious $\l \to \l^*$):
\be\label{QFTtoCFT}
\frac{-p^2}{\m^2} \to p^2\, .
\ee
For the conformal 3-point function the momentum space expression is not that simple 
as it involves in general the triple-K integrals \cite{Skenderis 3K, Skenderis CFT}.
What we will do instead is to take the QFT expression for the 3-point function and check whether it satisfies the conformal Ward identities,
after the substitution \eqref{QFTtoCFT}. The two relevant generators are those of the dilatation and special conformal transformation (SCT).
The fast way to determine the dilatation generator is to consider the general form of a renormalized 3-point fuction
\be
\braket{{\braket{\mathcal{O}_A(p_1) \mathcal{O}_B(p_2)\mathcal{O}_C(p_3)}}}= p_1^{m_1}p_2^{m_2} p_3^{m_3} f\left(\frac{p_1^2}{\m^2}, \frac{p_2^2}{\m^2},\frac{p_3^2}{\m^2} \right)\, ,
\ee
which implies, along with \eqref{sumruledims}, that the derivative with respect to the renormalization scale $\m$ is equivalent to
\be\label{derivativerule}
\m \frac{\partial }{\partial \m} = - \sum_{i=1}^{3} p_i \frac{\partial}{\partial p_i} +\left[\mathcal{O}_A \right] 
+\left[\mathcal{O}_b \right] +\left[\mathcal{O}_C \right] -2d\, .
\ee
Then, the Callan-Symanzik equation  takes the following form :
\be
 \left[- \sum_{i=1}^{3} p_i \frac{\partial}{\partial p_i}  + \b_\l \frac{\partial}{\partial \l} + \D_{\mathcal{O}_A} + \D_{\mathcal{O}_B} + \D_{\mathcal{O}_C}  -2d \right] \braket{{\braket{\mathcal{O}_A(p_1) \mathcal{O}_B(p_2)\mathcal{O}_C(p_3)}}}=0\, .
  \ee
Now, since we are interested in the conformal limit, we have to adjust to the $d$-dimensional version of the $\b$-function according to
\be
\b_\l \to \hat{\b}_\l
\ee
and set $\hat{\b}_\l^*=0$. What remains is the action of the dilatation generator on the conformal correlator
\be\label{3ptcftD}
 \left[- \sum_{i=1}^{3} p_i \frac{\partial}{\partial p_i}  + \D_{\mathcal{O}_A} + \D_{\mathcal{O}_B} + \D_{\mathcal{O}_C}  -2d \right] 
 \braket{{\braket{\mathcal{O}_A(p_1) \mathcal{O}_B(p_2)\mathcal{O}_C(p_3)}}}^*= 0\, .
  \ee
The momentum space SCT Ward identity reads (see \cite{Skenderis 3K, Claudio CFT}):
\be\label{3ptcftSCT}
p_2^\m (K_{p_2}-K_{p_1})  \braket{{\braket{\mathcal{O}_A(p_1) \mathcal{O}_B(p_2)\mathcal{O}_C(p_3)}}}^*
 + p_3^\m (K_{p_3}-K_{p_1}) \braket{{\braket{\mathcal{O}_A(p_1) \mathcal{O}_B(p_2)\mathcal{O}_C(p_3)}}}^*= 0
\ee
where
\be 
\bal
K_ {(p_2,p_3)}&= \frac{\partial^2}{\partial (p_2,p_3) \partial (p_2,p_3)} + \frac{d+1-2\D_\phi}{(p_2,p_3)} \frac{\partial}{\partial(p_2,p_3)} \\
K_p &= \frac{\partial^2}{\partial p_1 \partial p_1} + \frac{d+1-2\D_{\mathcal{O}_2}}{p_1} \frac{\partial}{\partial p_1}
\eal
\ee
and it is known that it is satisfied if each coefficient of the independent four-momenta $k^\m$ and $ q^\m$ is equal to zero:
\be 
\bal
    \left(\frac{\partial^2}{\partial p_2 \partial p_2} + \frac{d+1-2\D_\phi}{p_2} \frac{\partial}{\partial p_2}    
    -\frac{\partial^2}{\partial p_1 \partial p_1} - \frac{d+1-2\D_{\mathcal{O}_2}}{p_1} \frac{\partial}{\partial p_1} \right)\ \braket{{\braket{\mathcal{O}_A(p_1) \mathcal{O}_B(p_2)\mathcal{O}_C(p_3)}}}^*= 0\\
    \left(\frac{\partial^2}{\partial p_3\partial p_3} + \frac{d+1-2\D_\phi}{p_3} \frac{\partial}{\partial p_3}    
    -\frac{\partial^2}{\partial p_1 \partial p_1} - \frac{d+1-2\D_{\mathcal{O}_2}}{p_1} \frac{\partial}{\partial p_1} \right)  \braket{{\braket{\mathcal{O}_A(p_1) \mathcal{O}_B(p_2)\mathcal{O}_C(p_3)}}}^*= 0
\eal
\ee
To summarize, after computing the renormalized 2-point and 3-point functions, we will apply the rules $p^2/\m^2\to p^2$,
$\l\to\l^*$, $\b_\l\to\b_\l^*$ and check whether the resulting expression is of the form \eqref{2ptcft} for the 2-point functions and if it is 
invariant under \eqref{3ptcftD} and \eqref{3ptcftSCT} for the 3-point functions.

\section{The 3-point function $\braket{{\mathcal O}_2 \phi\phi}$}

The renormalization of QFT correlators needs a renormalization condition, which we must get out our way.
We would like it to be as general as possible.
The condition we would like to impose is that the 3-point function has a preferred form at the Symmetric Point. 
We choose this form to be the expression for the correlator near the free UV fixed point. 
The general form of the Poincaré invariant 3-point function is
\be \label{Poincare 3pt}
\left<\mathcal{O}_a(x_1)\mathcal{O}_b(x_2) \mathcal{O}_c(x_3)\right> = 
\frac{c_{abc}}{\left|x_{1,2}\right|^{\alpha} \left|x_{1,3}\right|^{\beta} \left|x_{2,3}\right|^{\gamma}} \;, \;  |x_{i,j}| =|x_i - x_j|\, ,
\ee
with the only restriction on  the coefficients $\a$,$\b$,$\g$ stemming from dimensional analysis:
\be \label{dim analysis 3pt}
\a +\b +\g =\sum_{i=a,b,c}\left[\mathcal{O}_i \right]\, .
\ee
The above constraint is similar to the one imposed by the Poincaré + Scale invariance, which is 
\be
\left[ \a +\b +\g \right]_{\text{scale invariance}} =\sum_{i=a,b,c} \Delta_{\mathcal{O}_i }\, ,
\ee
with $\Delta_{\mathcal{O}_i}$ now the scaling dimension $\Delta_{\mathcal{O}_i}=\left[ \mathcal{O}_i \right] + \Gamma_{\mathcal{O}_i}$ of the operator. 
Inspired by the conformal structure of the 3pt function we will make the following choice for the coefficients $\a$,$\b$,$\g$ :
  \be\label{coeffscft}
 \bal
 \a = \left[ \mathcal{O}_a\right]+ \left[ \mathcal{O}_b\right] - \left[ \mathcal{O}_c\right]\\
 \b= \left[ \mathcal{O}_a\right]+ \left[ \mathcal{O}_c\right] - \left[ \mathcal{O}_b\right]\\
 \g= \left[ \mathcal{O}_c\right]+ \left[ \mathcal{O}_b\right] - \left[ \mathcal{O}_a\right]
 \eal
 \ee 
For the $\braket{\mathcal{O}_2 \phi\phi}$ correlator we have $\mathcal{O}_a =\mathcal{O}_2$ and $\mathcal{O}_b=\mathcal{O}_c=\phi$
 and since
 $ \left[ \mathcal{O}_2\right] = 2\left[\phi \right] = 2 \frac{d-2}{2}=d-2$, the coefficients $\a$,$\b$,$\g$ get the values
$\a = d-2 $,  $\b =d-2 $ and  $\g =0$.
Then, 
\be
\braket{ \mathcal{O}_2 (x_1)\phi(x_2) \phi(x_3) } = \frac{c_{\mathcal{O}_2 \phi \phi}}{\left|x_{1,2}\right|^{d-2} \left|x_{1,3}\right|^{d-2} \underbrace{\left|x_{2,3}\right|^{0}}_{1}}\, .
\ee
With a Fourier transformation we move to momentum space
\be\bal
\left< \mathcal{O}_2 (p_1)\phi(p_2) \phi(p_3) \right> =(2\pi)^d {\d}^{(d)}(p_1 +p_2 +p_3)\left[\frac{4\pi^{d/2}}{\Gamma(d-2)}\right]^2 \frac{1}{p_2^2}\frac{1}{p_3^2}c_{\mathcal{O}_2 \phi \phi}\, .
\eal\ee
Since $\left[\frac{4\pi^{d/2}}{\Gamma(d-2)}\right]^{-2}$ is finite for every $d>2$ we define the constant as 
$c_{\mathcal{O}_2 \phi \phi}=2 \left[\frac{4\pi^{d/2}}{\Gamma(d-2)}\right]^{-2} i^2$ to arrive at
 \be 
 \braket{ \mathcal{O}_2 (p_1)\phi(p_2) \phi(p_3) } =2\frac{i}{p_2^2}\frac{i}{p_3^2}  \left(2\pi \right)^d {\d}^{(d)}(p_1 +p_2 +p_3)\, .
 \ee
 Employing the double bracket notation of \cite{Skenderis CFT} we can write
 \be
\braket{\cdots} = \braket{\braket{\cdots}} \left(2\pi \right)^d {\d}^{(d)}(p_1 +p_2 +p_3)
 \ee
and see that the expression can be seen to corresponds to the diagram 
 \be \label{renorm cond o2}
\braket{ \braket{\mathcal{O}_2 (p_1)\phi(p_2) \phi(p_3)}}  = \begin{gathered}
 \begin{tikzpicture}
  \begin{feynman}
  \vertex(x) {$p_2$};
  \vertex[ right = of x](c);
  \vertex[above = 1cm of c ](y);
  \vertex[right = of c](z){$p_3$};
  \vertex[above= 1cm of y](r){$p_1$};
  \diagram{(x)--(y)--(z);
  };
 \diagram{(r)--[dashed](y)};
\end{feynman}
\filldraw[fill =black](y) circle(5pt );
\end{tikzpicture}
\end{gathered}
\ee
This coincides of course with the renormalization condition implied in \cite{Peskin}.  
The black circle in the above diagram indicates the position of the ${\mathcal O}_2$ operator, usually called 'the insertion.

\subsection{$O(\l)$ renormalization of $\braket{\mathcal{O}_2 \phi \phi }$}

The calculation at leading order is just a review of the one in \cite{Peskin}, but it will help us to illustrate our method in a simple context.
Besides that, there are some novel steps since we compute the correlator itself, not just the counter-term and we also connect 
our expressions to their conformal limit.

The relation between the bare and the renormalized correlation function is:
\be 
\braket{ \phi_0 (x_1) \phi_0 (x_2) \mathcal{O}_{2(0)}(y)} = Z_\phi Z_{\mathcal{O}_2}\braket{ \phi (x_1) \phi (x_2) \mathcal{O}_{2}(y)} \, .
\ee
From the 1-loop renormalization of the fundamental field we know that $Z_\phi$=1, so the bare and renormalized fields $\phi$ 
are equivalent to $O(\l)$. Also the coupling constant $\l_0$ is finite to $O(\l)$ so the counterterm $\d_\l$ can be set to zero. 
This means that he have to renormalize only the operator $\mathcal{O}_2(y)$:
\be \label{ren ff o}
\braket{ \phi(x_1) \phi(x_2) \mathcal{O}_{2 (0)} (y)} - \d_{\mathcal{O}_2} \braket{ \phi(x_1) \phi(x_2) \mathcal{O}_{2 } (y)}  = \braket{ \phi(x_1) \phi(x_2) \mathcal{O}_{2} (y) }\, .
\ee
The right hand side of the previous equation is finite, since it is the renormalized 3-point function. 
By applying the Wick contractions up to $O(\l)$, the bare correlator gives the following diagrams: 
 \be
  \bal
  \left< \phi (x_1) \phi(x_2) \mathcal{O}_{2(0)} (y)\right> =  \lim_{y_1 \rightarrow y} \left< \phi (x_1) \phi(x_2)\phi (y_1) \phi(y)\right> =
  \begin{gathered}
  \begin{tikzpicture}
  \begin{feynman}
  \vertex(x) {$x_1$};
  \vertex[ right = 1cm of x](c);
  \vertex[above = 1cm of c ](y);
  \vertex[right = 1cm of c](z){$x_2$};
  \diagram{(x)--(y)--(z);
  };
    \end{feynman}
    \filldraw[fill =black](y) circle(5pt );
  \end{tikzpicture}
  \end{gathered}+
  \begin{gathered}
  \begin{tikzpicture}
  \begin{feynman}
  \vertex(x) {$x_1$};
  \vertex[ right =1cm  of x](c);
  \vertex[above =0.5cm of c ](u);
  \vertex[above =1cm  of u](y);
  \vertex[right = 1cm of c](z){$x_2$};
  \diagram{(x)--(u)--[half left](y)--[half left](u)--(z);
  };
    \end{feynman}
    \filldraw[fill =black](y) circle(5pt );
  \end{tikzpicture}
  \end{gathered}
  \eal
 \ee
 The second diagram is the divergent one and it is given by the following form (in momentum space): 
 \be
 \bal 
  \begin{gathered}
  \begin{tikzpicture}
  \begin{feynman}
  \vertex(x) ;
  \vertex[ right = 1cm  of x](c);
  \vertex[above =0.5 cm of c ](u);
  \vertex[above =1cm  of u](y);
  \vertex[right = 1cm  of c](z);
  \diagram{(x)--(u)--[half left](y)--[half left](u)--(z);
  };
    \end{feynman}
    \filldraw[fill =black](y) circle(5pt);
  \end{tikzpicture}
  \end{gathered} &= i \l \frac{i}{p_2^2} \frac{i}{p_3^2} \int \frac{\mathrm{d}^d k}{(2\pi)^d} \frac{1}{k^2 \left( k -p_1^2 \right)^2} \d(p_1 +p_2 +p_3)\\
  &= i \l L_1(p_1) \frac{i}{p_2^2} \frac{i}{p_3^2} (2\pi)^d \d(p_1 +p_2 +p_3)\, .
 \eal
 \ee
 The $L_1(p)$ integral is nothing but the $B_0(p)$ integral of the Passarino-Veltman language. Its definition as well as the definition and computation
 of all other diagrams to be encountered in this paper, can be found in the Appendix.
 So the bare  3 point function is given by:
 \be \label{bare O2phiphi oreder 1}
 \left< \mathcal{O}_{2,0}(p_1)\phi_0(p_2) \phi_0(p_3) \right> = \frac{i}{p_2^2} 
 \frac{i}{p_3^2} \left[2 + i \l L_1 (p_1^2) \right] (2\pi)^d\d(p_1 +p_2 +p_3)\, .
 \ee
For $d=4-\ve$ and in the context of $\ve$-expansion the loop integral takes the form
\be 
L_1(p^2)= \frac{i}{16\pi^2} \left[\frac{2}{\ve} - \ln \left(\frac{-p^2 e^\g}{4\pi}  \right) +2 \right]\, .
\ee
We now select the renormalization condition \eqref{renorm cond o2}:
\be 
\braket{ \braket{ \phi \phi \mathcal{O}_2 }} =  \begin{gathered}
  \begin{tikzpicture}
  \begin{feynman}
  \vertex(x) ;
  \vertex[ right =1cm of x](c);
  \vertex[above = 0.5cm of c ](y);
  \vertex[right = 1cm  of c](z);
  \diagram{(x)--(y)--(z);
  };
    \end{feynman}
    \filldraw[fill =black](y) circle(5pt );
  \end{tikzpicture}
  \end{gathered}= 2\frac{i}{p_2^2}\frac{i}{p_3^2} \,\,\, \text{at the $S.P.$}
\ee
Solving  \eqref{ren ff o} for the counterterm $\d_{\mathcal{O}_2}$ we get
 \be
\bal
\d_{\mathcal{O}_2}=i  \frac{\l}{2}L_1(-\m^2)\, .
\eal
\ee
Thus, the renormalized 3-point function is
\be\label{O2phiphi}
\braket{\braket{ \mathcal{O}_2(p_1) \phi(p_2) \phi(p_3) } }=2\frac{i}{p_2^2}\frac{i}{p_3^2}\left[1+ 
\frac{\l}{2 (4\pi)^2} \ln \left(\frac{-p_1^2}{\m^2} \right) + O(\l^2)\right] \, .
\ee
The Callan-Symanzik equation of the 3-point function is:
\be
\left[\frac{\partial}{\ln \m} + \b_\l \frac{\partial}{\partial \l} + \G_{\mathcal{O}_2} + 2\g_\phi \right]\braket{ \mathcal{O}_2(p_1) \phi(p_2) \phi(p_3) } =0
\ee
where $\b_\l = \frac{3\l^2}{16\pi^2}  + O(\l^3)$ and $\g_\phi = \frac{\l^2}{12 (4\pi)^4} + O(\l^3) $.
Since
\bea
\m \frac{\partial}{\partial \m} \left< \mathcal{O}_2(p_1) \phi(p_2) \phi(p_3) \right> &=&- \frac{i}{p_2^2} \frac{i}{p_3^2}\frac{\l^2}{16\pi^2} \nonumber \\
\b_\l \frac{\partial}{\partial \l} \left< \mathcal{O}_2(p_1) \phi(p_2) \phi(p_3) \right> &=& O(\l^3)
\eea
and solving the Callan-Symnazink equation to $O(\l)$ for $\G_{\mathcal{O}_2}$, we get
\be
\G_{\mathcal{O}_2} = \frac{\l}{16\pi^2} \, ,
\ee
in agreement with the known result.

Now we will study the conformal limit of the 3-point function, under the assumption that the system 
approaches the Wilson-Fisher fixed point in the IR.
Naively we should take the limit
\be
\braket{\braket{\mathcal{O}_2(p_1) \phi(p_2) \phi(p_3) }}^* = \lim_{\m \rightarrow 0 }2\frac{i}{p_2^2}\frac{i}{p_3^2}
\left[1+ \frac{\ve}{6} \ln \left(\frac{-p_1^2}{\m^2} \right) + O(\ve^2)\right] \, ,
\ee
which is of course singular. We need to use some sort of regularization.
Inspired by lattice QFT we use the regularisation scheme in order to absorb the infinity $\frac{-p_1^2}{\m^2} \rightarrow p_1^2$.
This is essentially a 'QFT to CFT' version of the argument \eqref{QFTtoCFT} presented previously.

With this rule, the presumed conformal 3-point function up to $O(\ve)$ is
\be\label{phi phi phi2 conformal corr}
\left<\left< \mathcal{O}_2(p_1) \phi(p_2) \phi(p_3) \right>\right>^* =
2\frac{i}{p_2^2}\frac{i}{p_3^2}\left[1+ \frac{\ve}{6} \ln \left(p_1^2 \right) + O(\ve^2)\right]\, .
\ee
We have to check this. The fast way is to recognize that the coefficient $\g$ in \eqref{coeffscft} is equal to zero and
the 3-point function in this case reduces effectively to a 2-point function. Then indeed the above is of the general form
\eqref{2ptcft}. But we also have to treat it in a way that can be generalized to less special 3-point functions.
As mentioned in the general discussion, if the above is indeed a conformal correlator, 
it must obey the Ward identities associated with the generators of the conformal transformations,
in particular of the dilatations and the special conformal transformations, given in sect. \ref{conlimsect}.
It is straightforward to check that the form \eqref{phi phi phi2 conformal corr} indeed satisfies both identities to order $O(\ve)$.
The reason we can be sure this is correct to $O(\l)$ is that $\beta_\l$, which carries the information for the breaking of scale invariance, is of order $O(\lambda^2)$.

\subsection{$O(\l^2)$ renormalization of $\left<\mathcal{O}_2 \phi \phi \right>$}

As a first step we will find the $O(\l^2)$ diagrams of the bare 3-point function. The Feynman diagrams 
that come up after the limiting procedure described above involve now also the coupling $\l_0$. There are two types of loop diagrams at this order 
that appear before the limits.
One is the Candy, the classic 1-loop diagram that renormalizes the coupling and the other is the sunset, the 2-loop diagram that yields the leading order
contribution to wave function renormalization. Of course, after the limit different diagrams are formed, often each limit adding one more loop to the original diagram,
as external legs are sewed together. In this case we have:
\be
\bal
\left<\mathcal{O}_{2,0} (x) \phi_0(y) \phi_0(z) \right>_{\text{(sunset)}}& = \lim_{x_1 \rightarrow x} \left[\begin{gathered} \begin{tikzpicture} 
     \begin{feynman}
     \vertex (x){$x$};
     \vertex[below =0.7cm of x](r1);
     \vertex[below =0.6cm of r1](r2);
     \vertex[below =2cm of x](u){$z$};
     \vertex[right=   0.5cm of x](y){$x_1$};
     \vertex[right = 0.5cm of  u](l){$y$};
     \diagram{
     (x)--(u)
     };
     \diagram{(r1)--[half left](r2)-- [half left](r1)};
     \diagram{
     (y)--(l)
     };
 \end{feynman} 
 \end{tikzpicture}
 \end{gathered} + \begin{gathered} \begin{tikzpicture} 
     \begin{feynman}
     \vertex (x){$x$};
     \vertex[below =0.7cm of x](r1);
     \vertex[below =0.6cm of r1](r2);
     \vertex[below =2cm of x](u){$y$};
     \vertex[right=   0.5cm of x](y){$x_1$};
     \vertex[right = 0.5cm of  u](l){$z$};
     \diagram{
     (x)--(u)
     };
     \diagram{(r1)--[half left](r2)-- [half left](r1)};
     \diagram{
     (y)--(l)
     };
 \end{feynman} 
 \end{tikzpicture}
 \end{gathered}  + \begin{gathered} \begin{tikzpicture} 
     \begin{feynman}
     \vertex (x){$x_1$};
     \vertex[below =0.7cm of x](r1);
     \vertex[below =0.6cm of r1](r2);
     \vertex[below =2cm of x](u){$y$};
     \vertex[right=   0.5cm of x](y){$x$};
     \vertex[right = 0.5cm of  u](l){$z$};
     \diagram{
     (x)--(u)
     };
     \diagram{(r1)--[half left](r2)-- [half left](r1)};
     \diagram{
     (y)--(l)
     };
 \end{feynman} 
 \end{tikzpicture}
 \end{gathered} + \begin{gathered} \begin{tikzpicture} 
     \begin{feynman}
     \vertex (x){$x_1$};
     \vertex[below =0.7cm of x](r1);
     \vertex[below =0.6cm of r1](r2);
     \vertex[below =2cm of x](u){$z$};
     \vertex[right=   0.5cm of x](y){$x$};
     \vertex[right = 0.5cm of  u](l){$y$};
     \diagram{
     (x)--(u)
     };
     \diagram{(r1)--[half left](r2)-- [half left](r1)};
     \diagram{
     (y)--(l)
     };
 \end{feynman} 
 \end{tikzpicture}
 \end{gathered}  \right] \\
 \left<\mathcal{O}_{2,0} (x) \phi_0(y) \phi_0(z) \right>_{\text{(candy)}}& = \lim_{x_1\rightarrow x} \left[ \begin{gathered} \begin{tikzpicture} 
     \begin{feynman}
     \vertex (x){$x$};
     \vertex[below = 1cm of x](z1);
     \vertex[right = 0.5cm of z1](z11);
     \vertex[below = 2cm of x](u){$x_1$};
     \vertex[right=  2 cm of x](y){$y$};
     \vertex[below=1cm of y](z2);
     \vertex[left= 0.5cm of z2](z22);
     \vertex[right =  2cm of  u](l){$z$};
     \diagram{
     (x)--(z11)--[half left](z22)--[half left](z11)--(u)
     };
     \diagram{
     (y)--(z22)--(l)
     };
 \end{feynman} 
 \end{tikzpicture}
 \end{gathered} + \begin{gathered} \begin{tikzpicture} 
     \begin{feynman}
     \vertex (x){$x$};
     \vertex[below = 1cm of x](z1);
     \vertex[right = 0.5cm of z1](z11);
     \vertex[below = 2cm of x](u){$y$};
     \vertex[right=  2 cm of x](y){$x_1$};
     \vertex[below=1cm of y](z2);
     \vertex[left= 0.5cm of z2](z22);
     \vertex[right =  2cm of  u](l){$z$};
     \diagram{
     (x)--(z11)--[half left](z22)--[half left](z11)--(u)
     };
     \diagram{
     (y)--(z22)--(l)
     };
 \end{feynman} 
 \end{tikzpicture}
 \end{gathered}+
  \begin{gathered} \begin{tikzpicture} 
     \begin{feynman}
     \vertex (x){$x$};
     \vertex[below = 1cm of x](z1);
     \vertex[right = 0.5cm of z1](z11);
     \vertex[below = 2cm of x](u){$z$};
     \vertex[right=  2 cm of x](y){$x_1$};
     \vertex[below=1cm of y](z2);
     \vertex[left= 0.5cm of z2](z22);
     \vertex[right =  2cm of  u](l){$y$};
     \diagram{
     (x)--(z11)--[half left](z22)--[half left](z11)--(u)
     };
     \diagram{
     (y)--(z22)--(l)
     };
 \end{feynman} 
 \end{tikzpicture}
 \end{gathered}
 \right]
 \eal 
\ee
The Sunset contribution does not add a loop in the limit but the Candy does. We now compute both.
All DR integrals that appear during the computation are defined in Appendix.
 
\textbf{Sunset contribution:}
We begin with the limits of the first class of diagrams above, in order to construct the bare 3-point function in position space.
\bea
&& \braket{\mathcal{O}_{2,0} (x) \phi_0(y) \phi_0(z) }_{\text{(sunset)}} = \nonumber\\
&& 2 \frac{(-i\l_0)^2}{6} i^6 \left[\int \frac{e^{i(k+q)x}e^{-iky}e^{-iqz}}{k^4q^2}S_1(k^2)  
+\int \frac{e^{i(k+q)x}e^{-iqy}e^{-ikz}}{k^4q^2}S_1(k^2)  \right] \, , 
\eea
where $S_1(k^2)$ denotes the sunset loop integral 
\be
S_1(k^2) = \int \frac{\mathrm{d}^d q \mathrm{d}^d l}{(2\pi)^{2d}} \frac{1}{l^2 q^2 (l+q-k)^2}\, .
\ee
We move to momentum space via a Fourier transformation. This contribution to the bare 3-point function in momentum space is then given by
\be
\bal
\left<\mathcal{O}_{2,0} (p_1) \phi_0(p_2) \phi_0(p_3) \right>_{\text{(sunset)}}=
-2\frac{\l_0^2}{6} \frac{i}{p_2^2} \frac{i}{p_3^2} \left[\frac{S_1(p_2^2)}{p_2^2} +  \frac{S_1(p_3^2)}{p_3^2} \right] (2\pi)^d\d(p_1+p_2+p_3)\, .
\eal
\ee

\textbf{Candy Contribution:} We follow the same steps for the limits of the second class of diagrams. 
In momentum space we obtain the following expression:
\bea
&& \braket{\mathcal{O}_{2,0} (p_1) \phi_0(p_2) \phi_0(p_3) }_{\text{(candy)}}= \nonumber\\
&& -\frac{\l_0^2}{2}\frac{i}{p_2^2}\frac{i}{p_3^2} \left[\left[ L_1(p_1^2)\right]^2 + \int \frac{L_1\left((k-p_2)^2\right)+L_1\left((k-p_3)^2 \right)}{k^2 \left(k+p_1\right)^2 } \right]\tilde{\d}(p_1+p_2+p_3)\, ,
\eea
where $\tilde{\d}(\cdots) = (2\pi)^d \d(\cdots)$.
Adding the two contributions, the total $O(\l^2)$ bare 3-point function therefore is:
 \bea\label{bare O2 phi phi order 2}
&& \braket{ \mathcal{O}_{2,0}(p_1) \phi_0(p_2)\phi_0(p_3)}_{O(\l^2)} = \nonumber\\
&& \frac{i}{p_2^2} \frac{i}{p_3^2}\left\{ -\frac{\l_0^2}{2} \left[\left[ L_1(p_1^2)\right]^2 + 
\int \frac{\mathrm{d}^dk}{(2\pi)^d} \frac{L_1\left((k-p_2)^2\right)+L_1\left((k-p_3)^2 \right)}{k^2 \left(k+p_1\right)^2 } \right] \right.\nonumber\\
&&\left.-2\frac{\l_0^2}{6}\left[\frac{S_1(p_2^2)}{p_2^2} +  \frac{S_1(p_3^2)}{p_3^2} \right] \right\}\tilde{\d}(p_1+p_2+p_3)\, .
\eea
We have to also take into account the (leading order) corrections to the coupling constant  
and to the field $\phi$, as they appear in $\l_0 = Z_\l Z_\phi^{-2}\l$ and $\phi_0 = Z_\phi^{1/2}\phi$,
with $Z_\phi = 1 +\d_\phi$ and $Z_\l = 1 +\d_\l$ which are known from the standard renormalization of the Lagrangean (see Appendix ):
\be
\d_\l^{(1)} = -i\frac{3 \l}{2} L_1(-\m^2) =\frac{3\l}{2} \frac{1}{16\pi^2} \left[\frac{2}{\ve} - \ln \left(\frac{\m^2 e^\g}{4\pi}  \right) +2 \right]
\ee
\be
\d_\phi = - \frac{1}{-\m^2} \frac{\l^2}{6}S_1(-\m^2) = - \frac{\l^2}{12 (4\pi)^4}\left[ \frac{2}{\ve} - \ln \left( \frac{\m^2e^\g}{4\pi} \right) + \frac{13}{4} \right]
\ee
We define the renormalized operator $\mathcal{O}_2$ as $\mathcal{O}_{2,0} = Z_{\mathcal{O}_2} \mathcal{O}_2$
and we expand $Z_{\mathcal{O}_2}$ as
\be
Z_{\mathcal{O}_2}= 1 + \d_{\mathcal{O}_2}^{(1)} + \d_{\mathcal{O}_2}^{(2)} + \cdots  
\ee
where $\d_{\mathcal{O}_2}^{(n)}$ is the counterterm multiplied by $\l^n$. The $\d_{\mathcal{O}_2}^{(1)}$ counterterm is already known 
from the $O(\l)$  renormalization procedure for the 3-point function:
\be \label{d1o2}
 \bal
\d_{\mathcal{O}_2}^{(1)}=i  \frac{\l}{2}L_1(-\m^2)\, .
 \eal
 \ee
The renormalized 3-point function is determined by the relation
\be\label{renormOphiphi}
\braket{\mathcal{O}_{2,0}(p_1) \phi_0(p_2)\phi_0(p_3)}  =  Z_{\mathcal{O}_2}Z_\phi \braket{\mathcal{O}_2(p_1) \phi(p_2)\phi(p_3)}
\ee
with the bare 3-point function, taking all contributions into account, being equal to
\be\label{zo2zphi renorm 2}
\braket{\mathcal{O}_{2,0}(p_1) \phi_0(p_2)\phi_0(p_3)}  =\frac{i}{p_2^2}\frac{i}{p_3^2}C_{\mathcal{O}_2 \phi \phi}^{\text{\textbf{bare}}}(p_1^2) \tilde{\d}(p_1+p_2+p_3)\, ,
\ee
with
\bea
C_{\mathcal{O}_2 \phi \phi}^{\text{\textbf{bare}}}(p_1^2) &=& 2 + i \l_0 L_1(p_1) -\frac{\l_0^2}{2} \left[\left( L_1(p_1)\right)^2 + \int \frac{L_1(k-p_2)+L_1(k-p_3)}{k^2 \left(k+p_1\right)^2 } \right] \nonumber\\
&-& 2\frac{\l_0^2}{6}\Bigl[\frac{S_1(p_2)}{p_2^2} + \frac{S_1(p_3)}{p_3^2} \Bigr]\, .
\eea
In terms of the renormalized coupling constant this becomes
\bea\label{co2phiphi bare}
C_{\mathcal{O}_2 \phi \phi}^{\text{\textbf{bare}}}(p_{1,2,3}^2) &=& 2 + i \l L_1(p_1)+i \d_\l \l L_1(p_1) -\frac{\l^2}{2} \left[\left( L_1(p_1)\right)^2 + I_4(p_1,p_2) + I_4(p_1,p_3) \right] \nonumber\\
&-&2\frac{\l^2}{6}\left[\frac{S_1(p_2)}{p_2^2} +  \frac{S_1(p_3)}{p_3^2} \right]\, ,
\eea
where we have used that $Z_\l Z_\phi^{-2}\l = \l + \d_\l^{(1)} \l $ and  $\left(Z_\l Z_\phi^{-2}\l \right)^2 = \l^2 + O(\l^3)$.
We can write the renormalized 3-point function in the form
\be
\left<\mathcal{O}_2(p_1) \phi(p_2) \phi(p_3) \right> = \frac{i}{p_2^2}\frac{i}{p_3^2} C_{\mathcal{O}_2 \phi \phi}^{\text{\textbf{R}}}(p_{1,2,3}) \tilde{\d}(p_1+p_2+p_3)\, .
\ee
Of course, the $O(\l)$ result of the renormalized expression will not be affected by the 
$O(\l^2)$ renormalization procedure. This allows us to write
\be
C_{\mathcal{O}_2 \phi \phi}^{\text{\textbf{R}}}(p_{1,2,3}) =2 + C_{\mathcal{O}_2 \phi \phi}^{\textbf{R} (1)}(p_{1,2,3}) +
C_{\mathcal{O}_2 \phi \phi}^{\textbf{R} (2)}(p_{1,2,3})
\ee
with $C_{\mathcal{O}_2 \phi \phi}^{\textbf{R} (1)}(p_{1,2,3}) = i\l \left[ L_1(p_1^2)- L_1(-\m^2)  \right]$, or
\be\label{zz02}
Z_{\mathcal{O}_2}Z_{\phi}C_{\mathcal{O}_2 \phi \phi}^{\text{\textbf{R}}}(p_{1,2,3}) = 2 + i\l L_1(p_1^2) +2\d_\phi + 2\d_{\mathcal{O}_2}^{(2)}  - \frac{\l^2}{2}L_1(p_1^2)L_1(-\m^2) + C_{\mathcal{O}_2 \phi \phi}^{\textbf{R} (2)}(p_{1,2,3})\, .
\ee
Next, we substitute \eqref{co2phiphi bare} and \eqref{zz02} in \eqref{renormOphiphi} to obtain
\be \bal
C_{\mathcal{O}_2 \phi \phi}^{\textbf{R} (2)}(p_{1,2,3}) =&- \frac{\l^2}{2} \left[ \left(L_1(p_1^2) - L_1(-\m^2)  \right)^2  - \left[L_1(-\m^2)\right]^2 \right] + \l^2 \left[ L_1(p_1^2)L_1(-\m^2) + D(p_1) \right]\\
& \; - 2 \left[ \frac{\l^2}{6}\frac{S_1(p_2^2)}{p_2^2} +\d_\phi   \right]    - 2 \left[ \frac{\l^2}{6}\frac{S_1(p_2^2)}{p_2^2} 
+ \d_{\mathcal{O}_2}^{(2)}  \right] \, .
\eal\ee
Recalling that:
\be
 \left[L_1(p_1^2) - L_1(-\m^2)  \right]^2  = - \ln^2 \left(\frac{-p_1^2}{\m^2} \right)
\ee
and
\be
\bal
\left[ L_1(p_1^2)L_1(-\m^2) + D(p_1) \right](4\pi)^4 &= -\frac{2}{\ve^2} + \frac{2\ln \left(\frac{\m^2 e^\g}{4\pi} \right)  -3}{\ve} \\
 & \; \; \; + \frac{1}{2} \ln^2 \left( \frac{-p_1^2}{\m^2} \right) - \ln \left( \frac{-p_1^2}{\m^2} \right)\\
 &\; \; \;- \frac{1}{2}\left[ G(p_1,p_2) + G(p_1,p_3) \right] \\
 &\; \; \; +\text{(momentum independent terms)} 
\eal
\ee
we arrive at
\be
\bal
C_{\mathcal{O}_2 \phi \phi}^{\textbf{R} (2)}(p_{1,2,3}^2) &= \frac{\l^2}{2(4\pi)^4} \ln^2 \left( \frac{-p_1^2}{\m^2} \right) + \frac{\l^2}{2(4\pi)^4} \ln^2 \left( \frac{-p_1^2}{\m^2} \right)  -\frac{\l^2}{(4\pi)^4} \ln \left( \frac{-p_1^2}{\m^2} \right)  + \frac{\l^2}{6(4\pi)^4}\ln\left( \frac{-p_2^2}{\m^2} \right) \\
& \;+ \frac{\l^2}{2}\left[L_1(-\m^2)\right]^2 - 2 \left[ \frac{\l^2}{6}\frac{S_1(p_2^2)}{p_2^2} + \d_{\mathcal{O}_2}^{(2)}  \right] - \frac{\l^2}{2(4\pi)^4} \left[G(p_1,p_2) + G(p_1,p_3) \right]
\eal
\ee
Using the renormalization condition, which is equivalent to the vanishing of $C_{\mathcal{O}_2 \phi \phi}^{\textbf{R} (2)}(p_{1,2,3} )$ at the symmetric point and solving for $\d_{\mathcal{O}_2}^{(2)}$, we get that
\be
\d_{\mathcal{O}_2}^{(2)} = - \frac{\l^2}{6}\frac{ S_1(-\m^2)}{-\m^2} + \frac{\l^2}{4} \left[L_1(-\m^2) \right]^2 - \frac{\l^2}{4(4\pi)^4}2G_{s.p.}
\ee
where
\be
\bal
2G_{s.p.} &= \left.G(p_1,p_2) \right|_{s.p.} + \left.G(p_1,p_3)\right|_{s.p.} \\
&= 2 \int_0^1 \mathrm{d}z \mathrm{dy}\frac{z}{1-z} \ln \left(\frac{z^2 y(1+y)-z(1+2y)}{y(y-1)} \right)\, .
\eal
\ee
therefore, the renormalized 3 point function will be given by: 
\be \label{CR l2}\bal 
C_{\mathcal{O}_2 \phi \phi}^{\textbf{R} }(p_{1,2,3}) &= 2  \left\{1 + \frac{\l}{2(4\pi)^2}  \ln \left( \frac{-p_1^2}{\m^2}\right)+ \frac{\l^2}{2(4\pi)^4}  \ln^2 \left( \frac{-p_1^2}{\m^2}\right)   \right.\\
& \left.
-\frac{\l^2}{2(4\pi)^4}  \ln \left( \frac{-p_1^2}{\m^2}\right) + \frac{\l^2}{12 (4\pi)^4} \left[  \ln \left( \frac{-p_2^2}{\m^2}\right)   +  \ln \left( \frac{-p_3^2}{\m^2}\right)\right] - \frac{\l^2}{2(4\pi)^4}\hat{G}(p_1,p_2,p_3)
\right\}\, ,
\eal \ee
where 
\bea\label{G hat}
\hat{G}(p_1,p_2,p_3) &=& G(p_1,p_2) + G(p_1,p_3) -2G_{s.p} \nonumber\\
&=&\int_0^1\mathrm{d}y \mathrm{d}z \frac{z }{1-z}\Bigl[\ln \left( \frac{-2 yz(1-z) p_1 
\cdot p_2 +yz(1-yz)p_1^2 + z(1-z)p_2^2}{p_1^2 \left[ z^2y(1+y)-z(1+2y) \right] }\right) \nonumber\\
&+& \left(p_2 \leftrightarrow p_3 \right) \Bigr]\, .
\eea
The next step is to determine the anomalous dimension $\Gamma_{\mathcal{O}_2}$ up to $O(\l^2)$ from the Callan-Symanzik equation.
We have
\be\label{cRo2phiphi}
\bal
\frac{\partial}{\partial \ln \m}C_{\mathcal{O}_2 \phi \phi}^{\textbf{R} }(p_{1,2,3}^2) &= 2 \left[ -\frac{\l}{(4\pi)^2} - 4\g_\phi +  \frac{\l^2}{(4\pi)^4} - 2\frac{\l^2}{(4\pi)^4} \ln \left( \frac{-p_1^2}{\m^2}\right)  \right] + O(\l^3) \\
\b_\l \partial_\l C_{\mathcal{O}_2 \phi \phi}^{\textbf{R} }(p_{1,2,3}^2)& = \frac{3\l^2}{(4\pi)^4}  \ln \left( \frac{-p_1^2}{\m^2}\right) + O(\l^3) \\
\left[\Gamma_{\mathcal{O}_2} + 2\g_\phi \right]  C_{\mathcal{O}_2 \phi \phi}^{\textbf{R} }(p_{1,2,3}^2)& = 2 \left[ \frac{\l}{(4\pi)^2} +\frac{\l^2}{2(4\pi)^4}  \ln \left( \frac{-p_1^2}{\m^2}\right)   + 2\g_\phi  + \G_{\mathcal{O}_2}^{(2)}\right] + O(\l^3)
\eal
\ee
and applying the Callan-Symanzik eqution yields the result
\be
{\G_{\mathcal{O}_2} = \frac{\l}{(4\pi)^2} - \frac{5}{6} \frac{\l^2}{(4\pi)^4}}\, .
\ee
This is in agreement with the result in \cite{Henriksson_2023}, by making the substitution 
\be
\bal
\hat{\b}_{\l} =0 \rightarrow -\ve \l + \frac{3\l^2}{(4\pi)^2} - \frac{17}{3} \frac{\l^3}{(4\pi)^4} =0 \\
\rightarrow  \ve =\frac{3\l}{(4\pi)^2} - \frac{17}{3} \frac{\l^2}{(4\pi)^4}
\eal
\ee
in the CFT expression.

\subsection{Callan-Symanzik equation and dilatation Ward identity}

To conformal limit of \eqref{CR l2} is more involved now, it being a genuine 3-point function. 
We start the discussion by writing down the Callan-Symanzik equation in terms of derivatives with respect to external momenta. 
The perturbative calculation just showed that the 3-point function has the general form
\be\label{o2 phi phi H G}
\left<\left< \mathcal{O}_2(p_1)\phi(p_2) \phi(p_3) \right> \right> = \frac{i}{p_2^2}\frac{i}{p_3^2} 
H\left( \frac{p_1^2}{\m^2}, \frac{p_2^2}{\m^2},\frac{p_3^2}{\m^2} ;\l \right)  -\frac{\l^2}{(4\pi)^4} \frac{i}{p_2^2}\frac{i}{p_3^2} \hat{G}(p_1,p_2,p_3)\, ,
\ee
where $ H\left( \frac{p_1^2}{\m^2}, \frac{p_2^2}{\m^2},\frac{p_3^2}{\m^2} ;\l \right)$ contains all the logarithmic 
parts of the correlation function and is defined (see \eqref{CR l2}) as:
\be
H\left( \frac{p_1^2}{\m^2}, \frac{p_2^2}{\m^2},\frac{p_3^2}{\m^2} ;\l \right)= C_{\mathcal{O}_2 \phi \phi}^{\textbf{R} }(p_{1,2,3})+  \frac{\l^2}{(4\pi)^4}\hat{G}(p_1,p_2,p_3)
\ee
with $\hat{G}(p_1,p_2,p_3)$ given by   \eqref{G hat}. 
Since the Callan Symanzik equation is linear, we can study the two parts of \eqref{o2 phi phi H G} separately. 

We will  begin with the term proportional to $\hat{G}(p_1,p_2,p_3)$. 
This terms is $\m$ independent and since it is of $O(\l^2)$, it can be neglected as a higher order contribution. In other words
\be
 \left[\b_\l \partial_\l +2\g_{\phi} + \G_{\mathcal{O}_2}\right] \l^2 \frac{i}{p_2^2}\frac{i}{p_3^2} \hat{G}(p_1,p_2,p_3)= O(\l^3)\, .
\ee
There is a simple reason why this term does not contribute to the Callan-Symanzik equation. In fact $\hat{G}(p_1,p_2,p_3)$ is scale invariant since
\be
 \hat{G}(p_1,p_2,p_3) = \hat{G}(\alpha p_1, \alpha p_2, \alpha p_3)\, .
\ee 
A more formal way to check the scale invariance of $\hat{G}(p_1,p_2,p_3)$ of course is to check as before the Dilatation Ward Identity, to $O(\l^2)$:
\be\label{Dilatation for G}
\left [-p_1 \frac{\partial}{\partial p_1} -p_2 \frac{\partial}{\partial p_2} -p_3 \frac{\partial}{\partial p_3} + 
2\D_\phi + \D_{\mathcal{O}_2} -2d \right]\l^2\frac{i}{p_2^2}\frac{i}{p_3^2}\hat{G}(p_1,p_2,p_3) = O(\l^3)
\ee
using $\D_{\mathcal{O}_2}= d-2 +\G_{\mathcal{O}_2}$ and $\D_\phi =\frac{d-2}{2}+ \g_\phi$ we can write the previous expression in a more familiar way:
\be
\left [-p_1 \frac{\partial}{\partial p_1} -p_2 \frac{\partial}{\partial p_2} -p_3 \frac{\partial}{\partial p_3}-4  + 2\g_\phi + 
\G_{\mathcal{O}_2}  \right]\l^2\frac{i}{p_2^2}\frac{i}{p_3^2}\hat{G}(p_1,p_2,p_3) = O(\l^3)\, .
\ee
This equation is identical to the Callan-Symazik equation with a vanishing $\beta$ function, 
by means of the connection \eqref{derivativerule} between the $\m$ derivative and the $p_i$ derivatives.

Next we turn to the term proportional to $H$. Since the ${\hat G}$ term is removed from the picture, the 
Callan-Symanzik equation for the term proportional to $H\left( \frac{p_1^2}{\m^2}, \frac{p_2^2}{\m^2},\frac{p_3^2}{\m^2} ;\l \right)$ is
\be
\bal
\left[ \m \frac{\partial}{\partial \m} + \b_\l \partial_\l +2\g_{\phi} + \G_{\mathcal{O}_2}\right]\frac{i}{p_2^2}\frac{i}{p_3^2} H\left( \frac{p_1^2}{\m^2}, \frac{p_2^2}{\m^2},\frac{p_3^2}{\m^2} ;\l \right)=0\, .
\eal
\ee
Using once more
\be
\bal
\m\frac{\partial}{\partial \m}H\left( \frac{p_1^2}{\m^2}, \frac{p_2^2}{\m^2},\frac{p_3^2}{\m^2} ;\l \right)& = -\sum_{j=1}^{3}p_i \frac{\partial}{\partial p_j}H\left( \frac{p_1^2}{\m^2}, \frac{p_2^2}{\m^2},\frac{p_3^2}{\m^2} ;\l \right) \\
\Rightarrow \m\frac{\partial}{\partial \m} \frac{i}{p_2^2} \frac{i}{p_3^2}H\left( \frac{p_1^2}{\m^2}, \frac{p_2^2}{\m^2},\frac{p_3^2}{\m^2} ;\l \right)& = \left[ -\sum_{j=1}^{3}p_i \frac{\partial}{\partial p_j} -4 \right]  \frac{i}{p_2^2} \frac{i}{p_3^2}H\left( \frac{p_1^2}{\m^2}, \frac{p_2^2}{\m^2},\frac{p_3^2}{\m^2} ;\l \right)
\eal
\ee
the equation takes the form
\be
 \left[ -\sum_{j=1}^{3}p_i \frac{\partial}{\partial p_j} -4 + \b_\l \partial_\l +2\g_{\phi} + \G_{\mathcal{O}_2} \right] 
 \frac{i}{p_2^2} \frac{i}{p_3^2}H\left( \frac{p_1^2}{\m^2}, \frac{p_2^2}{\m^2},\frac{p_3^2}{\m^2} ;\l \right)=0\, .
\ee
Using $\D_{\mathcal{O}_2}= d-2 +\G_{\mathcal{O}_2}$ and $\D_\phi =\frac{d-2}{2}+ \g_\phi$ we can rewrite this as 
\be\label{momenta CS for H}
\left[ -\sum_{j=1}^{3}p_i \frac{\partial}{\partial p_j}  + 2 \D_\phi + \D_{\mathcal{O}_2} -2d \right] \frac{i}{p_2^2} \frac{i}{p_3^2}H\left( \frac{p_1^2}{\m^2}, \frac{p_2^2}{\m^2},\frac{p_3^2}{\m^2} ;\l \right)=- \b_\l \partial_\l \frac{i}{p_2^2} \frac{i}{p_3^2}H\left( \frac{p_1^2}{\m^2}, \frac{p_2^2}{\m^2},\frac{p_3^2}{\m^2} ;\l \right)\, .
\ee
Combining finally \eqref{o2 phi phi H G},\eqref{Dilatation for G} and \eqref{momenta CS for H}, the Callan-Symanzik equation of the 3-point function takes the following form\cite{EFT anomalous, RG coef S-matrix}:
\be 
\left[ -\sum_{j=1}^{3}p_i \frac{\partial}{\partial p_j}  + 2 \D_\phi + \D_{\mathcal{O}_2} -2d \right]\left< \left< \mathcal{O}_2 (p_1)\phi(p_2) \phi(p_3) \right> \right> = - \b_\l \partial_\l \left< \left< \mathcal{O}_2 (p_1)\phi(p_2) \phi(p_3) \right> \right> 
\ee
Now we can clearly identify the $\beta_\l$-function as the source of the breaking of scale invariance on the right-hand side of the equation.
In this form, on the fixed point where we are instructed to perform the shift $\frac{-p_i^2}{\m^2}\to p_i^2$ and send $\b_\l\to 0$, we obtain a 
correlator that obeys the dilatation Ward identity.
It would be interesting to study also the breaking of Special Conformal Symmetry, but this is beyond the scope of this work.
\footnote {We just state that for this simple theory scale invariance is expected to imply full conformal invariance. In Section 4 of \cite{Rychkov_Trace} the authors present the Ward Identities expressing the breaking of scale and special conformal invariance by the running of the coupling (see equations (4.14) (4.15) in \cite{Rychkov_Trace}). 
They prove that when the system reaches the fixed point, it attains scale and conformal symmetry.
It is  interesting to note though that it is simple to verify that the $\hat{G}$ term of the correlation 
function $\left< \mathcal{O}_2 \phi \phi \right>$ 
obeys the SCT Ward identity for $p_1 \rightarrow 0$.
  }

\section{The 3-point function $\braket{\mathcal{O}_4\phi\phi}$ }

In this case, it is easier to find directly the bare diagrams because the contributions to this 3-point function can be visualised as insertions 
of the $\mathcal{O}_4$ operator in the propagator. So we can immediately find the bare  $O(\l)$ and $O(\l^2)$ diagrams, which are: 
\be
\bal
\left<\left< \mathcal{O}_{4,0}(p_1) \phi_0(p_2) \phi_0(p_3) \right> \right>&=\begin{gathered}
\begin{tikzpicture}
\begin{feynman}
\vertex (x);
\vertex[right =  0.5 cm of x](x1);
\vertex[right =  0.6 cm of x1](x2);
\vertex[right =  0.5 cm of x2](x3);
\diagram{(x)--(x1)--[half left](x2)--[half left](x1)--(x3)};
\end{feynman}
\filldraw[fill=black] (x1) ++(-0.07cm, -0.07cm) rectangle ++(0.2cm, 0.2cm);
\end{tikzpicture}
\end{gathered}+\begin{gathered}
\begin{tikzpicture}
\begin{feynman}
\vertex (x);
\vertex[right =  0.5 cm of x](x1);
\vertex[right =  0.6 cm of x1](x2);
\vertex[right =  0.5 cm of x2](x3);
\diagram{(x)--(x1)--[half left](x2)--[half left](x1)--(x3)};
\end{feynman}
\filldraw[fill=black] (x2) ++(-0.07cm, -0.07cm) rectangle ++(0.2cm, 0.2cm);
\end{tikzpicture}
\end{gathered}+\begin{gathered}
\begin{tikzpicture}
\begin{feynman}
\vertex(x);
\diagram [horizontal=a to b, layered layout] {
  a -- b --b--c
};
\path (b)--++(90:0.5) coordinate (A);
\path  (b)--++(90:1) coordinate (B);
\draw (A) circle(0.5);
\draw (B)circle(0.25);
\path (B)--++(1 :-0.2)coordinate (C);
\end{feynman}
\filldraw[fill=black] (C) ++(-0.13cm, -0.1cm) rectangle ++(0.15cm, 0.15cm);
 \end{tikzpicture}
\end{gathered} +
\begin{gathered}
\begin{tikzpicture}
\begin{feynman}
\vertex(x);
\diagram [horizontal=a to b, layered layout] {
  a -- b --b--c
};
\path (b)--++(90:0.5) coordinate (A);
\path  (b)--++(90:1) coordinate (B);
\draw (A) circle(0.5);
\draw (B)circle(0.25);
\path (B)--++(1 :0.2)coordinate (C);
\end{feynman}
\filldraw[fill=black] (C) ++(-0.04cm, -0.1cm) rectangle ++(0.15cm, 0.15cm);
 \end{tikzpicture}
\end{gathered} \\&+
\begin{gathered}
\begin{tikzpicture}
\begin{feynman}
\vertex (x);
\vertex[right= 0.5cm of x](x1);
\vertex[right= 1cm of x1](x2);
\vertex[right= 1cm of x2](x3);
\vertex[right= 0.5cm of x3](x4);
\vertex[above= 1cm of x2](x5);
\diagram{(x)--(x1)--[quarter left](x5)--[quarter left](x1)--(x3)--[quarter left](x5)--[quarter left](x3)--(x4)};
\end{feynman}
\filldraw[fill=black] (x5) ++(-0.1cm, -0.1cm) rectangle ++(0.2cm, 0.2cm);
\end{tikzpicture}
\end{gathered}
+\begin{gathered}
\begin{tikzpicture}
\begin{feynman}
\vertex (x);
\vertex[right= 0.5cm of x](x1);
\vertex[right= 1cm of x1](x2);
\vertex[right= 1cm of x2](x3);
\vertex[right= 0.5cm of x3](x4);
\vertex[above= 1cm of x2](x5);
\diagram{(x)--(x1)--[quarter left](x5)--[quarter left](x1)--(x3)--[quarter left](x5)--[quarter left](x3)--(x4)};
\end{feynman}
\filldraw[fill=black] (x1) ++(-0.1cm, -0.1cm) rectangle ++(0.2cm, 0.2cm);
\end{tikzpicture}
\end{gathered}+\begin{gathered}
\begin{tikzpicture}
\begin{feynman}
\vertex (x);
\vertex[right= 0.5cm of x](x1);
\vertex[right= 1cm of x1](x2);
\vertex[right= 1cm of x2](x3);
\vertex[right= 0.5cm of x3](x4);
\vertex[above= 1cm of x2](x5);
\diagram{(x)--(x1)--[quarter left](x5)--[quarter left](x1)--(x3)--[quarter left](x5)--[quarter left](x3)--(x4)};
\end{feynman}
\filldraw[fill=black] (x3) ++(-0.1cm, -0.1cm) rectangle ++(0.2cm, 0.2cm);
\end{tikzpicture}
\end{gathered}
\eal
\ee
with: 
\be
\begin{gathered}
\begin{tikzpicture}
\begin{feynman}
\vertex (x);
\vertex[right =  0.5 cm of x](x1);
\vertex[right =  0.6 cm of x1](x2);
\vertex[right =  0.5 cm of x2](x3);
\diagram{(x)--(x1)--[half left](x2)--[half left](x1)--(x3)};
\end{feynman}
\filldraw[fill=black] (x1) ++(-0.07cm, -0.07cm) rectangle ++(0.2cm, 0.2cm);
\end{tikzpicture}
\end{gathered}= -4\l_0 \frac{i}{p_2^2}\frac{i}{p_3^2}S_1(p_2^2)
\ee
\be
\begin{gathered}
\begin{tikzpicture}
\begin{feynman}
\vertex (x);
\vertex[right =  0.5 cm of x](x1);
\vertex[right =  0.6 cm of x1](x2);
\vertex[right =  0.5 cm of x2](x3);
\diagram{(x)--(x1)--[half left](x2)--[half left](x1)--(x3)};
\end{feynman}
\filldraw[fill=black] (x2) ++(-0.07cm, -0.07cm) rectangle ++(0.2cm, 0.2cm);
\end{tikzpicture}
\end{gathered}= - 4\l_0 \frac{i}{p_2^2}\frac{i}{p_3^2}S_1(p_3^2)
\ee
\be
\begin{gathered}
\begin{tikzpicture}
\begin{feynman}
\vertex(x);
\diagram [horizontal=a to b, layered layout] {
  a -- b --b--c
};
\path (b)--++(90:0.5) coordinate (A);
\path  (b)--++(90:1) coordinate (B);
\draw (A) circle(0.5);
\draw (B)circle(0.25);
\path (B)--++(1 :-0.2)coordinate (C);
\end{feynman}
\filldraw[fill=black] (C) ++(-0.13cm, -0.1cm) rectangle ++(0.15cm, 0.15cm);
 \end{tikzpicture}
\end{gathered} =\begin{gathered}
\begin{tikzpicture}
\begin{feynman}
\vertex(x);
\diagram [horizontal=a to b, layered layout] {
  a -- b --b--c
};
\path (b)--++(90:0.5) coordinate (A);
\path  (b)--++(90:1) coordinate (B);
\draw (A) circle(0.5);
\draw (B)circle(0.25);
\path (B)--++(1 :0.2)coordinate (C);
\end{feynman}
\filldraw[fill=black] (C) ++(-0.04cm, -0.1cm) rectangle ++(0.15cm, 0.15cm);
 \end{tikzpicture}
\end{gathered} = -i3\l_0^2\frac{i}{p_2^2}\frac{i}{p_3^2}ST({p_1}^2)
\ee
 \be
 \begin{gathered}
\begin{tikzpicture}
\begin{feynman}
\vertex (x);
\vertex[right= 0.5cm of x](x1);
\vertex[right= 1cm of x1](x2);
\vertex[right= 1cm of x2](x3);
\vertex[right= 0.5cm of x3](x4);
\vertex[above= 1cm of x2](x5);
\diagram{(x)--(x1)--[quarter left](x5)--[quarter left](x1)--(x3)--[quarter left](x5)--[quarter left](x3)--(x4)};
\end{feynman}
\filldraw[fill=black] (x5) ++(-0.1cm, -0.1cm) rectangle ++(0.2cm, 0.2cm);
\end{tikzpicture}
\end{gathered}=-i 6\l_0^2 \frac{i} {p_2^2} \frac{i}{p_3^2} \left[T(p_2,p_3) +\left( p_2 \leftrightarrow p_3  \right)\right]
 \ee
 \be
 \begin{gathered}
\begin{tikzpicture} 
\begin{feynman}
\vertex (x);
\vertex[right= 0.5cm of x](x1);
\vertex[right= 1cm of x1](x2);
\vertex[right= 1cm of x2](x3);
\vertex[right= 0.5cm of x3](x4);
\vertex[above= 1cm of x2](x5);
\diagram{(x)--(x1)--[quarter left](x5)--[quarter left](x1)--(x3)--[quarter left](x5)--[quarter left](x3)--(x4)};
\end{feynman}
\filldraw[fill=black] (x3) ++(-0.1cm, -0.1cm) rectangle ++(0.2cm, 0.2cm);
\end{tikzpicture}
\end{gathered} = -i\frac{3}{2}\l_0^2 TB(p_3^2)
 \ee
Here, $S_1(p^2)$ is the standard sunset integral, which has already been introduced (see \eqref{s1}, \eqref{s1 exp}) 
and the 'Sunset-Tadpole' $ST({p_1}^2)$ is given in \eqref{st general d} and \eqref{st e exp}.
 
To make it easier to follow the process, we give below the values of the integrals in the context of the $\epsilon$-expansion:
 \be
ST({p_1}^2) = i\frac{{p_1}^2}{(4\pi)^6} \left[ -\frac{1}{6\ve} + \frac{1}{4} \ln \left(\frac{-p_1^2e^\g}{4\pi} \right) - \frac{25}{24} \right]
 \ee
 \be
\bal
TB(p^2)&=\frac{ip^2}{(4\pi)^6} \left[ \frac{4}{3 \epsilon^2} -\frac{2\ln\left( \frac{-p^2e^\g}{4\pi} \right) -\frac{20}{3}}{\epsilon}  + 
\frac{3}{2} \ln^2 \left( \frac{-p^2e^\g}{4\pi}\right) -10 \ln \left(\frac{-p^2e^\g}{4\pi}\right)  -\frac{\pi^2}{12} +\frac{64}{3}\right]
\eal
\ee
\be\bal
T(p_2,p_3) = \frac{i}{(4\pi)^6} &\left\{ \frac{2}{3\epsilon^2} \left(p_2^2 +p_3^2 \right)  -p_2^2 \frac{\ln \left(\frac{-p_2^2 e^\g}{4\pi} \right)}{\epsilon} -
p_3^2 \frac{\ln \left(\frac{-p_3^2 e^\g}{4\pi} \right)}{\epsilon}  + \frac{\frac{11}{3}p_2^2 +\frac{23}{6} p_3^2}{\epsilon} \right.  \\
& \;+p_2^2\left[ \frac{18}{24} \left(\ln \left(\frac{-p_2^2 e^\g}{4\pi} \right) - \frac{11}{3}\right)^2 + \frac{11}{4} - \frac{\pi^2}{24} \right]\\
&\left. + p_3^2 \left[ \frac{54}{72} \left(\ln \left(\frac{-p_3^2 e^\g}{4\pi} \right) - \frac{23}{6}\right)^2 +\frac{157}{48} - \frac{7\pi^2}{72} \right]  \right\} + I_{f}(p_2,p_3)
\eal \ee
The $I_f$ term which appears in the  $T$ integral is a finite integral with respect to the Feynman parameters.
This term doesn't  contribute to the renormalization procedure, so we do not care of its exact form. 
The two types of 'tent' integrals $T$ and $TB$ are given in the Appendix.
In total, the bare 3-point function is given by 
\be\label{bare o4phiphi}
\bal
\left< \left< \mathcal{O}_{4,0}(p_1)\phi_0(p_2)\phi_0(p_3) \right> \right> = &-4\l_0 \frac{i}{p_2^2}\frac{i}{p_3^2} 
\left[S_1(p_2^2) + S_1(p_3^2) \right] - i6 \l_0^2 \frac{i}{p_2^2} \frac{i}{p_3^2} ST(p_1^2) \\
&- i\frac{3}{2} \l_0^2 \frac{i}{p_2^2}\frac{i}{p_3^2}  \left[TB(p_2^2) + TB(p_3^2) \right] -i 6\l_0^2 \frac{i} {p_2^2} \frac{i}{p_3^2} \left[T(p_1,p_2) +T(p_1,p_3) \right]
\eal
\ee
Looking at the powers of external momenta of each diagram we  observe that the bare 3-point function has the following form:
\be
\bal
\left< \left< \mathcal{O}_{4,0}(p_1)\phi_0(p_2)\phi_0(p_3) \right> \right> 
&=- i6 \l_0^2 p_1^2 \frac{i}{p_2^2} \frac{i}{p_3^2} \frac{ ST(p_1^2)}{p_1^2}\\
&+\frac{i}{p_2^2} \frac{i}{p_3^2} p_2^2 \left[-4\l_0  \frac{S_1(p_2^2)}{p_2^2}  - i\frac{3}{2} \l_0^2    \frac{TB(p_2^2)}{p_2^2}  -i 6\l_0^2  \frac{\left[T(p_1,p_2) +T(p_1,p_3) \right]}{p_2^2} \right] \\
&+ \left(p_2 \leftrightarrow p_3 \right)\, .
\eal
\ee
We observe that the bare correlation function has two different external momentum structures. As a result, we cannot define an overall $Z_{\mathcal{O}_4}$ that renormalizes all the divergences appearing in the bare correlation function.
This implies the presence of mixing of $\mathcal{O}_4$ under renormalization with an operator of mass dimension 4. 
As we will see in the next sections, this operator is $K_2=\Box \phi^2$. 

The operator mixing implies that :
\be
\mathcal{O}_{4,0} = Z_{ \mathcal{O}_4}  \mathcal{O}_4 + Z_{mixing} M \, ,
\ee
where $M$ is the (unknown for now) operator that mixes with $\mathcal{O}_4$.  
The mixing implies that the renormalized 3-point function of the operator $M$ will have the form
\be\label{M form}
\braket{M \phi \phi} = p_1^2 \frac{i}{p_2^2} \frac{i}{p_3^2} C_{M \phi \phi}\left(\frac{p_i^2}{ \m^2} ; \l \right) \, ,
\ee
where the function $C_{M \phi \phi}( \m ; \l)$ is determined by the loop diagrams of the corresponding bare correlation function. 
Furthermore, this function is equal to some constant $\mathcal{C}$ at the Symmetric Point of the energy scale $\m$, that is
\be
C_{M \phi \phi}( \m ; \l)= \mathcal{C}\, .
\ee
Using the above definitions we obtain the following relation between the renormalized and the  bare correllation functions:
\be
Z_{ \mathcal{O}_4} Z_\phi \braket{ \mathcal{O}_4 \phi \phi } = \braket{\mathcal{O}_{4,0}\phi_0 \phi_0} - Z_{mixing} Z_\phi \braket{M \phi \phi}\, .
\ee
The renormalization condition for$\braket {\mathcal{O}_4 \phi \phi }$  is
\be
\braket{\mathcal{O}_4 \phi \phi } = -4\l \frac{i}{p_2^2} \frac{i}{p_3^2} \left( p_2^2 +p_3^2 \right) \; \text{at $S.P.$}
\ee
Combining \eqref{M form} with the bare 3-point function of $\mathcal{O}_{4,0}$ in terms 
of the renormalized coupling constant $\l$, the above equality takes the following form:
 \be
 \bal
 Z_{ \mathcal{O}_4} Z_\phi \braket{ \mathcal{O}_4 \phi \phi } &=p_1^2 
 \frac{i}{p_2^2 }\frac{i}{p_3^2} \left[ Z_M Z_\phi C_{M \phi \phi}\left(\frac{p_i^2}{ \m^2}\right) - i6 \l^2 \frac{ST(p_1^2)}{ p_1^2} \right]\\
+\frac{i}{p_2^2}& \frac{i}{p_3^2} p_2^2 \left\{-4\l_0  \frac{S_1(p_2^2)}{p_2^2} +i6\l^2 L_1(-\m^2) \frac{S_1(p_2^2)}{p_2^2} \right.\\
& \, \, \, \, \;   \left. - i\frac{3}{2} \l^2    \frac{TB(p_2^2)}{p_2^2}  -i 6\l^2  \frac{\left[T(p_1,p_2) +T(p_1,p_3) \right]}{p_2^2} \right\} + \left(p_2 \leftrightarrow p_3 \right)
 \eal
 \ee
Imposing the renormalization condition we obtain the following set of equations:
\bea
&Z_{\mathcal{O}_4}Z_\phi=  \frac{-1}{4\l}\left[-4\l  \frac{S_1(p_2^2)}{p_2^2}  - i\frac{3}{2} \l^2    \frac{TB(p_2^2)}{p_2^2} +i6\l^2 L_1(-\m^2) \frac{S_1(p_2^2)}{p_2^2}  -i 6\l^2  \frac{\left[T(p_1,p_2) +T(p_1,p_3) \right]}{p_2^2} \right]_{S.P.}\\
&Z_M Z_\phi \mathcal{C} =\left. i6 \l^2 \frac{ST(p_1^2)}{ p_1^2}\right|_{S.P.}
\eea
The necessary counterterm for the divergences which are associated  with the external momentum structure 
$\frac{i}{p_2^2} \frac{i}{p_3^2}(p_2^2 + p_3^2)$ has a peculiarity because there does 
not exist any finite tree-level diagram in the bare correlation function. 
For this reason, we must take
\be
Z_{\mathcal{O}_4} = \d_{\mathcal{O}_4}^{(0)} + \d_{\mathcal{O}_4}^{(1)} +\d_{\mathcal{O}_4}^{(2)} + \cdots  
\ee
Taking into account the fact that $Z_\phi = 1 + O(\l^2)$, we can neglect it as a higher order term. 
We conclude that
\bea
&Z_{\mathcal{O}_4}=  -\left[-  \frac{S_1(p_2^2)}{p_2^2}  - i\frac{3}{8} \l    
\frac{TB(p_2^2)}{p_2^2} +i\frac{3}{2}\l L_1(-\m^2) \frac{S_1(p_2^2)}{p_2^2}  
-i \frac{3}{2}\l  \frac{\left[T(p_1,p_2) +T(p_1,p_3) \right]}{p_2^2} \right]_{S.P.}\\
&Z_M  =\left. i\frac{6}{\mathcal{C}} \l^2 \frac{ST(p_1^2)}{ p_1^2}\right|_{S.P.}
\eea
Now we can straightforwardly derive the renormalized expression of the 3-point function:
\bea \label{renormalized O4phiphi}
\braket{\braket{ \mathcal{O}_4(p_1) \phi (p_2) \phi(p_3) }} &=& -4\frac{i}{p_2^2}\frac{i}{p_3^2} 
\left\{p_2^2 \left[ \l + \frac{9}{2} \frac{\l^2}{16\pi^2} \ln \left( \frac{-p_2^2}{\m^2}\right) \right] + 
\left(p_2 \leftrightarrow p_3 \right)\right\} \nonumber\\
&+& p_1^2 \frac{i}{p_2^2}\frac{i}{p_3^2} \frac{\l^2}{4(4\pi)^6} \ln \left(\frac{-p_1^2}{\m^2}\right)
\eea
The above renormalized expression contains the information about the mixing of the operator $\mathcal{O}_4$. 
The Callan-Symanzik equation of the mixed operator has the following  form:
\be
\left[\m \frac{\partial}{\partial \m } + \b_\l \frac{\partial}{\partial \l} + \G_{\mathcal{O}_4} + 2\g_\phi \right] \braket{\braket{\mathcal{O}_4 \phi \phi}} + \G_{mix} \braket{\braket{M\phi \phi}}=0
\ee
Acting with derivatives on \eqref{renormalized O4phiphi} 
\be\bal
\m \frac{\partial}{\partial \m}\left< \left< \mathcal{O}_4(p_1) \phi (p_2) \phi(p_3) \right>\right>&=- \frac{9\l}{16\pi^2}\left[-4\l \frac{i}{p_2^2}\frac{i}{p_3^2}  (p_2^2 +p_3^2) \right] - p_1^2 \frac{i}{p_2^2}\frac{i}{p_3^2} \frac{\l^2}{2(4\pi)^6} + O(\l^3) \\
\b_\l \partial_\l\left< \left< \mathcal{O}_4(p_1) \phi (p_2) \phi(p_3) \right>\right>&= \frac{3\l}{16\pi^2} \frac{i}{p_2^2} \left[ -4 \l \frac{i}{p_3^2} (p_2^2 +p_3^2)  \right]+ O(\l^3)\\
(\G_{\mathcal{O}_4}+2 \g_\phi)\left< \left< \mathcal{O}_4(p_1) \phi (p_2) \phi(p_3) \right>\right>&=\G_{\mathcal{O}_4}\left[-4\l \frac{i}{p_2^2}\frac{i}{p_3^2} (p_2^2 +p_3^2) \right]  + O(\l^3)
\eal\ee
we can solve the Callan-Symazik equation for $\G_{\mathcal{O}_4}$ and $\G_{mix}$. We obtain
\be\label{Gamma mix o4}
{\G_{mix}} = \frac{\l^2}{\mathcal{C} 2(4\pi)^6} \, .
\ee
We remind that $\mathcal{C}$ is the constant in $\braket{M \phi \phi}$. In the next sections we will discuss further this mixing term.
 Form the Callan-Symanzik equation we also extract the value of the anomalous dimension of the operator $\mathcal{O}_4$:
\be
\G_{\mathcal{O}_4} = \frac{6\l}{16\pi^2} + O(\l^2)\, ,
\ee
which is in agreement with the known result presented in  \cite{Henriksson_2023} and 
in agreement with the definition of the anomalous dimension of an operator which is associated with a coupling constant  $\l$ as:
\be
\G_{\mathcal{O}_4} = \frac{\partial \b_\l}{\partial \l}\, .
\ee
We will not need to go to higher order for this correlator. The reason is that we will be eventually interested in the 
correlator $\braket{\l \phi^4 \phi\phi}$ which, at this order, is already $O(\l^3)$.

\section{The 3-point function $\braket{K_2 \phi \phi}$}

We begin the discussion of the spin zero derivative operators of dimension $d$ with the 
the simplest case, that of the operator $K_2$, defined as:
\be
K_{2,0}(x) = \Box_x \lim_{x_1 \rightarrow x} \phi_0(x) \phi_0(x_1)
\ee
Its bare 3-point function is given by:
\be
\left< K_{2,0}(x) \phi_0(y) \phi_0(z) \right> =\Box_x \lim_{x_1 \rightarrow x} \left< \phi_0(x) \phi_0(x_1) \phi_0(y) \phi_0(z) \right> - (\text{contact terms})
\ee

\subsection{$O(\l)$ renormalization of $\braket{ K_2 \phi \phi }$}

The $O(\l)$ contributions are extracted from the limit of the following diagrams:
\be 
\bal
\Box_x \braket{ \phi_0^2(x) \phi_0(y)\phi_0(z) }&= \Box_x \lim_{x_1 \rightarrow x}\left[   
 \begin{gathered} \begin{tikzpicture} 
     \begin{feynman}
     \vertex (x){$x$};
     \vertex[below =2cm of x](u){$y$};
     \vertex[right=  0.5cm of x](y){$x_1$};
     \vertex[right =  0.5cm of  u](l){$z$};
     \diagram{
     (x)--(u)
     };
     \diagram{
     (y)--(l)
     };
 \end{feynman} 
 \end{tikzpicture}
 \end{gathered}+
 \begin{gathered} \begin{tikzpicture} 
     \begin{feynman}
     \vertex (x){$x$};
     \vertex[below =2cm of x](u){$z$};
     \vertex[right=   0.5cm of x](y){$x_1$};
     \vertex[right = 0.5cm of  u](l){$y$};
     \diagram{
     (x)--(u)
     };
     \diagram{
     (y)--(l)
     };
 \end{feynman} 
 \end{tikzpicture}
 \end{gathered}  + \begin{gathered} \begin{tikzpicture} 
     \begin{feynman}
     \vertex (x){$x$};
     \vertex[below = 2cm of x](u){$x_1$};
     \vertex[right=  1cm of x](y){$y$};
     \vertex[right =  1cm of  u](l){$z$};
     \diagram{
     (x)--(l)
     };
     \diagram{
     (y)--(u)
     };
 \end{feynman} 
 \end{tikzpicture}
 \end{gathered}\right] \\
 &=-2 \int \frac{i^2(k_1+k_2)^2 e^{i(k_1 +k_2)x}e^{-ik_1y}e^{-k_2 z} }{k_1^2 k_2^2} \\
 & \; \; +i\l \int \frac{(k_1 +k_2)^2 e^{i(k_1 +k_2)x} e^{-iq_1y} e^{-iq_2 z}}{k_1^2 k_2^2 q_1^2 q_2^2} \tilde{\d}(k_1 + k_2 -q_1 -q_2)
 \eal
 \ee
where $\tilde{\d}(k_1 +k_2 -q_1 -q_2) = (2\pi)^d \d^{(d)}(k_1 +k_2 -q_1 -q_2)$.  
The fact that no terms analogous to delta functions appear in the above expression implies 
that there are no contact terms, so we can immediately calculate  $\braket{\braket{ K_{2,0}(p_1) \phi_0(p_2) \phi_0(p_3)}}$. 
Moving to momentum space we obtain the simple expression
\be
\braket{\braket{ K_{2,0}(p_1) \phi_0(p_2) \phi_0(p_3) }} = -p_1^2 \frac{i}{p_2^2} \frac{i}{p_3^2} \left[ 2 + i\l_0 L_1(p_1^2) \right]\, .
\ee
Comparing the above with \eqref{bare O2phiphi oreder 1} we observe that
\be
\braket{\braket{ K_{2,0}(p_1) \phi_0(p_2) \phi_0(p_3) }} = -p_1^2 \braket{\braket{ \mathcal{O}_{2,0}(p_1) \phi_0(p_2) \phi_0(p_3) }}\, .
\ee
The renormalization condition for this 3-point function is :
\be
\braket{\braket{ K_{2}(p_1) \phi(p_2) \phi(p_3) }} =-p_1^2 \frac{i}{p_2^2} \frac{i}{p_3^2} \; , \; 
\text{at the S.P.}
\ee
The procedure is exactly the same as in the case of $\mathcal{O}_2$ operator:
\be
\bal
Z_{K_2}Z_{\phi} \braket{\braket{ K_{2}(p_1) \phi(p_2) \phi(p_3) }} = \braket{\braket{ K_{2,0}(p_1) \phi_0(p_2) \phi_0(p_3) }} \\
\Rightarrow \d_{K_2}^{(1)} = i \frac{\l}{2} L_1(-\m^2)\, ,
\eal
\ee
so that the renormalized 3-point function is given by
\be
\braket{\braket{ K_{2}(p_1) \phi(p_2) \phi(p_3) }} = 
-p_1^2 \frac{i}{p_2^2}\frac{i}{p_3^2} \left[2 +\frac{\l}{(4\pi)^2}\ln \left( \frac{-p_1^2}{\m^2} \right) + O(\l^2) \right]\, .
\ee
From the Callan-Symanzik equation we can extract the anomalous dimension
\be
\G_{K_2} = \frac{\l}{16\pi^2} + O(\l^2)\, .
\ee

\subsection{$O(\l^2)$ renormalization of $\left< K_2 \phi \phi \right>$}

We begin with the evaluation of the $O(\l^2)$ contributions to the bare 3-point function.
\be \bal
\braket{K_{2,0}(x) \phi_{0}(y) \phi_0(z) }_{O(\l^2)} =& \;  \Box_x \lim_{x_1 \rightarrow x} \left[ \begin{gathered} \begin{tikzpicture} 
     \begin{feynman}
     \vertex (x){$x$};
     \vertex[below =0.7cm of x](r1);
     \vertex[below =0.6cm of r1](r2);
     \vertex[below =2cm of x](u){$z$};
     \vertex[right=   0.5cm of x](y){$x_1$};
     \vertex[right = 0.5cm of  u](l){$y$};
     \diagram{
     (x)--(u)
     };
     \diagram{(r1)--[half left](r2)-- [half left](r1)};
     \diagram{
     (y)--(l)
     };
 \end{feynman} 
 \end{tikzpicture}
 \end{gathered}+ \begin{gathered} \begin{tikzpicture} 
     \begin{feynman}
     \vertex (x){$x_1$};
     \vertex[below =0.7cm of x](r1);
     \vertex[below =0.6cm of r1](r2);
     \vertex[below =2cm of x](u){$y$};
     \vertex[right=   0.5cm of x](y){$x$};
     \vertex[right = 0.5cm of  u](l){$z$};
     \diagram{
     (x)--(u)
     };
     \diagram{(r1)--[half left](r2)-- [half left](r1)};
     \diagram{
     (y)--(l)
     };
 \end{feynman} 
 \end{tikzpicture}
 \end{gathered} + \begin{gathered} \begin{tikzpicture} 
     \begin{feynman}
     \vertex (x){$x_1$};
     \vertex[below =0.7cm of x](r1);
     \vertex[below =0.6cm of r1](r2);
     \vertex[below =2cm of x](u){$z$};
     \vertex[right=   0.5cm of x](y){$x$};
     \vertex[right = 0.5cm of  u](l){$y$};
     \diagram{
     (x)--(u)
     };
     \diagram{(r1)--[half left](r2)-- [half left](r1)};
     \diagram{
     (y)--(l)
     };
 \end{feynman} 
 \end{tikzpicture}
 \end{gathered} + \begin{gathered} \begin{tikzpicture} 
     \begin{feynman}
     \vertex (x){$x$};
     \vertex[below =0.7cm of x](r1);
     \vertex[below =0.6cm of r1](r2);
     \vertex[below =2cm of x](u){$y$};
     \vertex[right=   0.5cm of x](y){$x_1$};
     \vertex[right = 0.5cm of  u](l){$z$};
     \diagram{
     (x)--(u)
     };
     \diagram{(r1)--[half left](r2)-- [half left](r1)};
     \diagram{
     (y)--(l)
     }; 
 \end{feynman} 
 \end{tikzpicture}
 \end{gathered}\right] \\
 &+\Box_x \lim_{x_1 \rightarrow x} \left[  \begin{gathered} \begin{tikzpicture} 
     \begin{feynman}
     \vertex (x){$x$};
     \vertex[below = 1cm of x](z1);
     \vertex[right = 0.5cm of z1](z11);
     \vertex[below = 2cm of x](u){$x_1$};
     \vertex[right=  2 cm of x](y){$y$};
     \vertex[below=1cm of y](z2);
     \vertex[left= 0.5cm of z2](z22);
     \vertex[right =  2cm of  u](l){$z$};
     \diagram{
     (x)--(z11)--[half left](z22)--[half left](z11)--(u)
     };
     \diagram{
     (y)--(z22)--(l)
     };
 \end{feynman} 
 \end{tikzpicture} \end{gathered} +\begin{gathered} \begin{tikzpicture} 
     \begin{feynman}
     \vertex (x){$x$};
     \vertex[below = 1cm of x](z1);
     \vertex[right = 0.5cm of z1](z11);
     \vertex[below = 2cm of x](u){$y$};
     \vertex[right=  2 cm of x](y){$x_1$};
     \vertex[below=1cm of y](z2);
     \vertex[left= 0.5cm of z2](z22);
     \vertex[right =  2cm of  u](l){$z$};
     \diagram{
     (x)--(z11)--[half left](z22)--[half left](z11)--(u)
     };
     \diagram{
     (y)--(z22)--(l)
     };
 \end{feynman} 
 \end{tikzpicture}
 \end{gathered}+
  \begin{gathered} \begin{tikzpicture} 
     \begin{feynman}
     \vertex (x){$x$};
     \vertex[below = 1cm of x](z1);
     \vertex[right = 0.5cm of z1](z11);
     \vertex[below = 2cm of x](u){$z$};
     \vertex[right=  2 cm of x](y){$x_1$};
     \vertex[below=1cm of y](z2);
     \vertex[left= 0.5cm of z2](z22);
     \vertex[right =  2cm of  u](l){$y$};
     \diagram{
     (x)--(z11)--[half left](z22)--[half left](z11)--(u)
     };
     \diagram{
     (y)--(z22)--(l)
     };
 \end{feynman} 
 \end{tikzpicture}
 \end{gathered} \right] 
 \eal \ee
We can identify two types of contributions, defined by the limit of each diagram. 
The first kind, is the Sunset Contribution and the second is the Candy  Contribution.\\
\textbf{Sunset contribution}
\be \bal
&  \Box_x \lim_{x_1 \rightarrow x} \left[ \begin{gathered} \begin{tikzpicture} 
     \begin{feynman}
     \vertex (x){$x$};
     \vertex[below =0.7cm of x](r1);
     \vertex[below =0.6cm of r1](r2);
     \vertex[below =2cm of x](u){$z$};
     \vertex[right=   0.5cm of x](y){$x_1$};
     \vertex[right = 0.5cm of  u](l){$y$};
     \diagram{
     (x)--(u)
     };
     \diagram{(r1)--[half left](r2)-- [half left](r1)};
     \diagram{
     (y)--(l)
     };
 \end{feynman} 
 \end{tikzpicture}
 \end{gathered}+ \begin{gathered} \begin{tikzpicture} 
     \begin{feynman}
     \vertex (x){$x_1$};
     \vertex[below =0.7cm of x](r1);
     \vertex[below =0.6cm of r1](r2);
     \vertex[below =2cm of x](u){$y$};
     \vertex[right=   0.5cm of x](y){$x$};
     \vertex[right = 0.5cm of  u](l){$z$};
     \diagram{
     (x)--(u)
     };
     \diagram{(r1)--[half left](r2)-- [half left](r1)};
     \diagram{
     (y)--(l)
     };
 \end{feynman} 
 \end{tikzpicture}
 \end{gathered} + \begin{gathered} \begin{tikzpicture} 
     \begin{feynman}
     \vertex (x){$x_1$};
     \vertex[below =0.7cm of x](r1);
     \vertex[below =0.6cm of r1](r2);
     \vertex[below =2cm of x](u){$z$};
     \vertex[right=   0.5cm of x](y){$x$};
     \vertex[right = 0.5cm of  u](l){$y$};
     \diagram{
     (x)--(u)
     };
     \diagram{(r1)--[half left](r2)-- [half left](r1)};
     \diagram{
     (y)--(l)
     };
 \end{feynman} 
 \end{tikzpicture}
 \end{gathered} + \begin{gathered} \begin{tikzpicture} 
     \begin{feynman}
     \vertex (x){$x$};
     \vertex[below =0.7cm of x](r1);
     \vertex[below =0.6cm of r1](r2);
     \vertex[below =2cm of x](u){$y$};
     \vertex[right=   0.5cm of x](y){$x_1$};
     \vertex[right = 0.5cm of  u](l){$z$};
     \diagram{
     (x)--(u)
     };
     \diagram{(r1)--[half left](r2)-- [half left](r1)};
     \diagram{
     (y)--(l)
     }; 
 \end{feynman} 
 \end{tikzpicture}
 \end{gathered}\right] = 2\frac{\l_0^2}{6}  \int \frac{-(k+q)^2\; e^{i(k+q)x}e^{-iky}e^{-iqz}}{k^4 l_1^2 l_2^2 (l_1+l_2 -k^2)q^2} 
 + \left( y \leftrightarrow z \right)
\eal
\ee
The momentum space expression is given by
\be
\bal
 \left< K_{2,0}(p_1)\phi_0(p_2) \phi_0(p_3) \right>_{(\text{sun})}&= 2 \frac{\l_0^2}{6}
 \int \frac{-(k+q)^2\; \d(p_1-k-q)\d(p_2+k)\d(p_3+q)}{k^4 l_1^2 l_2^2 (l_1+l_2 -k^2)q^2} + \left( p_2 \leftrightarrow p_3 \right)\\
&=2 \frac{\l_0^2}{6}  \frac{i}{p_2^2} \frac{i}{p_3^2} (p_2+p_3)^2 \left[ \frac{S_1(p_2^2)}{p_2^2} 
+ \frac{S_1(p_3^2)}{p_3^2}  \right] \d(p_1+p_2+p_3)\\
&=2 \frac{\l_0^2}{6}  \frac{i}{p_2^2} \frac{i}{p_3^2} p_1^2 \left[ \frac{S_1(p_2^2)}{p_2^2} 
+ \frac{S_1(p_3^2)}{p_3^2}  \right] \d(p_1+p_2+p_3)
\eal \ee
\textbf{Candy contribution}

The Candy contribution is given by the following limit:
\be
\bal
&\Box_x \lim_{x_1 \rightarrow x}  \begin{gathered} \begin{tikzpicture} 
     \begin{feynman}
     \vertex (x){$x$};
     \vertex[below = 1cm of x](z1);
     \vertex[right = 0.5cm of z1](z11);
     \vertex[below = 2cm of x](u){$x_1$};
     \vertex[right=  2 cm of x](y){$y$};
     \vertex[below=1cm of y](z2);
     \vertex[left= 0.5cm of z2](z22);
     \vertex[right =  2cm of  u](l){$z$};
     \diagram{
     (x)--(z11)--[half left](z22)--[half left](z11)--(u)
     };
     \diagram{
     (y)--(z22)--(l)
     };
 \end{feynman} 
 \end{tikzpicture} \end{gathered} = \frac{\l_0^2}{2} \int \frac{-\left(k_1+k_2\right)^2 \; e^{i(k_1 +k_2)x} e^{-iq_1y}e^{-iq_2z}}{k_1^2 k_2^2 l^2 \left(l-k_1-k_2 \right)^2 q_1^2 q_2^2}\tilde{\d}(k_1+k_2-q_1-q_2)
\eal
\ee

\be\label{candy contr k2 cross} 
\bal
\Box_x \lim_{x_1\rightarrow x}\begin{gathered} \begin{tikzpicture} 
     \begin{feynman}
     \vertex (x){$x$};
     \vertex[below = 1cm of x](z1);
     \vertex[right = 0.5cm of z1](z11);
     \vertex[below = 2cm of x](u){$y$};
     \vertex[right=  2 cm of x](y){$x_1$};
     \vertex[below=1cm of y](z2);
     \vertex[left= 0.5cm of z2](z22);
     \vertex[right =  2cm of  u](l){$z$};
     \diagram{
     (x)--(z11)--[half left](z22)--[half left](z11)--(u)
     };
     \diagram{
     (y)--(z22)--(l)
     };
 \end{feynman} 
 \end{tikzpicture}
 \end{gathered} =\frac{\l_0^2}{2} \int \frac{-\left(k_1-q_1\right)^2 \; e^{i(k_1 -q_1)x} e^{ik_2y}
 e^{-iq_2z}}{k_1^2 k_2^2 l^2 \left(l-k_1-k_2 \right)^2 q_1^2 q_2^2}\tilde{\d}(k_1+k_2-q_1-q_2) 
 \eal
 \ee
  Moving to momentum space and taking into account the crossing symmetric contribution of \eqref{candy contr k2 cross} we obtain
 \be
 \bal
 \left< K_{2,0}(p_1) \phi_0(p_2) \phi_0(p_3) \right>_{(\text{candy})}=
 \frac{i}{p_2^2} \frac{i}{p_3^2} p_1^2 \frac{\l_0^2}{2}\left\{ \left[L_1(p_1^2) \right]^2  -2D(p_1^2)\right\} \tilde{\d}(p_1+p_2+p_3)\, ,
 \eal
 \ee
where $D(p_1^2)$ is 
\be
D(p_1^2) = -\frac{1}{2} \int \frac{L_1(k-p_2) + L_1(k-p_3)}{k^2 (k+p_1)^2} \, .
\ee
The total $O(\l_0^2)$ contribution is then given by:
\be
 \bal
\left<\left< K_{2,0}(p_1) \phi_0(p_2)\phi_0(p_3)\right>\right>_{O(\l^2)} = p_1^2\frac{i}{p_2^2} \frac{i}{p_3^2}\left\{ \frac{\l_0^2}{2} \left[\left( L_1(p_1)\right)^2 -2 D(p_1^2)\right] +2\frac{\l_0^2}{6}\left[\frac{S_1(p_2)}{p_2^2} +  \frac{S_1(p_3)}{p_3^2} \right] \right\}
\eal
\ee
Comparing this result with \eqref{bare O2 phi phi order 2} we observe again that: 
\be
\left<\left< K_{2,0}(p_1) \phi_0(p_2)\phi_0(p_3)\right>\right>_{O(\l^2)} =-p_1^2 \left<\left< \mathcal{O}_{2,0}(p_1) \phi_0(p_2)\phi_0(p_3)\right>\right>_{O(\l^2)} 
\ee
The renormalization procedure was already completed in the previous section, so we can use the expression \eqref{CR l2} to obtain the renormalized 3-point function:
\be
\left<\left< K_{2}(p_1) \phi_0(p_2)\phi_0(p_3)\right>\right> =
- p_1^2 \frac{i}{p_2^2}\frac{i}{p_3^2}C_{\mathcal{O}_2 \phi \phi}^{\textbf{R} }(p_{1,2,3})
\ee
Using this relation, we can argue in fact that the $O(\l^2)$ anomalous dimension is equal to 
$\Gamma_{\mathcal{O}_2}$.\footnote{The box in this case does not contribute to the anomalous dimension of the operator $\phi^2$
(the same happens with $\phi$ and as a matter of fact with any operator of the form $\Box {\mathcal O}$). 
We could have used the statement that $\Delta_{\Box \mathcal{O}} = 2 + \Delta_{\mathcal{O}}$, 
which is equivalent to $\Gamma_{\Box \mathcal{O}} = \Gamma_{\mathcal{O}}$, to avoid the previous analysis. 
Nevertheless, we presented an explicit calculation that confirms this statement.}
\be
\G_{K_2} =\frac{\l}{(4\pi)^2} - \frac{5}{6} \frac{\l^2}{(4\pi)^4} + O(\l^3)
\ee

\section{The 3-point function $\braket{K_3 \phi \phi}$ }

In this section, we will study the 3-point function of the operator $K_3$.
Additionally, we will provide a proof  that the classical equation of motion persists in the quantum system within the 3-point function, 
resulting in the equivalence between the $K_3$ and $K_4$ as operators. 
First, we will prove this equivalence using Feynman diagrams up to order $O(\lambda_0^2)$, which has not been done. 
Then, we will employ the Schwinger-Dyson equation to demonstrate the expected equivalence to all orders. 
In addition we will present the evaluation of the bare 3-point function of the $K_1$ operator, 
in order to confirm the $F=0$ identity at the quantum level. This will allow us  to reduce the operator mixing to a $2 \times 2$ system.
Finally, we will solve the mixing problem. The solution is almost identical to the one presented 
in a previous section for the $\mathcal{O}_4$ operator. Our result for the mixing factor is consistent with 
that in \cite{Collins_Brown}, differing only in numerical constants due to different conventions.

\subsection{The bare correlator and the equation of motion}

We recall that the $K_{3,0}$ operator is defined as
\be
K_{3,0}= \lim_{x_1 \rightarrow x} \phi_0(x_1)\Box_x \phi_0(x) ,
\ee
consequently the 3-point function is given by:
\be
\braket{K_{3,0}(x) \phi_0(y) \phi_0(z)} = \lim_{x_1 \rightarrow x} \Box_x \braket{\phi_0(x) \phi_0(x_1) \phi_0(y) \phi_0(z)} 
- (\text{contact terms})
\ee 
One can show that the contact terms in this case are given by
\be
-i \d(x-y)\braket{\phi_0(x) \phi_0(z)}-i \d(x-z) \braket{\phi_0(x)\phi_0(y)}\, .
\ee
As we will see, the first non-vanishing contribution to $\braket{ K_{3,0} \phi_0 \phi_0}$ is of order $O(\l_0^2)$.

We begin with the  $O(\l^0)$ diagrams:
\be
\bal
\lim_{x_1 \rightarrow x} \Box_x \left[  \begin{gathered} \begin{tikzpicture} 
     \begin{feynman}
     \vertex (x){$x$};
     \vertex[below =2cm of x](u){$y$};
     \vertex[right=  0.5cm of x](y){$x_1$};
     \vertex[right =  0.5cm of  u](l){$z$};
     \diagram{
     (x)--(u)
     };
     \diagram{
     (y)--(l)
     };
 \end{feynman} 
 \end{tikzpicture}
 \end{gathered}+
 \begin{gathered} \begin{tikzpicture} 
     \begin{feynman}
     \vertex (x){$x$};
     \vertex[below =2cm of x](u){$z$};
     \vertex[right=   0.5cm of x](y){$x_1$};
     \vertex[right = 0.5cm of  u](l){$y$};
     \diagram{
     (x)--(u)
     };
     \diagram{
     (y)--(l)
     };
 \end{feynman} 
 \end{tikzpicture}
 \end{gathered} \right] &=  \int \frac{\mathrm{d}^d k \mathrm{d}^dq}{(2\pi)^{2d}} \frac{i^2}{k^2 q^2} \left(- q^2 \right) \left(e^{i(k+q)x}e^{-ky}e^{-qz} + ( y \leftrightarrow z) \right) \\
 &=- i \d(x-y) \Braket{\phi_0(x) \phi_0(z) } -i \d(x-z) \braket{\phi_0(x) \phi_0(y) } 
 \eal
\ee
The above terms constitute the $O(\l_0^0)$ contribution to $\lim_{x_1 \to x}\Box_x \braket{\phi_0 (x) \phi_0(x_1) \phi_0 (y) \phi_0(z)}$. These are contact terms and, as such, get cancelled, as we can see from the definition of $\braket{K_{3,0} \phi_0 \phi_0}$.

The $O(\l_0)$ contibution is given by:
\be\bal
 \lim_{x_1 \rightarrow x} \Box_x \left[\begin{gathered} \begin{tikzpicture} 
     \begin{feynman}
     \vertex (x){$x$};
     \vertex[below = 2cm of x](u){$x_1$};
     \vertex[right=  1cm of x](y){$y$};
     \vertex[right =  1cm of  u](l){$z$};
     \diagram{
     (x)--(l)
     };
     \diagram{
     (y)--(u)
     };
 \end{feynman} 
 \end{tikzpicture}
 \end{gathered} \right] &=  -i \l_0 \int \frac{\mathrm{d}^d k_{1,2} \mathrm{d}^dq_{1,2}}{(2\pi)^{4d}} \frac{i^4 (-k_1^2)}{k_1^2 k_2^2 q_1^2 q_2^2} e^{i(k_1+k_2)x}e^{-iq_1y}e^{-iq_2z} \tilde{\d}(k_1 + k_2-q_1 -q_2)\\
 &=i\l_0 \int \frac{\mathrm{d}^dk}{(2\pi)^d} \frac{1}{k^2} \int \frac{\mathrm{d}^d q_{1,2}}{(2\pi)^{2d}}\frac{ie^{iq_1(x-y)}}{q_1^2} \frac{i e^{iq_2(x-z)}}{q_2^2} =0
 \eal
 \ee
The vanishing of the above expression is a consequence of the loop integral being a scaleless DR integral.

We proceed with the $O(\l_0^2)$ calculation of the bare 3-point function.
As we have already discussed there are two different types of contributions to the $O(\l_0^2)$ diagrams. 
We begin with the one based on the Candy diagrams.\\

 \textbf{ Candy contributions:}
 
We begin with the channel of the total 3, the one which is crossing symmetric by itself:
\be
\bal
&\lim_{x_1 \rightarrow x} \Box_x
   \begin{gathered} \begin{tikzpicture} 
     \begin{feynman}
     \vertex (x){$x$};
     \vertex[below = 1cm of x](z1);
     \vertex[right = 0.5cm of z1](z11);
     \vertex[below = 2cm of x](u){$x_1$};
     \vertex[right=  2 cm of x](y){$y$};
     \vertex[below=1cm of y](z2);
     \vertex[left= 0.5cm of z2](z22);
     \vertex[right =  2cm of  u](l){$z$};
     \diagram{
     (x)--(z11)--[half left](z22)--[half left](z11)--(u)
     };
     \diagram{
     (y)--(z22)--(l)
     };
 \end{feynman} 
 \end{tikzpicture}
 \end{gathered} =  \frac{(-i\l_0)^2}{2}i^6 \int \frac{-k_1^2 e^{i(k_1 +k_2)x } e^{-iq_1 y } e^{-q_2 z}}{k_1^2 k_2^2 l_1^2 l_2^2 q_1^2q_2^2}\tilde{ \d} (k_1 +k_2-l_1 -l_2) \tilde{\d} (l_1 +l_2 -q_1 -q_2 )  \, ,
 \eal
 \ee
 The above integral is with respect to $k_i , q_i, l_i$ with $i=1,2$.
 Moving to momentum space we obtain:
 \be
 \bal
 &  i^2 \frac{\l_0^2}{2} \int \frac{\tilde{\d}(p_1-k_1-k_2)\tilde{ \d}(p_2+q_1) \tilde{\d} (p_3+q_2) \tilde{\d} (k_1 +k_2-l_1 -l_2) \tilde{\d} (l_1 +l_2 -q_1 -q_2 )}{ k_2^2 l_1^2 l_2^2 q_1^2q_2^2}\\
 &=\frac{\l^2}{2} \frac{i}{p_2^2} \frac{i}{p_3^2} \int \frac{\mathrm{d}^d l}{(2\pi)^d} \frac {\tilde{\d}(p_1+p_2+p_3)}{l^2 (l -p_1)^2}  \int\frac{\mathrm{d}^d k}{(2\pi)^d} \frac{1}{k^2}\to0
\eal
\ee
This contribution vanishes, since it is a scaleless integral.
Now we turn to the other two channels, whose sum preserves crossing symmetry.
We have
\be
\bal
&\lim_{x_1 \rightarrow x} \Box_x \begin{gathered} \begin{tikzpicture} 
     \begin{feynman}
     \vertex (x){$x$};
     \vertex[below = 1cm of x](z1);
     \vertex[right = 0.5cm of z1](z11);
     \vertex[below = 2cm of x](u){$y$};
     \vertex[right=  2 cm of x](y){$x_1$};
     \vertex[below=1cm of y](z2);
     \vertex[left= 0.5cm of z2](z22);
     \vertex[right =  2cm of  u](l){$z$};
     \diagram{
     (x)--(z11)--[half left](z22)--[half left](z11)--(u)
     };
     \diagram{
     (y)--(z22)--(l)
     };
 \end{feynman} 
 \end{tikzpicture}
 \end{gathered}+
  \begin{gathered} \begin{tikzpicture} 
     \begin{feynman}
     \vertex (x){$x$};
     \vertex[below = 1cm of x](z1);
     \vertex[right = 0.5cm of z1](z11);
     \vertex[below = 2cm of x](u){$z$};
     \vertex[right=  2 cm of x](y){$x_1$};
     \vertex[below=1cm of y](z2);
     \vertex[left= 0.5cm of z2](z22);
     \vertex[right =  2cm of  u](l){$y$};
     \diagram{
     (x)--(z11)--[half left](z22)--[half left](z11)--(u)
     };
     \diagram{
     (y)--(z22)--(l)
     };
 \end{feynman} 
 \end{tikzpicture}
 \end{gathered} =\\
 &= \frac{(-i\l_0)^2}{2} i^6 \int \frac{-k_1^2 e^{(k_1-q_1)x}e^{ik_2 y} e^{-iq_2z}}{k_1^2 k_2^2 l_1^2 l_2^2 q_1^2 q_2^2}\tilde{\d} (k_1 +k_2-l_1 -l_2)\tilde{\d} (l_1 +l_2 -q_1 -q_2 ) +(y \leftrightarrow z) 
 \eal
 \ee
In momentum space the above limit is given by 
 \be
 \bal
&\frac{\l_0^2}{2} i^2 \int \frac{\tilde{\d}(p_1-k_1+q_1)\d(p_2-k_2) \tilde{\d}(p_3+q_2)\tilde{\d} (k_1 +k_2-l_1 -l_2)\tilde{ \d} (l_1 +l_2 -q_1 -q_2 ) +(p_2 \leftrightarrow p_3)}{k_2^2 l_1^2 l_2^2 q_1^2 q_2^2} \\
&= \frac{\l_0^2}{2} \frac{i}{p_2^2} \frac{i}{p_3^2} \int \frac{1}{l_1^2 l_2^2 \left( l_1 +l_2 +p_3 \right)^2} \tilde{\d}(p_1 +p_2 +p_3)  +(p_2 \leftrightarrow p_3) \\
&=\frac{\l_0^2}{2} \frac{i}{p_2^2} \frac{i}{p_3^2} \left[S_1(p_2^2) + S_1(p_3^2) \right] \tilde{\d}(p_1 +p_2 + p_3)
\eal
\ee
We see that this contribution does not produce any contact terms.

\textbf{Sunset contribution:}

Taking into account all the  remaining $O(\l_0^2)$ diagrams, the sunset contribution is given by:
\be
\bal
&\lim_{x_1 \rightarrow x}\Box_x  \left[ \begin{gathered} \begin{tikzpicture} 
     \begin{feynman}
     \vertex (x){$x$};
     \vertex[below =0.7cm of x](r1);
     \vertex[below =0.6cm of r1](r2);
     \vertex[below =2cm of x](u){$z$};
     \vertex[right=   0.5cm of x](y){$x_1$};
     \vertex[right = 0.5cm of  u](l){$y$};
     \diagram{
     (x)--(u)
     };
     \diagram{(r1)--[half left](r2)-- [half left](r1)};
     \diagram{
     (y)--(l)
     };
 \end{feynman} 
 \end{tikzpicture}
 \end{gathered}   + \begin{gathered} \begin{tikzpicture} 
     \begin{feynman}
     \vertex (x){$x_1$};
     \vertex[below =0.7cm of x](r1);
     \vertex[below =0.6cm of r1](r2);
     \vertex[below =2cm of x](u){$y$};
     \vertex[right=   0.5cm of x](y){$x$};
     \vertex[right = 0.5cm of  u](l){$z$};
     \diagram{
     (x)--(u)
     };
     \diagram{(r1)--[half left](r2)-- [half left](r1)};
     \diagram{
     (y)--(l)
     };
 \end{feynman} 
 \end{tikzpicture}
 \end{gathered} + \begin{gathered} \begin{tikzpicture} 
     \begin{feynman}
     \vertex (x){$x$};
     \vertex[below =0.7cm of x](r1);
     \vertex[below =0.6cm of r1](r2);
     \vertex[below =2cm of x](u){$y$};
     \vertex[right=   0.5cm of x](y){$x_1$};
     \vertex[right = 0.5cm of  u](l){$z$};
     \diagram{
     (x)--(u)
     };
     \diagram{(r1)--[half left](r2)-- [half left](r1)};
     \diagram{
     (y)--(l)
     };
 \end{feynman} 
 \end{tikzpicture}
 \end{gathered}+ \begin{gathered} \begin{tikzpicture} 
     \begin{feynman}
     \vertex (x){$x_1$};
     \vertex[below =0.7cm of x](r1);
     \vertex[below =0.6cm of r1](r2);
     \vertex[below =2cm of x](u){$z$};
     \vertex[right=   0.5cm of x](y){$x$};
     \vertex[right = 0.5cm of  u](l){$y$};
     \diagram{
     (x)--(u)
     };
     \diagram{(r1)--[half left](r2)-- [half left](r1)};
     \diagram{
     (y)--(l)
     };
 \end{feynman} 
 \end{tikzpicture}
 \end{gathered}  \right]
 \eal
 \ee
In momentum space the limit gives
\be
\frac{\l_0^2}{6}  \frac{i}{p_2^2} \frac{i}{p_3^2}   \left[S_1(p_2^2) + S_1(p_3^2)  \right] 
\tilde{\d}(p_1+p_2+p_3)   + \left[   \frac{i \l_0^2}{6}\frac{i}{p_2^2}S_1(p_2^2)  \frac{i}{p_2^2} 
+\frac{i \l_0^2}{6}\frac{i}{p_3^2}S_1(p_3^2)  \frac{i}{p_3^2} \right] \tilde{\d}(p_1+p_2 +p_3)
\ee
The terms in the squared brackets in the above expression must be contact terms, since they 
contain the $\l_0^2$ corrections to the propagators. We can always check it explicitly by applying a Fourier transformation:
\be
\bal
&\int \frac{\mathrm{d}^dp_1}{(2\pi)^d} \frac{\mathrm{d}^dp_2}{(2\pi)^d}\frac{\mathrm{d}^dp_3}{(2\pi)^d}e^{ip_1 x}e^{ip_2 y} e^{ip_3 z}\frac{i \l_0^2}{6}\frac{i}{p_2^2}S_1(p_2^2)  \frac{i}{p_2^2}  \tilde{\d}(p_1+p_2 +p_3) =\\
&\int \frac{\mathrm{d}^dp_3}{(2\pi)^d} e^{ip_3 (z-x)}\int \frac{\mathrm{d}^dp_2}{(2\pi)^d}e^{ip_2 (y-x)}\frac{i \l_0^2}{6}\frac{i}{p_2^2}S_1(p_2^2)  \frac{i}{p_2^2}  =-i \d(x-z) \braket{\phi_0(x) \phi_0(y)}_{O(\l_0^2)}
\eal
\ee
As a result, the total $O(\l_0^2)$ contribution is given by 
\be
\braket{\braket{K_{3,0}(p_1)\phi_0(p_2) \phi_0(p_3) }} = 
\frac{2 \l_0^2}{3}\frac{i}{p_2^2} \frac{i}{p_3^2} \left[S_1(p_2^2) + S_1(p_3^2) \right]
\ee
Comparing the above result with \eqref{bare o4phiphi} we observe that, at $O(\l_0)$
\be
\left< \left<K_{3,0}(p_1)\phi_0(p_2) \phi_0(p_3) \right> \right> = 
-\frac{\l_0}{6} \braket{\braket{\mathcal{O}_{4,0}(p_1)\phi_0(p_2) \phi_0(p_3) }}\, .
\ee
Therefore, using the definition \eqref{K4 def} of the $K_{4,0}$ operator we confirm that to leading order
\be \label{K3 K4 equality}
K_{3,0} = - \frac{1}{6}K_{4,0}\, .
\ee
Of course this is not a coincidence. The Schwinger-Dyson equation implies that $K_4$ and $K_3$  operators are equivalent:
\be
\bal
\Box_x \braket{ \phi_0(x) \phi_0(x_1) \cdots \phi_0(x_n) } = 
&\braket{ \frac{\d S_{int}\left[\phi_0(x) \right]}{\d \phi_0} \phi_0(x_1)\cdots \phi_0(x_n) }
\\&-i\sum_j \d_{x,x_j} \braket{\phi_0(x_1) \cdots \phi_0(x_{j-1}) \phi_0(x_{j+1}) \cdots \phi_0(x_n) }\, ,
\eal
\ee
where the sum on the right hand side of the Schwinger-Dyson equation gives the contact terms. Considering the limit $x_1 \to x$ and using the definition of $K_{3,0}$ we conclude that
\be
K_{3,0}(x) = \lim_{x_1 \to x}  \frac{\d S_{int}\left[\phi_0(x) \right]}{\d \phi_0} \phi_0(x_1)\, ,
\ee 
which for the case of the $\l\phi^4$ theory becomes 
\be
K_{3,0}(x) = - \frac{1}{6} \l_0 \phi_0^4 \equiv- \frac{1}{6}K_{4,0}
\ee
to all orders. 
Thus, at the operator level we can safely use
\be\label{Eid}
E = K_3+\frac{1}{6} K_4 = 0\, .
\ee
This allows us to eliminate $K_4$ from the basis in favour of $K_3$.

\subsection{Bare 3-point function $\braket{K_{1,0} \phi_0\phi_0}$ and the $F$-identity}

Before proceeding to the renormalization of the 3-point function of $K_3$ we should clarify what is 
the basis of the independent operators with mass dimension equal to $d$, which participate in the mixing under renormalization. 
The one sure candidate  is the  operator $K_2$, that we already know that does not require any mixing for its renormalization. 
Although, this operator can participate in the renormalization of other $d$-dimensional operators. 
This means that the mixing matrix will have the following form
\be
Z_{IJ}=
\begin{bmatrix}
Z_{ K_2}  & \vec{0}  \\
\vdots & \ddots
\end{bmatrix}
\ee
The other candidate could be $K_1$, but this operator is not linearly independent form $K_2$ and $K_3$. 
This can be visualised with the use of the $F$-identity
\be
F_0 \equiv ( \partial_\m \phi_0)^2 + \phi_0 \Box \phi_0 - \frac{1}{2} \Box \phi_0^2 =0
\ee
whose classical version was shown earlier.
We will now confirm this identity diagrammatically up to $O(\l_0^2)$.

We will have to be careful with the limits and the derivatives. The 3-point function of $K_{1,0}$ is given by
\be
 \braket{ K_{1,0}(x) \phi_0(y) \phi_0 (z) } = \lim_{x_1 \rightarrow x} 
 \left[\left(\partial^{(x_1)} \cdot \partial^{(x)}  \right)\braket{ \phi_0(x_1)\phi_0(x) \phi_0(y) \phi_0 (z) }  \right] - ( \text{contact terms})
\ee
We begin with the $O(\l_0)$ contributions
\be
\bal
 &\lim_{x_1 \rightarrow x} \left(\partial^{(x_1)} \cdot \partial^{(x)}  \right) \left[   
 \begin{gathered} \begin{tikzpicture} 
     \begin{feynman}
     \vertex (x){$x$};
     \vertex[below =2cm of x](u){$y$};
     \vertex[right=  0.5cm of x](y){$x_1$};
     \vertex[right =  0.5cm of  u](l){$z$};
     \diagram{
     (x)--(u)
     };
     \diagram{
     (y)--(l)
     };
 \end{feynman} 
 \end{tikzpicture}
 \end{gathered}+
 \begin{gathered} \begin{tikzpicture} 
     \begin{feynman}
     \vertex (x){$x$};
     \vertex[below =2cm of x](u){$z$};
     \vertex[right=   0.5cm of x](y){$x_1$};
     \vertex[right = 0.5cm of  u](l){$y$};
     \diagram{
     (x)--(u)
     };
     \diagram{
     (y)--(l)
     };
 \end{feynman} 
 \end{tikzpicture}
 \end{gathered}  + \begin{gathered} \begin{tikzpicture} 
     \begin{feynman}
     \vertex (x){$x$};
     \vertex[below = 2cm of x](u){$x_1$};
     \vertex[right=  1cm of x](y){$y$};
     \vertex[right =  1cm of  u](l){$z$};
     \diagram{
     (x)--(l)
     };
     \diagram{
     (y)--(u)
     };
 \end{feynman} 
 \end{tikzpicture}
 \end{gathered}\right]=  \\
 &=\int\frac{i^2\left(-2 k\cdot q e^{i(k+q)x}e^{-iky}e^{-iqz} \right)}{k^2 q^2}  -i\l \int \frac{-k_1\cdot k_2 e^{i(k_1 +k_2)x} e^{-iq_1y}e^{-iq_2z}}{k_1^2 k_2^2 q_1^2 q_2^2}\tilde{\d}(k_1+k_2-q_1-q_2) 
 \eal
 \ee
We apply a Fourier transformation to obtain
 \be
 \bal
&  -2 (p_2 \cdot p_3) \frac{i}{p_2^2} \frac{i}{p_3^2}\tilde{ \d}(p_1+p_2+p_3)  - 
i \l \frac{p_1^2}{2} L_1(p_1) \frac{i}{p_2^2}\frac{i}{p_3^2} \tilde{\d}(p_1+p_2+p_3)\\&= 
- \frac{p_1^2}{2} \left[ 2\frac{i}{p_2^2} \frac{i}{p_3^2} + i\l L_1(p_1)\frac{i}{p_2^2} \frac{i}{p_3^2}  
\right]\tilde{\d}(p_1+p_2 +p_3)+ i \left[ \frac{i}{p_2^2} + \frac{i}{p_3^2} \right ] \tilde{\d}(p_1+p_2+p_3)
\eal
\ee
The last term in the above expression is actually a contact term. We can perform an inverse Fourier transformation to check it:
\be
\bal
\int\frac{\mathrm{d}^dp_{1,2,3}}{(2\pi)^{3d}} e^{ip_1x}e^{ip_2y} e^{ip_3z} & i \left[ \frac{i}{p_2^2} + \frac{i}{p_3^2} \right ] \tilde{\d}(p_1+p_2+p_3)= \\
  &=i \d(y-x) \left< \phi_0(z)\phi_0(x)\right>  +i \d(z-x) \braket{ \phi_0(y)\phi_0(x)}\, .
\eal
\ee
So we conclude that
\be
\bal
\left<K_{1,0}(p_1) \phi_0(p_2) \phi_0 (p_3) \right>_{O(\l)}  &= - \frac{p_1^2}{2}  \left[ 2\frac{i}{p_2^2} \frac{i}{p_3^2} + i\l L_1(p_1)\frac{i}{p_2^2} \frac{i}{p_3^2}  \right]\d(p_1+p_2 +p_3)\\
&= \frac{1}{2} \braket{ K_{2,0}(p_1) \phi_0(p_2) \phi_0(p_3)}_{O(\l_0)}\, .
\eal
\ee
This result is consistent with the $F$-identity, since the first non-vanishing contribution 
to $\braket{ K_{3,0}\phi_0 \phi_0 }$ is of order $O(\l_0^2)$.

The $O(\l_0^2)$ contribution to the bare 3-point function is given by the following limits:
\be
\bal
\left<K_{1,0}(x) \phi_0(y) \phi_0(z) \right>_{O(\l^2)} = &\lim_{x_1 \rightarrow x} \partial^{(x_1)}\cdot \partial^{(x)} \left[ \begin{gathered} \begin{tikzpicture} 
     \begin{feynman}
     \vertex (x){$x$};
     \vertex[below =0.7cm of x](r1);
     \vertex[below =0.6cm of r1](r2);
     \vertex[below =2cm of x](u){$z$};
     \vertex[right=   0.5cm of x](y){$x_1$};
     \vertex[right = 0.5cm of  u](l){$y$};
     \diagram{
     (x)--(u)
     };
     \diagram{(r1)--[half left](r2)-- [half left](r1)};
     \diagram{
     (y)--(l)
     };
 \end{feynman} 
 \end{tikzpicture}
 \end{gathered} + \begin{gathered} \begin{tikzpicture} 
     \begin{feynman}
     \vertex (x){$x$};
     \vertex[below =0.7cm of x](r1);
     \vertex[below =0.6cm of r1](r2);
     \vertex[below =2cm of x](u){$y$};
     \vertex[right=   0.5cm of x](y){$x_1$};
     \vertex[right = 0.5cm of  u](l){$z$};
     \diagram{
     (x)--(u)
     };
     \diagram{(r1)--[half left](r2)-- [half left](r1)};
     \diagram{
     (y)--(l)
     };
 \end{feynman} 
 \end{tikzpicture}
 \end{gathered}  + \begin{gathered} \begin{tikzpicture} 
     \begin{feynman}
     \vertex (x){$x_1$};
     \vertex[below =0.7cm of x](r1);
     \vertex[below =0.6cm of r1](r2);
     \vertex[below =2cm of x](u){$y$};
     \vertex[right=   0.5cm of x](y){$x$};
     \vertex[right = 0.5cm of  u](l){$z$};
     \diagram{
     (x)--(u)
     };
     \diagram{(r1)--[half left](r2)-- [half left](r1)};
     \diagram{
     (y)--(l)
     };
 \end{feynman} 
 \end{tikzpicture}
 \end{gathered} + \begin{gathered} \begin{tikzpicture} 
     \begin{feynman}
     \vertex (x){$x_1$};
     \vertex[below =0.7cm of x](r1);
     \vertex[below =0.6cm of r1](r2);
     \vertex[below =2cm of x](u){$z$};
     \vertex[right=   0.5cm of x](y){$x$};
     \vertex[right = 0.5cm of  u](l){$y$};
     \diagram{
     (x)--(u)
     };
     \diagram{(r1)--[half left](r2)-- [half left](r1)};
     \diagram{
     (y)--(l)
     };
 \end{feynman} 
 \end{tikzpicture}
 \end{gathered} 
 \right] \\
 +&\lim_{x_1 \rightarrow x} \partial^{(x_1)}\cdot \partial^{(x)} \left[  \begin{gathered} \begin{tikzpicture} 
     \begin{feynman}
     \vertex (x){$x$};
     \vertex[below = 1cm of x](z1);
     \vertex[right = 0.5cm of z1](z11);
     \vertex[below = 2cm of x](u){$x_1$};
     \vertex[right=  2 cm of x](y){$y$};
     \vertex[below=1cm of y](z2);
     \vertex[left= 0.5cm of z2](z22);
     \vertex[right =  2cm of  u](l){$z$};
     \diagram{
     (x)--(z11)--[half left](z22)--[half left](z11)--(u)
     };
     \diagram{
     (y)--(z22)--(l)
     };
 \end{feynman} 
 \end{tikzpicture} \end{gathered} +\begin{gathered} \begin{tikzpicture} 
     \begin{feynman}
     \vertex (x){$x$};
     \vertex[below = 1cm of x](z1);
     \vertex[right = 0.5cm of z1](z11);
     \vertex[below = 2cm of x](u){$y$};
     \vertex[right=  2 cm of x](y){$x_1$};
     \vertex[below=1cm of y](z2);
     \vertex[left= 0.5cm of z2](z22);
     \vertex[right =  2cm of  u](l){$z$};
     \diagram{
     (x)--(z11)--[half left](z22)--[half left](z11)--(u)
     };
     \diagram{
     (y)--(z22)--(l)
     };
 \end{feynman} 
 \end{tikzpicture}
 \end{gathered}+
  \begin{gathered} \begin{tikzpicture} 
     \begin{feynman}
     \vertex (x){$x$};
     \vertex[below = 1cm of x](z1);
     \vertex[right = 0.5cm of z1](z11);
     \vertex[below = 2cm of x](u){$z$};
     \vertex[right=  2 cm of x](y){$x_1$};
     \vertex[below=1cm of y](z2);
     \vertex[left= 0.5cm of z2](z22);
     \vertex[right =  2cm of  u](l){$y$};
     \diagram{
     (x)--(z11)--[half left](z22)--[half left](z11)--(u)
     };
     \diagram{
     (y)--(z22)--(l)
     };
 \end{feynman} 
 \end{tikzpicture}
 \end{gathered} \right] \\
 &- \left(\text{contact terms} \right)
\eal
\ee
Again there are the Sunset and Candy Contributions.\\

\textbf{Sunset contribution}

The sunset contribution is given by the following diagrams: 
\be
\bal 
&\lim_{x_1 \rightarrow x} \partial^{(x_1)}\cdot \partial^{(x)} \left[ \begin{gathered} \begin{tikzpicture} 
     \begin{feynman}
     \vertex (x){$x$};
     \vertex[below =0.7cm of x](r1);
     \vertex[below =0.6cm of r1](r2);
     \vertex[below =2cm of x](u){$z$};
     \vertex[right=   0.5cm of x](y){$x_1$};
     \vertex[right = 0.5cm of  u](l){$y$};
     \diagram{
     (x)--(u)
     };
     \diagram{(r1)--[half left](r2)-- [half left](r1)};
     \diagram{
     (y)--(l)
     };
 \end{feynman} 
 \end{tikzpicture}
 \end{gathered}+ \begin{gathered} \begin{tikzpicture} 
     \begin{feynman}
     \vertex (x){$x_1$};
     \vertex[below =0.7cm of x](r1);
     \vertex[below =0.6cm of r1](r2);
     \vertex[below =2cm of x](u){$y$};
     \vertex[right=   0.5cm of x](y){$x$};
     \vertex[right = 0.5cm of  u](l){$z$};
     \diagram{
     (x)--(u)
     };
     \diagram{(r1)--[half left](r2)-- [half left](r1)};
     \diagram{
     (y)--(l)
     };
 \end{feynman} 
 \end{tikzpicture}
 \end{gathered} + \begin{gathered} \begin{tikzpicture} 
     \begin{feynman}
     \vertex (x){$x_1$};
     \vertex[below =0.7cm of x](r1);
     \vertex[below =0.6cm of r1](r2);
     \vertex[below =2cm of x](u){$z$};
     \vertex[right=   0.5cm of x](y){$x$};
     \vertex[right = 0.5cm of  u](l){$y$};
     \diagram{
     (x)--(u)
     };
     \diagram{(r1)--[half left](r2)-- [half left](r1)};
     \diagram{
     (y)--(l)
     };
 \end{feynman} 
 \end{tikzpicture}
 \end{gathered} + \begin{gathered} \begin{tikzpicture} 
     \begin{feynman}
     \vertex (x){$x$};
     \vertex[below =0.7cm of x](r1);
     \vertex[below =0.6cm of r1](r2);
     \vertex[below =2cm of x](u){$y$};
     \vertex[right=   0.5cm of x](y){$x_1$};
     \vertex[right = 0.5cm of  u](l){$z$};
     \diagram{
     (x)--(u)
     };
     \diagram{(r1)--[half left](r2)-- [half left](r1)};
     \diagram{
     (y)--(l)
     }; 
 \end{feynman} 
 \end{tikzpicture}
 \end{gathered}\right] = \\ 
 &= \frac{(-i\l)^2}{6}i^6\left[ \int \frac{- (k\cdot q  )e^{i(k+q)x} e^{-iky}e^{-iqz}}{k^4 q^2 l_1^2 l_2^2 \left(l_1 +l_2-k\right)^2} +\int \frac{- (k\cdot q  )e^{i(k+q)x e^{-ikz}e^{-iqy}}}{k^4 q^2 l_1^2 l_2^2 \left(l_1 +l_2-k\right)^2} + \left( y \leftrightarrow z \right)\right] \eal \ee
We move to momentum and we get:
 \be
 \bal
 \left< K_{1,0}(p_1)\phi_0(p_2) \phi_0p_3)\right>_{sun}& =2\frac{(-i\l_0)^2}{6}i^6 \int \frac{- (k\cdot q )\tilde{\d}(p_1-k-q) \tilde{\d}(p_2+k)\tilde{\d}(p_3+q)}{k^4 q^2 l_1^2 l_2^2 \left(l_1 +l_2-k\right)^2} + \left( p_2 \leftrightarrow p_2 \right)\\
 &=  \frac{\l_0^2}{6} \left\{ p_1^2 \frac{i}{p_2^2}\frac{i}{p_3^2} \left[ \frac{S_1(p_2^2)}{p_2^2} + \frac{S_1(p_3^2)}{p_3^2}\right]  \right. \\
 & \left. -   \frac{i}{p_2^2}\frac{i}{p_3^2}  \left[S_1(p_2) + S_1(p_3) \right] - \frac{i}{p_2^2}S_1(p_2^2) \frac{i}{p_2^2} - \frac{i}{p_3^2}S_1(p_3^2) \frac{i}{p_3^2} \right\}\tilde{\d}(p_1 +p_2 +p_3 )
 \eal
 \ee
The last two terms are contact terms.To check this we perform a Fourier transformation back to position space:
\bea
&& -\frac{\l_0^2}{6}\int \mathrm{d}^d p_{1,2,3} \tilde{\d}(p_1 + p_2 +p_3)e^{ip_1x}e^{ip_2y} e^{ip_3z} 
\frac{i}{p_3^2}S_1(p_3) \frac{i}{p_3^2} \nonumber\\
&=& \frac{\l_0^2}{6}\int \mathrm{d}^dp_{2,3} e^{ip_2(y-x)} e^{ip_3(z-x)} 
\frac{i}{p_3^2}S_1(p_3) \frac{i}{p_3^2} = i\d(y-x) \left< \phi_0(z) \phi_0(x)\right>_{O(\l^2)}\, .
\eea
\\
\textbf{Candy contribution}

The first Candy contribution comes from :
\be
\bal
&\lim_{x_1 \rightarrow x} \partial^{(x_1)}\cdot \partial^{(x)} \left[  \begin{gathered} \begin{tikzpicture} 
     \begin{feynman}
     \vertex (x){$x$};
     \vertex[below = 1cm of x](z1);
     \vertex[right = 0.5cm of z1](z11);
     \vertex[below = 2cm of x](u){$x_1$};
     \vertex[right=  2 cm of x](y){$y$};
     \vertex[below=1cm of y](z2);
     \vertex[left= 0.5cm of z2](z22);
     \vertex[right =  2cm of  u](l){$z$};
     \diagram{
     (x)--(z11)--[half left](z22)--[half left](z11)--(u)
     };
     \diagram{
     (y)--(z22)--(l) 
     };
 \end{feynman} 
 \end{tikzpicture} \end{gathered} \right] = \\
 &= \frac{(-i\l_0)^2}{2}i^6\int \frac{-(k_1 \cdot k_2)e^{i(k_1 + k_2)x} e^{-iq_1y} e^{-iq_2z} \d(k_1+k_2 -q_1 -q_2)}{k_1^2 k_2^2 l^2 \left(k-k_1 -k_2 \right)^2q_1^2 q_2^2}
 \eal
 \ee
 After the Fourier transformation we arrive at
 \be
 \bal
 & \int \frac{\l^2}{2} \int \frac{-(k_1 \cdot k_2)\tilde{\d}(p_1-k_1-k_2) \d(p_2+q_1) \tilde{\d}(p_3 -q_2) \tilde{\d}(k_1 +k_2 -q_1 -q_2)}{k_1^2 k_2^2 l^2 \left(k-k_1 -k_2 \right)^2q_1^2 q_2^2}\\ 
 &=\frac{\l_0^2}{2} \frac{i}{p_2^2}\frac{i}{p_3^2} L_1(p_1) \int \frac{k \cdot(p_1-k)}{k^2 \left(p_1-k \right)^2} \tilde{\d}(p_1 +p_2 +p_3) \\ 
&= \frac{\l_0^2}{2} \frac{i}{p_2^2}\frac{i}{p_3^2} \frac{p_1^2}{2} \left[L_1(p_1) \right]^2 \tilde{\d}(p_1+p_2+p_3)
\eal
\ee  
In the last step we have used the 1-loop integral
\be
 \int \frac{k \cdot(p_1-k)}{k^2 \left(p_1-k \right)^2} =  \frac{p_1^2}{2} \int \frac{1}{k^2\left(k-p_1 \right)^2} = \frac{p_1^2}{2}L_1(p_1)
\ee
Now we consider the other two Candy contributions
\be\bal
&\lim_{x_1 \rightarrow x} \partial^{(x_1)}\cdot \partial^{(x)} \left[  \begin{gathered} \begin{tikzpicture} 
     \begin{feynman}
     \vertex (x){$x$};
     \vertex[below = 1cm of x](z1);
     \vertex[right = 0.5cm of z1](z11);
     \vertex[below = 2cm of x](u){$y$};
     \vertex[right=  2 cm of x](y){$x_1$};
     \vertex[below=1cm of y](z2);
     \vertex[left= 0.5cm of z2](z22);
     \vertex[right =  2cm of  u](l){$z$};
     \diagram{
     (x)--(z11)--[half left](z22)--[half left](z11)--(u)
     };
     \diagram{
     (y)--(z22)--(l)
     };
 \end{feynman} 
 \end{tikzpicture}
 \end{gathered}+
  \begin{gathered} \begin{tikzpicture} 
     \begin{feynman}
     \vertex (x){$x$};
     \vertex[below = 1cm of x](z1);
     \vertex[right = 0.5cm of z1](z11);
     \vertex[below = 2cm of x](u){$z$};
     \vertex[right=  2 cm of x](y){$x_1$};
     \vertex[below=1cm of y](z2);
     \vertex[left= 0.5cm of z2](z22);
     \vertex[right =  2cm of  u](l){$y$};
     \diagram{
     (x)--(z11)--[half left](z22)--[half left](z11)--(u)
     };
     \diagram{
     (y)--(z22)--(l)
     };
 \end{feynman} 
 \end{tikzpicture}
 \end{gathered} \right]= \\
 &= \frac{\l_0^2}{2} \int \frac{(k_1 \cdot q_1) e^{i(k_1-q_1)x} e^{ik_2 y}e^{-iq_2z} \tilde{\d}(k_1+k_2-q_1-q_2)}{k_1^2 k_2^2 q_1^2 q_2^2 l^2(l-k_1 -k_2)^2} +\left( y \leftrightarrow z \right) \eal \ee
Fourier transforming gives
 \be
 \bal
 & \frac{\l_0^2}{2} \frac{i}{p_2^2} \frac{i}{p_3^2} \int \frac{k\cdot(p_1-k) \d(p_1 +p_2 +p_3)}{k_1^2(p_1-k)^2 l^2 (l-k-p_2)^2} + \left( p_2 \leftrightarrow p_3 \right) = \\
  &= \frac{\l_0^2}{2} \frac{i}{p_2^2} \frac{i}{p_3^2} \left[-p_1^2 D(p_1^2) -\left(S_1(p_2)+S_1(p_3)\right) \right] \d(p_1+p_2+p_3)  
\eal \ee
where $D(p_1^2)$ has already been introduced in the renormalization of the coupling constant:
\be
D(p_1^2) = -\frac{1}{2} \int \frac{L_1(k-p_2) + L_1(k-p_3)}{k^2 (k+p_1)^2} \, .
\ee
The full $O(\l_0^2)$ contribution is given by :
\be
\bal
\left<\left<K_{1,0}(p_1) \phi_0(p_2) \phi_0(p_3) \right>\right>_{O(\l_0^2)}  = &p_1^2 \frac{i}{p_2^2}\frac{i}{p_3^2} \left[\frac{\l_0^2}{2} \frac{\left[L_1(p_1^2)\right]^2}{2} - \frac{\l_0^2}{2}D(p_1^2) + \frac{\l_0^2}{6} \left(\frac{S_1(p_2^2)}{p_2^2} + \frac{S_1(p_2^3)}{p_2^3} \right)  \right]\\
&- \frac{2\l_0^2}{3} \frac{i}{p_2^2}\frac{i}{p_3^2} \left[ S_1(p_2) + S_1(p_3)\right]
\eal
\ee
and the full bare correlation function up to $O(\l_0^2)$ is given by:
\be
\bal
\braket{\braket{K_{1,0}(p_1) \phi_0(p_2) \phi_0(p_3) }} = &p_1^2 \frac{i}{p_2^2}\frac{i}{p_3^2} 
\left[-1 -i\frac{\l_0}{2}L_1(p_1^2) + \frac{\l_0^2}{2} \frac{\left[L_1(p_1^2)\right]^2}{2} - 
\frac{\l_0^2}{2}D(p_1^2) \right. \\
 & \left.+ \frac{\l_0^2}{6} \left(\frac{S_1(p_2^2)}{p_2^2} + \frac{S_1(p_2^3)}{p_2^3} \right)  \right]
- \frac{2\l_0^2}{3} \frac{i}{p_2^2}\frac{i}{p_3^2} \left[ S_1(p_2) + S_1(p_3)\right] \\
=& \frac{1}{2} \braket{\braket{K_{2,0}(p_1) \phi_0(p_2) \phi_0(p_3) }} - \braket{\braket{K_{3,0}(p_1) \phi_0(p_2) \phi_0(p_3) }}\, .
\eal
\ee
This allows us to promote the identity to the operator level:
\be
K_1 = \frac{1}{2}K_2 - K_3
\ee
and eliminate $K_1$ from the basis. Together with \eqref{Eid} this leaves as the independent basis operators
$K_2$ and $K_3$. 
Moreover we have seen that $K_2$ does not mix with any other operator. It remains to see whether
$K_3$ mixes. If it does, the above analysis ensures that it can be only with $K_2$.

\subsection{Renormalization of $\braket{K_3\phi\phi}$ and mixing}

Using the expression of the 3-point function of $\mathcal{O}_4$ in terms of the renormalized coupling constant $\l$   
and the relation \eqref{K3 K4 equality} we directly obtain the expression of the bare 3-point function 
in terms of the renormalized coupling constant: 
\be \label{bare K3 phi phi}\bal
\left< \left< K_{3,0} (p_1)\phi_0(p_2) \phi_0 (p_3) \right> \right>=-\frac{1}{6} \frac{i}{p_2^2} \frac{i}{p_3^2}& \left\{-4 \l^2  \left[S_1(p_2^2) + S_1(p_3^2)\right] \right.\\
&+i6 \l^3 L_1(-\m^2) \left[S_1(p_2^2) + S_1(p_3^2)\right]\\
& -i6\l^3 ST(p_1^2) -i\frac{3}{2}\l^3 \left[ TB(p_2^2) + TB(p_3^2)\right]\\
&\left. -i 6\l^3 \left[T(p_2,p_3) +\left( p_2 \leftrightarrow p_3  \right)\right]
  \right\}
\eal
\ee
We note here that in the above expression the term $ST(p_1^2)$ is proportional 
to $p_1^2$ for $d \rightarrow 4$. Using \eqref{st general d} we can write

\be \label{K3 phi phi sep} \bal
&\braket{\braket {K_{3,0} (p_1)\phi_0(p_2) \phi_0 (p_3) }} =  
p_1^2 \frac{i}{p_2^2} \frac{i}{p_3^2} i \l^3 \frac{ST(p_1^2)}{p_1^2}\\
 &+i \l^2 \frac{i}{p_3^2} \left\{ \frac{2}{3}  \frac{ S_1(p_2^2)}{p_2^2} - 
 i \l L_1(-\m^2)\frac{S_1(p_2^2)}{p_2^2} +i\frac{1}{4}\l  \frac{TB(p_2^2)}{p_2^2} +
 i \l \frac{T(p_2,p_3)}{p_2^2}\right\} +\left( p_2 \leftrightarrow p_3  \right)
\eal
\ee
As in the case of $\mathcal{O}_4$ operator, the first term of the above expression is responsible 
for the mixing under renormalization of $K_3$ and we know that it can happen only with $K_2$. 
As the independent bare basis we define the vector (in the reduced basis we use the same indices but now $I,J=1,2$)
 \be
 Q_{I,0} = \begin{bmatrix}
 K_{2,0}\\
 K_{3,0}
 \end{bmatrix}
 \ee
 The renormalized basis is defined accordingly as 
 \be
 Q _{J}= \begin{bmatrix}
 K_{2}\\
 K_{3}
 \end{bmatrix}
 \ee
 and the relation between the bare and the renormalized vector is dictated by the renormalization matrix
 \be
 Q_{I,0} = Z_{IJ}Q_J\, .
 \ee
Taking into account that $K_2$ is renormalized by itself, the mixing matrix has the following form:
 \be
 Z_{IJ} = \begin{bmatrix}
 Z_{K_2} & 0 \\
 Z_{32} & Z_{K_3}\, .
 \end{bmatrix}
 \ee
Our goal is to determine $Z_{K_3}$ and $Z_{32}$ through consistency relations.
We write the Callan-Symanzik equation in the following form:
\be
\left[ \m \frac{\partial}{\partial \m} + \b_\l \frac{\partial}{\partial \l} + 2 \g_\phi \right] \braket{\braket{K_I \phi \phi}} 
+ \G_{IJ} \braket{\braket{K_J \phi \phi}} =0 \; ,
\ee
where $\G_{IJ}$ is the anomalous dimension matrix.  
Using the form of the $Z_{IJ }$ matrix we obtain the following relation between the bare 3-point 
function of $K_3$ and the renormalized correlation functions of $K_2$ and $K_3$.
\be
Z_{K_3}Z_{\phi} \braket{K_3 \phi \phi} = \braket{K_{3,0} \phi_0 \phi_0} - Z_{32}Z_{\phi}\braket{K_2 \phi \phi}
\ee
The renormalization condition for the 3-point function  of $K_3$ is :
\be \label{K3 renorm cond}
\braket{\braket {K_3(p_1) \phi(p_2) \phi (p_3) }} = \frac{2}{3} \l^2\frac{i}{p_2^2} 
\frac{i}{p_3^2}(p_2^2 + p_3^2) \; ,\;  \text{at $S.P$ ($p_1^2 =p_2^2 =p_3^2= -\m^2$)}
\ee
Recalling the renormalized 3-point function of $K_2$
\be
\left< \left< K_2(p_1) \phi(p_2) \phi(p_3) \right> \right> = 
-p_1^2 \frac{i}{p_2^2} \frac{i}{p_3^2} C^{\mathbf{R}}_{\mathcal{O}_2\phi \phi}(p_{1,2,3}) \; ,
\ee
with  $ C^{\mathbf{R}}_{\mathcal{O}_2 \phi \phi}$ given by \eqref{CR l2}, we get
\bea
\braket{\braket{ K_2(p_1) \phi(p_2) \phi(p_3) }} &= &-2p_1^2 \frac{i}{p_2^2}\frac{i}{p_3^2}  
\Bigl\{1 + \frac{\l}{2(4\pi)^2}  \ln \left( \frac{-p_1^2}{\m^2}\right)+ \frac{\l^2}{2(4\pi)^4}  \ln^2 \left( \frac{-p_1^2}{\m^2}\right) \nonumber\\
&-& \frac{\l^2}{2(4\pi)^4}  \ln \left( \frac{-p_1^2}{\m^2}\right) + \frac{\l^2}{12 (4\pi)^4} 
\left[  \ln \left( \frac{-p_2^2}{\m^2}\right)   +  \ln \left( \frac{-p_3^2}{\m^2}\right)\right] \nonumber\\
 &-& \frac{\l^2}{2(4\pi)^4}\hat{G}(p_1,p_2,p_3) \; \, \Bigr\}
\eea
The renormalization condition \eqref{K3 renorm cond} implies the following set of equations:
\begin{eqnarray}
\left. - Z_{32} C^{\mathbf{R}}_{\mathcal{O}_2 \phi \phi} (p_{1,2,3}) \right|_{\text{S.P.}}   + \left. i\l^3 \frac{ST(p_1^2)}{p_1^2}  \right|_{\text{S.P.}}  =0 \\
\label{Zk3 eq}   \left\{ \frac{2}{3}  \frac{ S_1(p_2^2)}{p_2^2} - i \l L_1(-\m^2)\frac{S_1(p_2^2)}{p_2^2} +i\frac{1}{4}\l  \frac{TB(p_2^2)}{p_2^2} +i \l \frac{T(p_2,p_3)}{p_2^2}\right\} _{\text{S.P.}} = \frac{2}{3}Z_{K_3}Z_\phi
\end{eqnarray}
Recalling also that $\left. C^{\mathbf{R}}_{\mathcal{O}_2 \phi \phi} (p_{1,2,3}) \right|_{\text{S.P.}}  = 2$  we can solve for $Z_{32}$
and obtain
\be
Z_{32} = \frac{ \l^3}{2} \frac{ST(-\m^2)}{-\m^2}\, .
\ee
Taking into account that $Z_\phi =1 + O(\l^2)$ we can solve \eqref{Zk3 eq} for $Z_{K_3}$ and we obtain the following expression:
\be
Z_{K_3} = \left\{   \frac{ S_1(p_2^2)}{p_2^2} - i \frac{3}{2} \l L_1(-\m^2)\frac{S_1(p_2^2)}{p_2^2} 
+i\frac{3}{8}\l  \frac{TB(p_2^2)}{p_2^2} +i \frac{3}{2} \l \frac{T(p_2,p_3)}{p_2^2}\right\} _{\text{S.P.}} + O(\l^2)
\ee
Summing all the contributions we have
\be \bal
\left< \left< K_{3}(p_1) \phi(p_2) \phi(p_3) \right> \right> &=  \frac{2\l}{3}\frac{i}{p_2^2}\frac{i}{p_3^2} \left\{p_2^2 
\left[ \l + \frac{9}{2} \frac{\l^2}{16\pi^2} \ln \left( \frac{-p_2^2}{\m^2}\right) \right] + 
\left(p_2 \leftrightarrow p_3 \right) \right\}\\
& - p_1^2 \frac{i}{p_2^2} \frac{i}{p_3^2} \frac{\l^3}{4 (4\pi)^6} \ln \left( \frac{-p_1^2}{\m^2} \right)
\eal \ee
Now we check the Callan-Symanzik equation:
\be\bal
\m \frac{\partial}{\partial \m}\left< \left< K_3(p_1) \phi (p_2) \phi(p_3) \right>\right>&=- \frac{9\l^2}{16\pi^2} \left[ \frac{2\l}{3} \frac{i}{p_2^2}\frac{i}{p_3^2}  (p_2^2 +p_3^2) \right]  + \frac{\l^3}{4 (4\pi)^6} p_1^2 \frac{i}{p_2^2} \frac{i}{p_3^2}+ O(\l^4) \\
\b_\l \partial_\l\left< \left< K_3(p_1) \phi (p_2) \phi(p_3) \right>\right>&=\left[ \frac{3\l}{16\pi^2} + \frac{\b_\l}{\l}\right] \left[  \frac{2 \l}{3}\frac{i}{p_2^2}\frac{i}{p_3^2} (p_2^2 +p_3^2) \right] + O(\l^4)\\
2\g_\phi \left< \left< K_3(p_1) \phi (p_2) \phi(p_3) \right>\right> &= O(\l^4)
\eal \ee
Using the relation
\be
 \frac{\l^3}{4 (4\pi)^6} p_1^2 \frac{i}{p_2^2} \frac{i}{p_3^2} = - \frac{\l^3}{4 (4\pi)^6} \braket{\braket{ K_2(p_1) \phi(p_2) \phi(p_3) }}
\ee
we have that
\be
\bal
\left[\m \frac{\partial}{\partial \m} + \b_\l \partial_\l \right]\braket{\braket{ K_3(p_1) \phi (p_2) \phi(p_3) }} 
&=\left( - \frac{6\l}{(4\pi)^2} + \frac{\b_\l}{\l}\right) \braket{\braket{ K_3(p_1) \phi (p_2) \phi(p_3) }} \\
& -\frac{\l^3}{4 (4\pi)^6} \braket{\braket{ K_2(p_1) \phi(p_2) \phi(p_3) }} 
\eal
\ee
This fixes  $\mathcal{C} =2$ in the expression \eqref{Gamma mix o4} for the $\G_{mix}$  of $\mathcal{O}_4$ operator.
Using also $\G_{\mathcal{O}_4} =\frac{6\l}{(4\pi)^2} + O(\l^2)$ the above  Callan-Symanzik equation can be written as
\be
\left[\m \frac{\partial}{\partial \m} + \b_\l \partial_\l +\left( \G_{\mathcal{O}_4}- \frac{\b_\l}{\l}\right) 
+ 2\g_\phi \right]\braket{\braket{ K_3(p_1) \phi (p_2) \phi(p_3) }}  + 
\frac{\l^3}{4 (4\pi)^6} \braket{\braket{ K_2(p_1) \phi(p_2) \phi(p_3) }} =0\, .
\ee
The above expression determines the elements of the $\G$-matrix as:
\be \label{G-matrix}
\G_{ij}= \begin{bmatrix}
\G_{K_2} & 0 \\
\frac{\l^3}{4(4\pi)^6} & \G_{\mathcal{O}_4} - \frac{\b_\l}{\l}
\end{bmatrix} = \begin{bmatrix}
\G_{K_2} & 0 \\
\frac{\l^3}{4(4\pi)^6} & \l \frac{\partial}{\partial \l} \left( \frac{\b_\l}{\l} \right)
\end{bmatrix}\, .
\ee
Τhis result is in agreement with the one presented in \cite{Collins_Brown}.

 \section{Construction of $\Theta$ and $\braket{\Theta \phi \phi}$}
 
 In the previous sections we computed the 3-point functions of the operators of dimension $d$
 and showed that the basis of such operators consists of $K_2$ and $K_3$. 
 As stated in the introduction, one of the final goals of this work
 is to define in a proper way the renormalized operator of the trace $\Theta$ of the EMT. 
 The constraints on the construction of this operators are the following:
 \begin{enumerate}
 \item The EM trace operator should vanish when the system reaches the fixed point.  This implies that
\be
\Theta \sim \b_\l
\ee
 \item $\Theta$ is an operator of mass-dimension $d$. 
 Combined with the above constraint we  can write
 \be \label{linear combination def}
 \Theta = \b_\l \left[ a(\l) K_2 +  b(\l) K_3  \right]
 \ee
 where $a(\l)$ and $b(\l)$ are some functions of $\l$ to be determined.
 \item The operator $\Theta$ has (to all oders) vanishing anomalous dimension, which implies that the 3-point function 
 of this operator should obey the Callan-Symanzik equation
  \be \label{Theta 3pt CS}
  \left[\m \frac{\partial}{\partial \m} + \b_\l \frac{\partial}{\partial \l} + 
  2\g_\phi  \right] \braket{\braket{ \Theta (p_1) \phi (p_2) \phi(p_3) }} \equiv \hat{R}\braket{\braket{ \Theta (p_1) \phi (p_2) \phi(p_3)}} =0
  \ee
 \end{enumerate}
Recalling the Callan-Symanzik equations obeyed by the 3-point functions of $K_2$ and $K_3$
  \begin{align}
  \hat{R} \braket{\braket{K_2(p_1) \phi(p_2) \phi(p_3) }}& = - \G_{K_2}\braket{\braket{ K_2(p_1) \phi(p_2) \phi(p_3) }} \\
  \hat{R} \braket{\braket{ K_3(p_1) \phi(p_2) \phi(p_3) }} &= - \l \frac{\partial}{\partial \l} \left( \frac{\b_\l}{\l} \right) \braket{\braket{ K_3(p_1) \phi(p_2) \phi(p_3)}}  - \G_{32}\braket{\braket{ K_2(p_1) \phi(p_2) \phi(p_3) }}
 \end{align}
 and from the definition \eqref{linear combination def}, we obtain
  \be \bal
  \hat{R} \braket{\braket{\Theta (p_1) \phi (p_2) \phi(p_3)}} = 
 &\left[ -\b_\l b(\l) \l \frac{\partial}{\partial \l} \left( \frac{\b_\l}{\l} \right) + 
 \b_\l \frac{\partial}{\partial \l } \left(\b_\l b(\l) \right)  \right]\braket{\braket{ K_3 (p_1) \phi (p_2) \phi(p_3) }} \\
&+ \left[\b_\l \frac{\partial}{\partial \l} \left(\b_\l a(\l) \right) - \b_\l b(\l)\G_{32} - 
\b_\l a(\l)\G_{K_2}  \right]\braket{\braket{ K_2 (p_1) \phi (p_2)\phi(p_3)}}
  \eal\ee
whose right hand side should vanish, by \eqref{Theta 3pt CS}. 
From the vanishing of the bracket multiplying the 3-point function of the $K_3$ operator we have
\be
\bal
-\b_\l b(\l) \l \frac{\partial}{\partial \l} \left( \frac{\b_\l}{\l} \right) + \b_\l \frac{\partial}{\partial \l } \left(\b_\l b(\l) \right) =0
\\ \Rightarrow \frac{1}{b(\l)} \frac{\partial b(\l)}{\partial \l} = - \frac{1}{\l} \\
\Rightarrow  { b(\l) = \frac{c}{\l}}
\eal
\ee
where $c$ is an integration constant.
Having this solution we proceed with the bracket multiplying the 3-point function $K_2$.
The condition is 
 \be
 \bal
 \b_\l \frac{\partial}{\partial \l} \left(\b_\l a(\l) \right) - \b_\l b(\l)\G_{32} - \b_\l a(\l)\G_{K_2} =0 \\
 \Rightarrow \b_\l \frac{\partial a(\l)}{\partial \l} + \left[ \frac{\partial \b_\l}{\partial \l} - \G_{K_2}\right] a(\l) - \frac{c \G_{32}}{\l} =0
\eal
 \ee
 and we will solve it by an oder by order calculation. We write
 \begin{align}
 & \b_\l = \sum_{n=2}^{\infty} b_n \l^n \rightarrow b_2 = \frac{3}{(4\pi)^2} \; , \; b_3 = -\frac{17}{3(4\pi)^6}  \\
  &\G_{K_2}= \sum_{m=1}^{\infty} \g_{m} \l^m \rightarrow \g_1 = \frac{1}{(4\pi)^2} \; , \; \g_2 = \frac{5}{6(4\pi)^4} \\ 
  &\G_{32}= \sum_{r=3}^{\infty} g_{r} \l^r \rightarrow g_3 = \frac{1}{4 (4\pi)^6}  \\ 
  &a(\l) = \sum_{\xi =0}^{\infty} a_{\xi} \l^\xi
  \end{align} 
 and then the differential equation becomes
  \be
   \sum_{n=2}^{\infty} b_n \l^n \sum_{\xi=0}^\infty \xi a_\xi \l^{\xi-1}  + 
   \left\{\sum_{n=2}^\infty n b_n \l^{n-1} -  \sum_{m=1}^{\infty} \g_{m} \l^m \right\}\sum_{\xi =0}^{\infty} 
   a_{\xi} \l^\xi - c \sum_{r=3}^{\infty} g_{r} \l^{r-1} =0
  \ee
The above relation should hold at each order of $\l$. Then,
\be
O(\l):  \; 2b_2 \l a_0 =0\Rightarrow a_0=0\, .
\ee
We use this result and proceed to $O(\l^2)$, where we have
\be
  \bal
  O(\l^2): \; & b_2 \l^2 a_1 + 2b_2 a_1 \l^2 - \g_1 a_1 \l^2 -c g_3 \l^2 =0 \\
  &\Rightarrow a_1 (3 b_2 -\g_1) =c  g_3\\
  &\Rightarrow a_1 =c \frac{1}{32 (4\pi)^4}
  \eal
\ee
We stop at this order because, for the $O(\lambda^3)$ term, we would need the value of $g_4$, which we do not have.
The conclusion of the analysis is that the trace operator which obeys a "diagonal" (in the sense of no mixing) Callan-Symanzik equation is:
\bea \label{Trace zero AD}
\Theta &=& c\frac{\b_\l}{\l} K_3 + c \b_\l \left[\frac{\l}{32 (4\pi)^4}  + O(\l^2) \right] K_2 \nonumber\\
           &=& -c \frac{1}{6}\b_\l \phi^4 + c \b_\l \left[\frac{\l}{32 (4\pi)^4}  + O(\l^2) \right] \Box \phi^2\, .
\eea 
For the rest of the analysis we will take the integration constant $c$ equal to 1. 
The 3-point function of $\Theta$ can be now easily obtained from those of $K_2$ and $K_3$. It is given by:  
  \be\label{Thetaphi2}
  \bal
\braket{\braket{ \Theta (p_1) \phi(p_2) \phi(p_3) }} = 
-& \b_\l p_1^2\frac{i}{p_2^2}\frac{i}{p_3^2} \left[\frac{\l}{16(4\pi)^4} - 
\frac{17}{3} \frac{\l^2}{48 (4\pi)^6} + \frac{27}{3} \frac{\l^2}{32 (4\pi)^6} \ln \left(\frac{-p_1^2}{\m^2} \right) \right] \\
   &+ i \b_\l \frac{i}{p_3^2} \left[ \frac{2}{3} \frac{\l}{(4\pi)^2} - \frac{32}{3} \frac{\l^2}{9 (4\pi)^2} +
   \frac{3 \l^2}{(4\pi)^2} \ln \left( \frac{-p_2^2}{\m^2} \right) \right] + (p_2 \leftrightarrow p_3)
  \eal
  \ee
 Equations \eqref{Trace zero AD} and \eqref{Thetaphi2} are two of the main results of this paper.
 Up to an irrelevant overall constant the expression in \eqref{Trace zero AD} agrees in the first term 
 with that in \cite{Collins_Brown} but we find a difference in the second term, which in our case is 
 proportional to $\b_\l$.
  Both definitions of the trace \eqref{Trace zero AD} and the one in \cite{Collins_Brown} 
 (with the component along $K_2=\Box \phi^2$ missing, i.e. with $\Theta \sim \beta_\l K_3$) 
 are consistent with a vanishing anomalous 
 dimension $\Gamma_{\Theta}$, with the latter 
 however satisfying a "non-diagonal" Callan-Symazik equation.
By non-diagonal it is meant that in the Callan-Symanzik equation of the correlator 
$\braket{\Theta\Theta}$ a contribution proportional to a non-diagonal entry of the $K_2-K_3$ mixing matrix appears.
In the basis of \eqref{Trace zero AD} this mixing term is absent thus making the Callan-Symanzik equation 
diagonal. In the following we will see that in both bases the eigenvalue $e_\Theta$ 
remains the same to leading order in $\l$.
This is due to the fact that the component along $K_2$ gives a contribution to $e_\Theta$ that is one order higher 
than what $K_3$ contributes. 
In fact, the leading order value of $e_\Theta$ is determined entirely by the $K_3$ component.

\section{The 2-point function $\braket{\Theta\Theta}$}

The goal of this section is to derive an expression for the 2-point function $\braket{\Theta\Theta}$ and 
a value for the eigenvalue $e_\Theta$ in \eqref{eTheta}.
In order to achieve this we have to compute first the 2-point functions involving the basis operators:
\be
\bal
\left<K_2 K_2\right>,\,\,
\left<K_3 K_3 \right>,\,\,
\left< K_3 K_2 \right>
\eal
\ee
The analysis of these correlation functions will be similar to one of the 3-point functions. 
Of course, the road to the result involves mainly the derivatives and the limits for the 4-$\phi$ system, 
as we are interested only in the $K_2$ and $K_3$ operators, however
the equivalence between $K_3$ and $K_4$ will bring in the discussion also the correlators 
$\braket{\phi^4 \phi^4}$ and $\braket{\phi^4 \phi^2}$.

In the analysis of the 3-point function, we solved the mixing problem between the 
$K_2$ and $K_3$ operators. We reproduced the anomalous dimension matrix, 
which encodes the information of the scaling dimensions of the operators through its eigenvalues. 
In fact, the eigenvalues of the anomalous dimension matrix $\G$ is independent of the type of correlation 
functions from which it is derived. On the other hand, the mixing matrix of the counterterms thus also generic elements 
of $\G$ can depend on the type of correlation function being analyzed, as its elements are evaluated using the 
loop diagrams for that specific $n$-point function. 

The Callan-Symanzik equation of a general 2-point function between two $K$-operators can be written as:
\be
\left[ \m \frac{\partial}{\partial \m} + \b_\l \frac{\partial}{\partial \l}    \right] \braket{ K_I K_J } +
\sum_{M} \G_{IM}\braket{ K_{M}K_J} + \sum_M \G_{JM} \braket{K_I K_M } = 0\, .
\ee
Using the form of the anomalous dimension matrix \eqref{G-matrix} we obtain the 
Callan-Symanzik equation for each two point function:
\begin{align}
\label{k2k2 cs}&\left[ \m \frac{\partial}{\partial \m} + \b_\l \frac{\partial}{\partial \l}    \right] \braket{ K_2 K_2 } + 2\G_{K_2} \braket{K_2 K_2 }=0\\
\label{k2k3 cs}&\left[ \m \frac{\partial}{\partial \m} + \b_\l \frac{\partial}{\partial \l}    \right] \braket{ K_3 K_2 }  + \G_{K_2} \braket{ K_3 K_2 }+ \G_{32} \braket{ K_2 K_2 } + \G_{K_3}  \braket{ K_3 K_2 } =0\\
\label{k3k3 cs}&\left[ \m \frac{\partial}{\partial \m} + \b_\l \frac{\partial}{\partial \l}    \right] \braket{ K_3 K_3 }  +2 \G_{32}\braket{ K_3 K_2 } + 2\G_{K_3}\braket{ K_3 K_3 }=0
\end{align}
We would like to first check if the definition of the trace operator $\Theta$ \eqref{Trace zero AD} 
is consistent with the Callan-Symanzik equation of the two-point function. We will repeat the same 
analysis as in the case of the 3-point function and we will check if the coefficients $a(\l)$ and $b(\l)$ 
are in the agreement with the ones found in the previous section.
As before, we set $\Theta = \b_\l \left[a(\l) K_2 + b(\l) K_3 \right]$ and write its 2-point function:
\be
\braket{ \Theta \Theta }  = \b_\l^2 a(\l)^2 \braket{K_2 K_2 } + \b_\l^2 b(\l)^2 \braket{ K_3 K_3} + 2a(\l) b(\l) \b_\l^2 \braket{ K_3 K_2}
\ee
Using the Callan-Symanzik equations for the two point functions \eqref{k2k2 cs},\eqref{k2k3 cs} and \eqref{k3k3 cs}, we get: 
\be
\bal
\left[ \m \frac{\partial}{\partial \m} + \b_\l \partial_\l \right]\braket {\Theta \Theta } &= \left\{\b_\l \partial_\l  \left[\b_\l^2 a^2(\l)\right] -2\b_\l^2 a^2(\l) \G_{K_2}  - 2\b_\l^2 a(\l) b(\l) \G_{32} \right\} \braket{ K_2 K_2 }\\
&+\left\{ \b_\l \partial_\l \left[\b_\l^2 b^2(\l) \right] -2\b_\l^2 b^2(\l) \G_{K_3} \right\} \braket {K_3 K_3 } \\
&+ \left\{2 \b_\l \partial_\l \left[a(\l) b(\l) \b_\l^2 \right] -2a(\l) b(\l) \b_\l^2 \left[\G_{K_2} 
+\G_{K_3}\right] -2\b_\l^2 b^2(\l) \G_{32} \right\} \braket{K_3 K_2 }
\eal
\ee
Demanding that the 2-point function of $\Theta$ obeys its Callan-Symanzik \eqref{CSTheta} we obtain the following set of equations:
\begin{align}
\label{1st of 2pt}&\partial_\l \left[ \b_\l a(\l) \right] -a(\l)\G_{K_2} - b(\l)\G_{32}=0 \\
\label{2nd of 2pt}&\partial_\l \left[ \b_\l b(\l) \right] - b(\l) \G_{K_3}=0 \\
\label{3rd of 2pt}&a(\l) \left\{ \partial_\l \left[ \b_\l b(\l) \right] - b(\l) \G_{K_3} \right\} + b(\l) \left\{\partial_\l \left[ \b_\l a(\l) \right] -a(\l)\G_{K_2} - b(\l)\G_{32} \right\} =0
\end{align}
We observe that \eqref{3rd of 2pt} is automatically satisfied if \eqref{1st of 2pt} and  
\eqref{2nd of 2pt} are satisfied. So we are left with a $2\times2$ system which is identical to the one that we solved 
in the case of the 3-point function. As a result, we obtain that the trace operator within the $K_I$ basis is uniquely defined as:
\be\label{ThetaTheta2}
\Theta = -\frac{1}{6}\b_\l \phi^4 + \b_\l \left[\frac{\l}{32 (4\pi)^4}  + O(\l^2) \right] \Box \phi^2\, .
\ee

\subsection{The two point function $\braket{K_2K_2}$}

From the definition of the operator $K_2$ we can evaluate its 2-point function as
\be
\braket{ K_{2,0} (x) K_{2,0}(y) } = \Box_x \Box_y \lim_{x_1 \rightarrow x} 
\lim_{y_1 \rightarrow y} \braket{ \phi_0 (x) \phi_0(x_1) \phi_0(y)\phi_0(y_1) }
\ee
that is, as
\be
\braket{ K_{2,0} (x) K_{2,0}(y) } = \Box_x \Box_y\braket{ \mathcal{O}_{2,0}(x) \mathcal{O}_{2,0}(y) }\, .
\ee
So we have to evaluate first the 2-point function of $\mathcal{O}_{2,0}$ operator
$\braket{\mathcal{O}_{2,0}(x) \mathcal{O}_{2,0}(y)}$ and then act with the boxes.
In this analysis we should be careful on how we consider the Wick contractions. 
Since $y_1$ and $y$ are identified, we cannot generate two separate contractions with $x$, 
as in the case of the 3-point function, which preserved crossing symmetry.
This will alter the symmetry factor of certain diagrams by a factor of 1/2. 
Taking this into account we get that: 
\be
\left< \mathcal{O}_{2,0} (x) \mathcal{O}_{2,0}(y) \right> = \begin{gathered} 
\begin{tikzpicture}
\begin{feynman}
\vertex(x);
\vertex [right= 2cm of x](y) ;
\diagram{(x)--[quarter left](y)--[quarter left](x)};
\end{feynman}
\filldraw[fill =black](y) circle(3pt );
\filldraw[fill =black](x) circle(3pt );
\end{tikzpicture}
\end{gathered} +
\begin{gathered} 
\begin{tikzpicture}
\begin{feynman}
\vertex(x);
\vertex [right= 1cm of x](z) ;
\vertex [right= 1cm of z](y) ;
\diagram{(x)--[quarter left](z)--[quarter left](y)--[quarter left](z)--[quarter left](x)};
\end{feynman}
\filldraw[fill =black](y) circle(3pt );
\filldraw[fill =black](x) circle(3pt );
\end{tikzpicture}
\end{gathered}
\ee
In momentum space this is
\be
\braket{ \mathcal{O}_{2,0}(p) \mathcal{O}_{2,0}(-p) } = \left[L_1(p^2) + i \l_0\left[ L_1(p^2)\right]^2  \right]\, .
\ee
Imposing the renormalization condition
\be
\braket{\mathcal{O}_{2}(p) \mathcal{O}_{2}(-p)} = 1 \; , p^2=-\m^2\, 
\ee
and using the definition of the renormalized correlator
\be
\mathcal{O}_{2,0} = Z_{\mathcal{O}_2} \mathcal{O}_2\, ,
\ee
we obtain
\be
Z_{\mathcal{O}_2} = \left[L_1(-\m^2) + i \l \left[ L_1(-\m^2)\right]^2  \right]^{1/2}\, .
\ee
The renormalized expression is fiven by:
\be
\braket{\mathcal{O}_2 (p) \mathcal{O}_2(-p) } =  \left[1 + \frac{\l}{(4\pi)^2} \ln \left( \frac{-p^2}{\m^2} \right)\right]\, .
\ee
As a result the renormalized two point function of the $K_2$ operator is given by:
\be \label{renormalized K2K2}
\braket{ K_2(p)K_2(-p) } = p^4\left[1 + \frac{\l}{(4\pi)^2} \ln \left( \frac{-p^2}{\m^2} \right) + O(\l^2)\right]\, .
\ee
This two point function obeys the Callan-Symanzik equation \eqref{k2k2 cs}.

\subsection{ The 2-point functions $\braket{K_3K_3}$ and $\braket{K_3K_2}$}

In order to compute the renormalized expression for the 2-point functions in the 23 sector,
we will use the Callan-Symanzik equations \eqref{k2k3 cs} and \eqref{k3k3 cs}, 
which means that we have to solve a $2\times 2$ system with unknowns the  
$\braket{ K_3 K_3}$ and $\braket{ K_3 K_2}$ correlation functions. For this, 
we will employ perturbation theory in order to find the logarithmic dependence of the correlation function on the energy scale $\m$.

\subsubsection{The $\braket{K_3 K_3}$ correlator}

The first contribution to the 2-point function of $K_3$ will be of $O(\l_0^2)$.
It is defined
\be
\braket {K_{3,0}(x) K_{3,0}(y) } = \lim_{z\rightarrow y} \Box_y\braket{K_{3,0}(x)\phi_0(y) \phi_0(z) }
\ee
from which we can find the first non-vanshing contribution to the 2-point function:
\be 
\bal
\left< K_{3,0}(x) \phi_0(y) \phi_0(z)\right> &= \int \mathrm{d}^dp_{1,2,3} e^{i p_1x} e^{ip_2y} e^{ip_3z} \left< K_{3,0}(p_1) \phi_0(p_2) \phi_0(p_3) \right> \\
\Rightarrow \Box_y\left< K_{3,0}(x) \phi_0(y) \phi_0(z)\right> &= \int \mathrm{d}^dp_{1,2,3}e^{i p_1x} e^{ip_2y} e^{ip_3z} \frac{-2\l^2}{3} \frac{i^2}{p_3^2}\left[S_1(p_2)+S_1(p_3) \right]\d(p_1+p_2+p_3) \\
&=\int \mathrm{d}^dp_{2,3}e^{-i(p_2+p_3) x} e^{ip_2y} e^{ip_3z} \frac{2\l^2}{3} \frac{1}{p_3^2}\left[S_1(p_2)+S_1(p_3) \right]\eal \ee
Considering the limit $z\to y$ we obtain
\be \bal
\left<K_{3,0}(x) K_{3,0}(y) \right>&=\int \mathrm{d}^dp_{2,3}e^{-i(p_2+p_3) (x-y)}  \frac{2\l_0^2}{3} \frac{1}{p_3^2}\left[S_1(p_2)+S_1(p_3)\right]\\
\eal
\ee
In momentum space the above expression is five by:
\be
\bal
 \left<K_{3,0}(p) K_{3,0}(-p) \right>&=\frac{2\l_0^2}{3}  \int \mathrm{d}^dk \frac{1}{k^2} \left[S_1(k)+ S_1(p-k) \right] = \frac{2\l_0^2}{3} \int \mathrm{d}^dk \frac{1}{k^2}S_1(p-k)
 \eal
\ee
The integral on the right hand side of the above expression can be recognized to originate from the 2-point function of the $\mathcal{O}_{4,0}$ operator and specifically from the Watermelon diagram
\be
\left<\mathcal{O}_{4,0}(p) \mathcal{O}_{4,0}(-p)\right> = \begin{gathered}
\begin{tikzpicture}
\begin{feynman}
    \vertex (x);
    \vertex[below = 1cm of x] (y);
    
    \diagram{
        (x) -- [half left] (y) -- [half left] (x) -- [quarter left] (y) -- [quarter left] (x)
    };
\end{feynman}
\filldraw[fill = black] (x) ++(-0.1cm, -0.1cm) rectangle ++(0.2cm, 0.2cm);
\filldraw[fill = black] (y) ++(-0.1cm, -0.1cm) rectangle ++(0.2cm, 0.2cm);
\end{tikzpicture}
\end{gathered} = 4! \int \mathrm{d}^dk \frac{1}{k^2}S_1(p-k) = 4! W(p^2)\, .,
\ee
Using this result we can see that
\be
\braket{\left(-\frac{\l_0} {6} \mathcal{O}_4 (p) \right)\left(-\frac{\l_0} {6} \mathcal{O}_4 (-p) \right) } = 
\frac{\l_0^2}{36}4! \int \mathrm{d}^dk \frac{1}{k^2}S_1(p-k)=\braket{ K_{3,0}(p) K_{3,0}(-p) }
\ee
which confirms again the $E$-identity
\be
K_{3,0} = -\frac{\l_0}{6} \mathcal{O}_{4,0} \equiv -\frac{1}{6}K_{4,0}\, .
\ee

{\bf Diagrammatic evaluation of $\braket{K_3 K_3}$}

The bare diagrams that contribute to leading and next to leading order to $\braket{K_{3,0}K_{3,0}}$ are:
\be
\braket{\mathcal{O}_{4,0}(p) \mathcal{O}_{4,0}(-p)} =\begin{gathered} \begin{tikzpicture}
\begin{feynman}
    \vertex (x);
    \vertex[below = 1cm of x] (y);
    
    \diagram{
        (x) -- [half left] (y) -- [half left] (x) -- [quarter left] (y) -- [quarter left] (x)
    };
\end{feynman}

\filldraw[fill = black] (x) ++(-0.1cm, -0.1cm) rectangle ++(0.2cm, 0.2cm);
\filldraw[fill = black] (y) ++(-0.1cm, -0.1cm) rectangle ++(0.2cm, 0.2cm);

\end{tikzpicture}
\end{gathered} +
\begin{gathered}
\begin{tikzpicture}
\begin{feynman}
    \vertex (x);
    \vertex[below = 0.5cm of x](z);
    \vertex[below = 0.5cm of z] (y);
    \diagram{
        (x) -- [quarter left] (z) -- [quarter left] (y) -- [quarter left] (z) -- [quarter left] (x)--[half left](y)--[half left](x)
    };
\end{feynman}

\filldraw[fill = black] (x) ++(-0.1cm, -0.1cm) rectangle ++(0.2cm, 0.2cm);
\filldraw[fill = black] (y) ++(-0.1cm, -0.1cm) rectangle ++(0.2cm, 0.2cm);

\end{tikzpicture}
\end{gathered}
\ee
with
\be
\begin{gathered}
\begin{tikzpicture}
\begin{feynman}
    \vertex (x);
    \vertex[below = 0.5cm of x](z);
    \vertex[below = 0.5cm of z] (y);
    \diagram{
        (x) -- [quarter left] (z) -- [quarter left] (y) -- [quarter left] (z) -- [quarter left] (x)--[half left](y)--[half left](x)
    };
\end{feynman}

\filldraw[fill = black] (x) ++(-0.1cm, -0.1cm) rectangle ++(0.2cm, 0.2cm);
\filldraw[fill = black] (y) ++(-0.1cm, -0.1cm) rectangle ++(0.2cm, 0.2cm);

\end{tikzpicture}
\end{gathered} = 4 \times 4! i \l_0 \int  \frac{\mathrm{d}^d k } {(2\pi)^{d}}\frac{1}{(k+p)^2 } TB(k^2) =4 \times 4! i\l_0Q(p^2)\, 
\ee
one order higher in $\l$ than the Watermelon.
We present in detail the evaluation of the related loop integrals in the Appendix.
Setting the renormalization condition for $\braket{\mathcal{O}_4 \mathcal{O}_4}$ 
\be
\braket{\mathcal{O}_4 \mathcal{O}_4} = 4!  p^4 \, , \text{at} \, p^2=-\m^2 \, , 
\ee
we obtain  the renormalized expression
\be
\braket{\mathcal{O}_4 \mathcal{O}_4}  = 4! p^4 \left[ 1 + \frac{6\l}{(4\pi)^2}  \ln \left(\frac{-p^2}{\m^2} \right)  + O(\l^2) \right]
\ee
It is easy to see that the above expression obeys the Callan-Symanzik equation
\be
\left[ \m \frac{\partial}{\partial \m} + \b_\l \frac{\partial}{\partial \l} + 2 \G_{\mathcal{O}_4} \right]
\braket{\mathcal{O}_4 \mathcal{O}_4}   = 0 + O(\l^2)
\ee
Where are the mixing terms? The answer is that the mixing terms are at least of order $O(\lambda^2)$. 
Recalling the analysis of the mixing of $\mathcal{O}_4$ for $d=4$, we found that 
$\Gamma_{\text{mixing}} \sim O(\lambda^2)$. Therefore, the mixing effects cannot be obtained 
through the leading-order 2-point function. It would be very interesting to proceed with the 
$O(\lambda^2)$ analysis of the 2-point function $\braket{\mathcal{O}_4 \mathcal{O}_4}$. 
However, this task is technically demanding and beyond the scope of this work.

Now it is trivial to write down the 2-point function of $K_3$,
using the result for the renormalized $\braket{\mathcal{O}_4 \mathcal{O}_4}$ and the $E$-identity: 
\be
\braket{K_3 K_3} =p^4 \frac{2\l^2}{3} \left[ 1 + \frac{6\l}{(4\pi)^2}  \ln \left(\frac{-p^2}{\m^2} \right)  + O(\l^2) \right]\, .
\ee

\subsubsection{The $\braket{K_3 K_2}$ correlator}

The bare 2-point function is defined in this case as
\be\label{32corr}
\braket{K_{3,0}(x) K_{2,0}(y) } =  \Box_ y\lim_{z\rightarrow y}\braket{K_{3,0}(x) \phi_0(y) \phi_0(z)}
\ee
and as a consequence the first non-vanishing contribution is of order $O(\l_0^2)$. 
We can schematically  write:
\be
\braket{ K_{3,0}(p) K_{2,0}(-p) }  =\s \l_0^2 \left[\text{Loop integral} \right] + O(\l_0^3)\, ,
\ee
where $\s$ is the symmetry factor of the loop integral. 
The above expression implies that the renormalized $\braket{K_{3,0} K_{2,0}}$ start at $O(\lambda^2)$.
In addition it has to vanish on the fixed point, since $\D_{K_2} \neq \D_{K_3}$. 
Previous experience instructs us that the renormalized 2-point function has to be proportional to the $\b$-function: 
$\braket{K_3(p) K_2(p)} \sim \b_\l$.
This leads us to the following ansatz for the renormalized correlator
\be
\braket{K_3(p) K_2(-p)} = c \b_\l \left[1 + c_{32} \l \ln \left(\frac{-p^2}{\m^2} \right) + \cdots \right]
\ee
where $c$ and $c_{32}$ are constants to be determined either by the leading order loop corrections or 
as a consistency condition from the CS equation
\be \label{K3K2 Cs equation}
\left[ \m \frac{\partial}{\partial \m} + \b_\l \frac{\partial}{\partial \l}    \right] \braket{ K_3 K_2 }  + 
\G_{K_2} \braket{ K_3 K_2 }+ \G_{32} \braket{ K_2 K_2 } + \G_{K_3}  \braket{ K_3 K_2 } =0\, .
\ee

{\bf Diagrammatic evaluation of $\braket{ K_3 K_2}$}

Fourier transforming the right hand side of \eqref{32corr} we have that
\be
\bal
\braket{K_{3,0}(x) \phi_0(y) \phi_0(z)} &= \int \frac{\mathrm{d}^dp_{1,2,3}}{(2\pi)^{3d}} e^{ip_1x}e^{ip_2y} e^{i(p_3)z}
\braket{\braket{K_{3,0}(p_1) \phi_0(p_2) \phi_0(p_3)}} (2\pi)^d \d(p_1 +p_2 +p_3) \\
 &=\int \frac{\mathrm{d}^dp_{1,2}}{(2\pi)^{2d}}e^{ip_1 x}e^{ip_2y} e^{-i(p_1+p_2)z}\braket{\braket{K_{3,0}(p_1) \phi_0(p_2) \phi_0(-p_1 -p_2)}}
\eal
\ee
But the form of $\braket{\braket{K_{3,0}(p_1) \phi_0(p_2) \phi_0(p_3) }}$ in terms of the renormalized 
coupling constant is already known from the previous section, see \eqref{bare K3 phi phi}. 
Therefore the two point function is
\be
\braket{K_{3,0}(x) K_{2,0}(y)} =\int \frac{ \mathrm{d}^dp_1}{(2\pi)^d} e^{ip_1(x-y)} 
\int \frac{\mathrm{d}^d p_2}{(2\pi)^d}(-p_1^2)\braket{\braket{K_{3,0}(p_1) \phi_0(p_2) \phi_0(-p_1 -p_2) }}\, .
\ee
In momentum space
\be
\braket{K_{3,0}(p) K_{2,0}(-p)} = (-p^2)\int \frac{\mathrm{d}^dk}{(2\pi)^d} \braket{\braket{K_{3,0}(p) \phi_0(k) \phi_0(-p -k)}}
\ee
The first term is of order $O(\l^2)$ and is given by: 
\be
\braket{K_{3,0}(p) K_{2,0}(-p)} = (-p^2) \frac{2\l^2}{3} \int  \frac{\mathrm{d}^dk}{(2\pi)^d}  
\frac{i}{k^2} \frac{i}{(k+p)^2}S_1\left(k^2 \right)  = p^2 \frac{2\l^2}{3}ST(p^2) 
\ee
The evaluation of $\braket{K_3 K_2}$ to $O(\l_0^3)$ becomes easier if it is reduced to the calculation of the  
two point function $\braket{\phi^4\phi^2}$:
\be
\braket{K_{3,0}(p) K_{2,0}(-p)} = p^2 \frac{\l_0}{6} \braket{\phi_0^4 \phi_0^2}\, .
\ee
The bare two point function  $\braket{\phi_0^4 \phi_0^2}$  up to $O(\l_0^2)$ is given by the following diagrams:
\be
\braket{\phi_0^4 \phi_0^2} =\begin{gathered}
\begin{tikzpicture}
\begin{feynman}
\vertex(x);
\vertex[right = 1cm of x](z);
\vertex[above =0.8cm of z](x1);
\vertex[below =0.8cm of z](x2);
\vertex[right = 1 cm of z](y);
\diagram{(x)--(x1)--[quarter left](x)--[quarter left](x1)-- (y)--[quarter left](x)};
\end{feynman}
\filldraw[fill=black] (x) ++(-0.1cm, -0.1cm) rectangle ++(0.2cm, 0.2cm);
\filldraw[fill =black](y) circle(3pt );
\end{tikzpicture}
\end{gathered} + \begin{gathered}
\begin{tikzpicture}
\begin{feynman}
\vertex(x);
\vertex[right = 1cm of x](z);
\vertex[above =0.8cm of z](x1);
\vertex[below =0.8cm of z](x2);
\vertex[right = 1 cm of z](y1);
\vertex[right =0.8 cm of y1](y);
\diagram{(x)--(x1)--[quarter left](x)--[quarter left](x1)-- (y1)--[quarter left](x)};
\diagram{(y1)--[half left](y)--[half left](y1)};
\end{feynman}
\filldraw[fill=black] (x) ++(-0.1cm, -0.1cm) rectangle ++(0.2cm, 0.2cm);
\filldraw[fill =black](y) circle(3pt );
\end{tikzpicture}
\end{gathered} + 
\begin{gathered}
\begin{tikzpicture}
\begin{feynman}
\vertex(x);
\vertex[right = 1cm of x](z);
\vertex[above =0.8cm of z](x1);
\vertex[below =0.8cm of z](x2);
\vertex[right = 1 cm of z](y);
\diagram{(x)--(x1)--[quarter left](y)--[quarter left](x1)--(x2)--[quarter left](y)--[quarter left](x2) -- (x)};
\end{feynman}
\filldraw[fill=black] (y) ++(-0.1cm, -0.1cm) rectangle ++(0.2cm, 0.2cm);
\filldraw[fill =black](x) circle(3pt );
\end{tikzpicture}
\end{gathered} +
\begin{gathered}
\begin{tikzpicture}
\begin{feynman}
\vertex(x);
\vertex[right = 1cm of x](z);
\vertex[right =1cm of z](z1);
\vertex[right = 1 cm of z1](y);
\diagram{(x)--(z)--[quarter left](z1)--[quarter left](z)--[quarter left](y)--[quarter left](z1) --[quarter left] (y)--[quarter left](x)};
\end{feynman}
\filldraw[fill=black] (y) ++(-0.1cm, -0.1cm) rectangle ++(0.2cm, 0.2cm);
\filldraw[fill =black](x) circle(3pt );
\end{tikzpicture}
\end{gathered}
\ee
Such a calculation has already been done in \cite{Rychkov_Trace}. Nevertheless, we would like to proceed 
with the renormalization of this two point function as we will focus on different aspects.
This a demanding task, since we have to deal with the mixing between the $K_3$ and $K_2$ operators at the 4-loop level.
The bare two point function $\braket{K_{3,0} K_{2,0}}$ can be expressed in terms 
of the renormalized two point functions $\braket{K_2 K_2}$ and $\braket{K_3 K_2}$:
\be\label{K3K2 bare mixing}
\braket{K_{3,0} K_{2,0}} =Z_{K_2}Z_{32} \braket{K_{2} K_{2}} + Z_{K_2}Z_{K_3}\braket{K_{3} K_{2}}\, .
\ee
What we already know from this relation is $\braket{K_2 K_2}$ given by \eqref{renormalized K2K2}, as it does not involve any mixing.  
In addition we know the expression for $Z_{K_2}$, also needed for the renormalization of $\braket{K_2 K_2}$:
 \be\bal
 Z_{K_2}& = \sqrt{L_1(-\m^2) + i \l \left[L_1(-\m^2) \right]^2}   + O(\l^2) \\
 &= \sqrt{L_1(-\m^2)} + \frac{i}{2}\left[L_1(-\m^2) \right]^{3/2} + O(\l^2) 
 \eal
 \ee
 Of course we know the bare expression of the two point function $\braket{K_{3,0}K_{2,0}}$ in terms 
 of the renormalized coupling constant $\l$. These are the 4-loop diagrams above. The result is
\be\bal
\braket{K_{3,0}K_{2,0}}= p^4   \frac{1}{6}&\left[ 4\l^2 \frac{ST(p^2)}{p^2}  - i 6  \l^3 L_1(-\m^2)\frac{ST(p^2)}{p^2}\right.  \\
 & \, \left.  + 3 i \l^3 \frac{SC(p^2)}{p^2}+i \frac{3}{2} \l^3  \frac{QC(p^2)}{p^2} + i6\l^3 \frac{ LT(p^2)}{p^2}        \right]\, ,
\eal \ee
where the corresponding loop integrals have been evaluated in the Appendix:
\be\bal
ST({p_1}^2)& = i\frac{p_1^2}{(4\pi)^6} \left[ -\frac{1}{6\epsilon} + \frac{1}{4} \ln \left(\frac{-p_1^2e^\g}{4\pi} \right) - \frac{25}{24} \right]\\
SC(p^2)&=-\frac{p^2}{(4\pi)^8} 
\left[ \frac{1}{16\epsilon} - \frac{\ln \left(\frac{-p^2e^\g}{4\pi} \right) - 5}{8} \right]
 \\
QC(p^2) &= \frac{p^2}{(4\pi)^8} \left[\frac{1}{2 \epsilon^2} - \frac{8 \ln \left(\frac{-p^2e^\g}{4\pi} \right) -35}{8 \epsilon} + \cdots  \right]\\
LT(p^2)&= -\frac{p^2}{(4\pi)^8} \left[  \frac{2}{\epsilon^2} - \frac{4 \ln(-p^2)}{\epsilon}  + \cdots \right]\, .
\eal\ee
From the analysis of the 3-point functions we know that the order $Z_{32} \sim O(\l^3)$.
This implies that:
\be
Z_{K_2}Z_{32} \braket{K_{2} K_{2}} =p^4 \sqrt{L_1(-\m^2)}Z_{32} + O(\l^4)
\ee
In addition we have to use the renormalization of $\braket{K_3 K_3}$:
\be
\braket{K_{3,0} K_{3,0}} = Z_{K_3}^2 \braket{K_3 K_3} + 2Z_{32}Z_{K_3}\braket{K_3 K_2} + Z_{32}^2\braket{K_2 K_2}
\ee
Taking into account that we work up to $O(\l^3)$ and the fact that $\braket{K_3 K_2}$ 
starts at $ \sim O(\l^2) $, the last two terms of the above expression can be neglected as higher order terms:
\be\bal
2Z_{32}Z_{K_3}\braket{K_3 K_2}  \sim O(\l^5)\\
Z_{32}^2\braket{K_2 K_2} \sim O(\l^6)
\eal\ee
As a result, the renormalization of $\braket{K_3 K_3}$  in not affected by any mixing effect up to $O(\l^3)$. 
This has already been discussed in the previous subsection, where we used the equations 
of motion in order to obtain the renormalized expression of $\braket{K_3 K_3}$
\be
\braket{K_{3,0} K_{3,0}} = Z_{K_3}^2 \braket{K_3 K_3} + O(\l^4)
\ee
Recalling that the renormalization condition of $\braket{K_3 K_3}$  is $\braket{K_3 K_3}=\frac{2\l^2}{3}p^4$ at $p^2 =-\m^2$ we obtain:
\be \bal
Z_{K_3}^2 &=\frac{\left.\braket{K_{3,0} K_{3,0}}\right|_{p^2 = -\m^2} }{\m^4} \frac{3}{2\l^2} + O(\l^4) \\
\eal\ee 
The bare $\braket{K_{3,0} K_{3,0}}$ in terms of the renormalized coupling constant $\l$ is given by:
\be\bal
\braket{K_{3,0} K_{3,0}} &= \frac{2\l^2}{3} \left[W(p^2) + 2 \d_\l W(p^2) +i 4 \l Q(p^2)  \right] + O(\l^4) \\
&=\frac{2\l^2}{3} \left[W(p^2) -i3 \l L_1(-\m^2) W(p^2) +i 4 \l Q(p^2)  \right] + O(\l^4)
\eal\ee
Using the above form of the bare two-point function we obtain:
\be
Z_{K_3}^2=\frac{1}{\m^4}\left[W(-\m^2) -i3 \l L_1(-\m^2) W(-\m^2) +i 4 \l Q(-\m^2)  \right] + O(\l^4)
\ee
One can check that the  above result reproduces the anomalous dimension of $K_3$ operator:
\be
\G_{K_3} = \frac{1}{Z_{K_3}} \m \frac{\mathrm{d}}{\mathrm{d} \m} Z_{K_3} = \frac{3\l}{(4\pi)^2} + O(\l^2)
\ee
This confirms that the assumption that the first contribution to $Z_{32} $ is of order $O(\l^3)$ is consistent.
Now, we have everything needed to obtain the expression for $Z_{32}$ and then the renormalized $\braket{K_3 K_2}$.
Considering the relation \eqref{K3K2 bare mixing} at $p^2 = -\m^2$ and solving for $Z_{32}$ we obtain:
\be
Z_{32} = \frac{1}{Z_{K_2}} \frac{\braket{K_{3,0} K_{2,0}}|_{p^2 = - \m^2}}{\m^4} - Z_{K_3} \frac{\braket{K_3 K_2}|_{p^2=-\m^2}}{\m^4}
\ee
where  $\left. \braket{K_{3} K_{2}} \right|_{p^2 =- \m^2}$ is defined by the renormalization condition.
Since $Z_{32} \sim O(\l^3)$ we may assume that:
\be
\frac{1}{Z_{K_2}} \frac{\braket{K_{3,0} K_{2,0}}|_{p^2 = - \m^2}}{\m^4} - Z_{K_3} \frac{\braket{K_3 K_2}|_{p^2=-\m^2}}{\m^4} = 0 + O(\l^3)
\ee
Solving the above equation we obtain the renormalization condition
\be
\frac{\braket{K_3 K_2}|_{p^2=-\m^2}}{\m^4}  = -i \frac{\l^2}{48\pi^2} + O(\l^3)
\ee
and observe that only the $O(\l^2)$ contribution of the renormalization condition is constrained.
As a result there is an $O(\l^3)$ freedom in defining the renormalization condition. 
Taking also into account that this 2-point correlation function has to vanish at the conformal limit  
we can consider the renormalization condition
\be
\braket{K_3 K_2} = -p^4\frac{i\b_\l}{9} + O(\l^3) \; , \, \text{at} \, p^2=-\m^2
\ee
Of course, this is not a direct proof that the renormalized correlation function vanishes at the WF fixed point,
it is rather a reasonable ansatz. 
So the off-diagonal element of the mixing matrix $Z_{32}$ is given by:
\be
Z_{32} = \frac{1}{Z_{K_2}} \frac{\braket{K_{3,0} K_{2,0}}|_{p^2 = - \m^2}}{\m^4} +i\frac{\b_\l}{9} Z_{K_3} 
\ee
Plugging this into \eqref{K3K2 bare mixing} we can evaluate the renormalized correlation function from the relation
\be
\braket{K_3 K_2} = \frac{1}{Z_{K_3} Z_{K_2}}  \left[\braket{K_{3,0}K_{2,0}} -  
\frac{\braket{K_{3,0} K_{2,0}}|_{p^2 = - \m^2}}{\m^4} \braket{K_2 K_2} \right] - \frac{i \b_\l}{9} \braket{K_2 K_2}\, .
\ee
Putting everything together, expanding in powers of $\l$ and taking the limit $\epsilon \to 0$ 
we obtain the form of the renormalized correlation function:
\be\bal
\braket{K_3 K_2} &=- \frac{i}{9}p^4 \left[ \frac{3\l^2}{(4\pi)^2} - \frac{17}{3}\frac{\l^3}{(4\pi)^4} - \frac{729}{4} \frac{\l^3}{(4\pi)^4} \ln \left( \frac{-p^2}{\m^2} \right) \right] + O(\l^4) \\
&=- \frac{i\b_\l}{9}p^4 \left[ 1 - \frac{243}{4}\frac{\l}{(4\pi)^2} \ln \left( \frac{-p^2}{\m^2} \right) \right] + O(\l^4)
\eal\ee
By applying the Callan-Symanzik equation \eqref{K3K2 Cs equation} we obtain the value of $\G_{32}$
\be
\G_{32} = i \frac{263}{6 } \frac{\l^3}{(4\pi)^2}\, .
\ee
Thus, the anomalous dimension matrix for the case of the 2-point functions is given by:
\be
\G_{IJ} = \begin{bmatrix}
\G_{K_2} & 0 \\
i \frac{263}{6 } \frac{\l^3}{(4\pi)^2} & \G_{K3}
\end{bmatrix}
\ee  
Although the value of $\G_{32}$ differs from the one obtained in the analysis of the 3-point functions
it is important to emphasize that the eigenvalues of the anomalous dimension matrix
which are associated with observables (the critical exponents), remain the same. 
The underlying reason for the discrepancy in the off-diagonal element of the anomalous dimension 
matrix is that, in the renormalization scheme used in this work, the $Z_\mathcal{O}$ 
factors depend on the type of correlation function under consideration. 

\subsection{$\braket{ \Theta \Theta }$ and the eigenvalue $e_\Theta$} 

Finally we are ready to compute $\braket{ \Theta \Theta }$. From \eqref{ThetaTheta2} we have
\be
\braket{ \Theta \Theta }  = \b_\l^2 \left(\frac{\l}{32 (4\pi)^4}\right)^2 \braket{K_2 K_2 } + 
\frac{\b_\l^2}{\l^2} \braket{ K_3 K_3} + 2\frac{1}{32 (4\pi)^4} \b_\l^2 \braket{K_3 K_2 }\, .
\ee
The orders of each contribution is
 \begin{align}\label{CScontr23}
& \b_\l^2 \left(\frac{\l}{32 (4\pi)^4}\right)^2 \braket{K_2 K_2 }  \sim O(\l^6)\\
  &\frac{\b_\l^2}{\l^2}  \braket{ K_3 K_3} =p^4\b_\l^2 \frac{2}{3} \left[ 1 + 
  \frac{6\l}{(4\pi)^2}  \ln \left(\frac{-p^2}{\m^2} \right)  + O(\l^2) \right]\\
  &2\frac{1}{32 (4\pi)^4} \b_\l^2 \braket{ K_3 K_2 }  \sim O(\l^6)
 \end{align}
So up to leading order the expression for the 2-point function is given by:
\be
\braket{ \Theta \Theta } =p^4\b_\l^2 \frac{2}{3} \left[ 1 + \frac{6\l}{(4\pi)^2}  \ln \left(\frac{-p^2}{\m^2} \right)  
+ O(\l^2) \right] + O(\l^6)\, .
\ee
This is the other main result of this paper.
We check the Callan-Symanzik equation:
\begin{align}
&\m\frac{\partial}{\partial \m}\braket{\Theta \Theta}  = -\frac{12 \l}{(4\pi)^2} \frac{6\l^4}{(4\pi)^4} 
+ O(\l^6)= -\frac{12 \l}{(4\pi)^2}\braket{ \Theta \Theta } + O(\l^6)\\
&\b_\l \partial_\l \braket{ \Theta \Theta } =\frac{12 \l}{(4\pi)^2} \frac{6\l^4}{(4\pi)^4} + O(\l^6) 
=\frac{12 \l}{(4\pi)^2}\braket{ \Theta \Theta } + O(\l^6)
\end{align}
It is obvious that it is satisfied. Moreover, the eigenvalue equation in \eqref{eTheta} is:
\be\label{eigbetalambda}
\b_\l \frac{\partial}{\partial\l} \braket{ \Theta \Theta } = \left[\frac{12 \l}{(4\pi)^2}  + O(\l^2) \right] \braket{ \Theta \Theta }
\ee
Thus, the eigenvalue we are after is simply
{
\be
e_\Theta = 2\Gamma_{\phi^4} + O(\l^2) \, .
\ee
}
We find it impressive that such a simple result emerges after the calculation that was needed to derive it.
However no matter how simple it is, we do not see a simple and safe way to just guess it. 
We note that it is not guaranteed that the eigenvalue $e_\Theta$ 
will keep on reproducing the anomalous dimension of the $\phi^4$ operator to all orders. For example, 
if we extend our analysis so that the corrections $O(\l^2)$ are included, 
the $O(\b_\l^2\cdot \l^2)\sim O(\l^6)$ contributions to the $\braket{K_3 K_2}$ and $\braket{K_2 K_2}$ correlators will
have to be taken into account.  
In addition, we observe from \eqref{CScontr23} that the leading order contribution to 
$\braket{\Theta\Theta}$ is entirely due to $\braket{K_3 K_3}$. This shows that we would have obtained the same
expression for it, in the minimal basis. 
Of course, this is true only up to the leading order. 
Therefore, we can argue that the leading-order eigenvalue is invariant under the choice 
of either the minimal or non-minimal definition of the trace operator.
 This provides an extra argument for the claim that $e_\Theta$
 encodes non-trivial information about the internal structure of the EMT.

It is interesting to ask if it is possible to form some other linear combination ${\mathcal O}$ of the basis operators such that the 
eigenvalue $e_{\mathcal O}$ of the new operator is of higher order than the order of the eigenvalues 
of $\braket{K_2K_2}$ and $\braket{K_3K_3}$. 
The leading order version of this question in perturbation theory is whether one can construct a linear combination that 
has $e_{\mathcal O}\sim \l^2$. Such a linear combination must have of course non-zero anomalous dimension.
Combining the results of this section and the observation made in the Introduction \eqref{eOO}, we can 
see that this is not possible, as one can not reduce in magnitude the anomalous dimension of an operator by forming
linear combinations, at least in not perturbation theory. 
A related fact is that even though the anomalous dimension is not additive in the sense that
$\Gamma_{{\mathcal O_1}+{\mathcal O_2}}\ne \Gamma_{\mathcal O_1}+\Gamma_{\mathcal O_2}$ ,
the eigenvalue in \eqref{eOO} is:
 \be
 e_{{\mathcal O_1}+{\mathcal O_2}} = e_{\mathcal O_1}+e_{\mathcal O_2}\, ,
 \ee
given the fact that the mixing effect is of higher order. The statement follows form the fact that $\Gamma_{\mathcal O_{1,2}}\ge 0$.

We conclude by defining a "charge", based on the generic form of the correlator:
\be\label{ctheta}
\braket{\Theta\Theta} =  p^4 c_\Theta\, , \hskip 1cm c_\Theta= \b_\l^2 \frac{2}{3} \left[ 1 + \frac{6\l}{(4\pi)^2}  \ln \left(\frac{-p^2}{\m^2} \right)  
+ O(\l^2) \right]\, .
\ee
This charge enters the conservation equation of the renormalized EMT, which will be the topic of a future work.

\section{Conclusions}

In this work, we presented in a self contained way the renormalization of correlation functions 
of certain composite operators in the four-dimensional $\phi^4$ theory. 
In a preparatory step we computed correlators of the form $\braket{{\mathcal O}\phi\phi}$, with ${\mathcal O}=\phi^2, \phi^4$.
A consistency check of the derived renormalized expressions is that they should reproduce the known results 
for the anomalous dimension functions of the operators, $\Gamma_{\mathcal{O}}$. 
Another consistency check we used was that under the simple substitution $\frac{-p^2}{\m^2}\to p^2$ the
expressions for the correlators coincide
with their corresponding expressions in conformal field theory, expanded in $\epsilon$ around the Wilson-Fisher fixed point.

We then defined a basis of operators with mass dimension equal to $d$. 
We solved the mixing problem in this basis and confirmed the anomalous dimension matrix presented in \cite{Collins_Brown}.
Using these results and the Callan-Symanzik equation satisfied by the 3-point function $\braket{\Theta \phi \phi}$, we determined the trace of the 
renormalized Energy-Momentum Tensor operator $\Theta$, to be of the form
\be
\Theta = -\frac{1}{6}\b_\l \phi^4 + \b_\l \left[\frac{\l}{32 (4\pi)^4}  + \l^2) \right] \Box \phi^2\, .
\ee
This expression constitutes the non-minimal definition of the trace of the EMT, 
which satisfies a Callan-Symanzik equation without any mixing term.
Based on this solution for $\Theta$, we proceeded to study its two-point function and showed that in the basis of 
dimension $d$ operators it is of the form $\braket{\Theta\Theta}= p^4 c_\Theta$, with $c_\Theta$ given in \eqref{ctheta}.
To leading order, we used only the two-point function $\braket{K_3 K_3}$, since the other two correlation functions $\braket{K_2 K_2}$ and $\braket{K_3K_3}$ contribute higher order terms. 
Consequently, we obtained the form of $\braket{\Theta\Theta}$
that was expected to be obtained by the minimal definition of $\Theta$.
Finally, we found that the $c_\Theta$ function determines an eigenvalue equation 
of the form \eqref{eigbetalambda}, with $e_\Theta = 2\G_{\phi^4} + \cdots$.
The leading order  eigenvalue is invariant under the definition of the trace. As we already stated, this invariance  implies that the eigenvalue $e_\Theta$ is a physical quantity contained in the two-point function $\braket{\Theta \Theta} $.
 In principle, such an analysis could be performed for any massless renormalizable field theory.  
 
 We would like to end with an important observation. Perhaps the most striking characteristic of the renormalization of
 the trace of the EMT in the massless theory is that in perturbation theory it is not implemented via the standard method where 
 the bare quantum operator is identified with the fields in the classical expression promoted to quantum fields and 
 then writing $\Theta_0=Z_\Theta \Theta$. This is of course due to the vanishing of the classical trace and
 it is the origin of a large freedom in the non-perturbative definition of a renormalized $\Theta$. 
 In general $\Theta$ is a quantum operator of dimension $d$ 
 that obeys the two constraints of its vanishing on the fixed point and of the vanishing of its anomalous dimension
 and there seem to be many ways to define such an operator.
 In order to give a physical meaning to any of these definitions, they must be associated with an observable. 
 Such an observable may be extracted from the eigenvalue $e_\Theta$. 
 A concrete way to obtain this eigenvalue is to project the operator on the basis
 comprised of $K_2$ and $K_3$. In this basis, it is contained in the charge function $c_\Theta$ as 
 showed in great detail. The solution however we presented is just
 one of infinitely many RG trajectories along which the operator $\Theta$ could approach the fixed point:
 it is simply the perturbative RG trajectory associated with the basis of dimension $d$ operators.
 
As discussed in the Introduction, alternatively we could have defined another trajectory, for example such as the one in \cite{Ising}, where it was defined 
directly non-perturbatively as
\be
\Theta_0 = z_\Theta^{1/2} \Theta\hskip .5 cm {\rm with}\hskip .5 cm \m \frac{\partial}{\partial \m} \braket{\Theta_0\Theta_0} = 0
\ee
with no reference to the classical trace or to a basis of operators.   
Here $0=\G_\Theta = \g_\Theta - 2\g_\phi$ and $\g_\Theta = \m \frac{\partial}{\partial \m} z_\Theta$.
This definition results in the eigenvalue equation 
\be\label{IsingEig}
\left(\m \frac{\partial}{\partial \m} + 2\g_\phi \right) \braket{\Theta\Theta} = 0
\ee
and a valid definition of $\Theta$ since its vanishing on the fixed point is guaranteed by the "other half" of the
CS equation $\left(\b_\l \frac{\partial}{\partial \l} - 2\g_\phi \right) \braket{\Theta\Theta} = 0$ and the form
$\braket{\Theta\Theta} = c_\Theta/|x|^{2d}$.
The eigenvalue $e_\Theta=2\g_\phi$ characterizes this RG flow as one where the twist does not play a role
and in \cite{Ising} it was linked to inflationary observables.  Similarly many other RG flows could be constructed. 
It would be interesting to understand better whether 
physical systems that are presumably described by $\phi^4$ theory or certain purely theoretical approaches have a preference for one of these trajectories. 
We summarize the situation in the following picture:
\begin{figure}[H]
    \centering
    \includegraphics[width=0.35\textwidth]{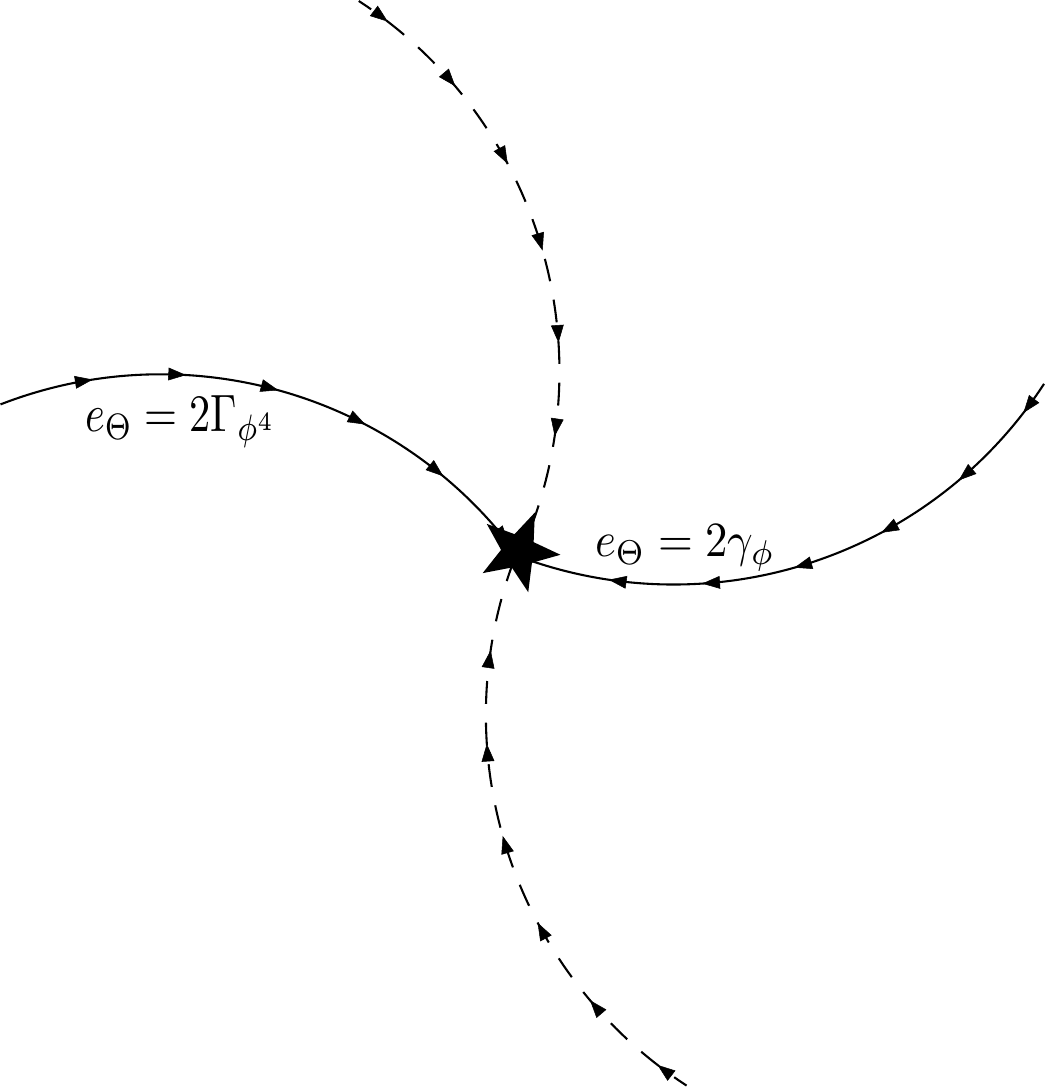}
    \caption{\label{RGTheta} Various RG trajectories defining a renormalized $\Theta$. The one labelled by $2\Gamma_{\phi^4}$ was 
    computed in this work.}
\end{figure}

\section*{Acknowledgments}

L.K. would like to thank S. Rychkov for  discussions during the CERN–SEENET-MTP–ICTP PhD School 2024 in Thessaloniki.

\newpage
\begin{appendix}
\newpage
\numberwithin{equation}{section}

\begin{center}{\bf\Huge Appendices}\end{center}
   
\section{Renormalization of the $\l \phi^4$-theory}

We review the standard renormalization of the $\lambda \phi^4$ theory and reproduce the results of the RG functions, used extensively throughout this work. 

\subsection{The classical field theory}

We are interested in the massless $\l \phi^4$-theory with bare action
\be \label{flat space action}
S ^{(0)}[\phi_0 ; \l_0]=  \int \mathrm{d}^dx \mathcal{L} =\int \mathrm{d}^dx \left[ \frac{1}{2} \partial_\n \phi_0 \partial^\n \phi_0 - \frac{\l_0}{4!} \phi_0^4  \right]
\ee
The equation of motion, the Energy-Momentum Tensor (EMT) and its trace are:
\begin{align}
\frac{\partial \mathcal{L}}{\partial \phi_0}- \partial_\m \frac{\partial \mathcal{L}}{\partial (\partial_\m \phi_0)} =0 \Rightarrow  \Box \phi_0 + \frac{\l}{6} \phi_0^3 =0 \\
 \label{canonical Tmn}  T^{(0)}_{\m \n}= \frac{\partial \mathcal{L}}{\partial (\partial^\m \phi_0)}\partial_\n \phi_0 -g_{\m \n} \mathcal{L} = \partial_\m \phi_0 \partial_\n \phi_0  + \frac{\l_0}{4!} g_{\m \n}\phi_0^4 - \frac{1}{2}g_{\m \n} \left(\partial \phi_0 \right)^2 \\
 \label{canonical Trace} \Theta^{(0)} =g^{\m \n} T^{(0)}_{\m \n}= \left(1 -\frac{d}{2}\right)\left(\partial \phi_0 \right)^2+ \frac{\l_0 d}{4!} \phi_0^4
\end{align}
Using the equations of motion we can check that the above EMT is conserved:
\be
\bal
\partial^\m T^{(0)}_{\m \n}&= \Box \phi_0 \partial_\n \phi_0 + \partial^\m \phi_0 \partial_\n (\partial_\m \phi_0) + \frac{\l_0}{6} \phi_0^4 \partial_{\n}\phi_0 - \frac{1}{2} \partial_\n (\partial \phi_0)^2 \\
&=\partial_\n \phi_0 \left[\Box \phi_0 + \frac{\l_0}{6} \phi_0^3 \right] +\partial^\m \phi_0 \partial_\n (\partial_\m \phi_0)  - \partial_{\n}(\partial_\m \phi_0)\partial^\m \phi_0 =0
\eal
\ee
With partial differentiation, we can write the trace of the EMT tensor also as 
\be
\Theta^{(0)} = \left(1 -\frac{d}{2}\right) \left[ \frac{1}{2} \Box \phi_0^2  - \phi_0 \Box \phi_0 \right]+ \frac{\l_0 d}{4!} \phi_0^4
\ee
For $d=4$ we obtain:
\be
\Theta^{(0)} = - \frac{1}{2}\Box \phi_0^2 +\phi_0 \left( \Box \phi_0 + \frac{\l_0}{6} \phi_0^3 \right)
\ee
and with the use of  the equations of motion, we conclude that:
\be
\Theta^{(0)} = - \frac{1}{2}\Box \phi_0^2\, ,
\ee
which means that the trace is equal to a surface term, that can be neglected. 

The $\l \phi_0^4$-theory is a Classical CFT.
The action \eqref{flat space action} can be thought of as the flat-space limit of the  action of a scalar field in curved space
\be
S^{(0)}_{\text{curved}} [\phi_0 ; \l_0] = \int \mathrm{d}^d x \sqrt{|g|} \left[\frac{1}{2}g^{\m \n} \partial_\m \phi_0 \partial_\n \phi_0  - \frac{\l_0}{4!} \phi_0^2 + \xi_0 R \phi_0^2 \right] \, , 
\ee
where $\xi_0$ is the dimensionless coupling constant of the conformally coupled scalar field
\be
\xi_0 = \frac{1}{2} \frac{d-2}{4(d-1)} \underbrace{\rightarrow}_{d=4} \frac{1}{12}\, .
\ee
The $R \phi_0^2$ term  contributes to the energy-momentum tensor
\be
T^{(0)\text{-curved}}_{\m \n} =  \nabla_\m \phi_0 \nabla_\n \phi_0  + \frac{\l_0}{4!} g_{\m \n}\phi_0^4 - \frac{1}{2}g_{\m \n}  \nabla_\r \phi_0 \nabla^\r \phi_0  + 2\xi_0 \left( G_{\m \n} -\nabla_\m \nabla_\n + g_{\m \n} \Box_{\text{(curved)}} \right) \phi_0^2
\ee
where $\nabla_{\m}$ is the covariant derivative, $\Box_{\text{(curved)}}= g^{\m \n} \nabla_{\m} \nabla_\n$ and $G_{\m\n}$ the Einstein tensor.
$G_{\m \n} = R_{\m \n} - \frac{1}{2}R g_{\m \n}$.
The flat space limit of the EMT is:
\be\bal
T^{(0)\text{-flat}}_{\m \n} &= \partial_\m \phi_0 \partial_\n \phi_0  + \frac{\l_0}{4!} g_{\m \n}\phi_0^4 - \frac{1}{2}g_{\m \n} \left(\partial \phi_0 \right)^2 +2 \xi_0 \left( g_{\m \n} \Box - \partial_\m \partial_\n \right) \phi_0^2 \\
&= T^{(0)}_{\m \n} +2 \xi_0 \left( g_{\m \n} \Box - \partial_\m \partial_\n \right) \phi_0^2 
\eal
\ee
Because of the $R\phi_0^2$ term, the flat space EMT gets "improved" by a tensorial term which is conserved:
\be
\partial^{\m}\left( g_{\m \n} \Box - \partial_\m \partial_\n \right) \phi_0^2 =0
\ee
This ensures that the the improved EMT is still conserved.
As a consequence of the above improvement,  the trace of EMT also gets improved and vanishes for $d=4$ as it should:
\be \bal
\Theta^{(0)\text{-flat}}& = \Theta^{(0)} + 2\xi_0 (d-1)\Box \phi_0^2 \\
&= - \frac{1}{2}\Box \phi_0^2 +6 \xi_0  \Box \phi_0^2
\eal
\ee
If we use the classical value of $\xi_0$ for $d=4$ we get:
\be
\Theta^{(0)\text{-flat}} =- \frac{1}{2}\Box \phi_0^2 +\frac{1}{2}   \Box \phi_0^2 =0
\ee
For the above result, we used the "classical" value of $\xi_0$ and obtained a vanishing trace. In QFT, $\xi$ receives corrections, 
leading to the breaking of conformal symmetry, which results in a non-vanishing quantum operator for the trace $\Theta \sim \b_\l$.

\subsection{1-loop renormalization}

For the propagator the renormalization condtions are
\be\label{prop renorm cond}
\bal
\left.\Pi (p^2)\right|_ {p^2 = - \m^2} &=0 \\
\left.\frac{\mathrm{d}\Pi (p^2)}{\mathrm{d}p^2} \right|_{p^2 =-\m^2}&=0 
\eal
\ee
with $\Pi(p^2)$ denoting its loop correction.
For the renormalization of the coupling constant $\l$ we define the condition:

\be\label{amp ren con}
\mathcal{M} = -i \l \; ,\; \text{at $S.P.$ with $s^2=t^2=u^2=-\m^2$} \ee
where $S.P. (\m)$ is the Symmetric Point \cite{Amit:1984ms} of the four external momenta $\left\{ p_i \right\}$, with $p_i \cdot p_j = -\m^2 \left( \d_{ij} - \frac{1}{4} \right)$
and the 3 channels are defined as $s = (p_1 +p_2)^2$, $t=(p_1+p_3)^2$ and $u= (p_1+p_4)^2$.
The renormalization condition \eqref{amp ren con} is equivalent to the definition of $\l$ as the magnitude of the $\phi \phi \rightarrow \phi \phi$ scattering amplitude at the $S.P.$, 
where
\be
\braket{ \phi(p_1) \phi(p_2) \phi(p_3) \phi(p_4) } = \mathcal{M} \frac{i}{p_1^2}\frac{i}{p_2^2}\frac{i}{p_3^2}\frac{i}{p_4^2} (2\pi)^d \d \left( \sum p_{in }- \sum p_{out} \right)
\ee
The amplitude $\mathcal{M}$  in the massless case is given by:
\be 
   \bal
   \mathcal{M} &=
   \begin{gathered}
   \begin{tikzpicture} 
     \begin{feynman}
     \vertex (x);
     \vertex[below= 1cm of x](z);
     \vertex[below = 1cm of z](u);
     \vertex[right = 1cm of z](v);
     \vertex[right=  2cm of x](y);
     \vertex[right = 2cm of  u](l);
     \diagram{
     (x)--(v) --(l)
     };
     \diagram{
     (y)--(v)--(u)
     };
 \end{feynman} 
 \end{tikzpicture} 
 \end{gathered} +
 \begin{gathered}
 \begin{tikzpicture} 
     \begin{feynman}
     \vertex (x);
     \vertex[below= 1cmof x](z);
     \vertex[below = 1cm of z](u);
     \vertex[right = 1cm of z](v);
     \vertex[right=  2cm of x](y);
     \vertex[right = 2cm of  u](l);
     \diagram{
     (x)--(v) --(l)
     };
     \diagram{
     (y)--(v)--(u)
     };
 \end{feynman} 
 \filldraw[fill=black] (v) ++(-0.1cm, -0.1cm) rectangle ++(0.2cm, 0.2cm);
  \draw[thick, white] (v) ++(-0.1cm, 0) -- ++(0.2cm, 0);
  \draw[thick, white] (v) ++(0, -0.1cm) -- ++(0, 0.2cm);
   \end{tikzpicture}
 \end{gathered} +
\begin{gathered}
\begin{tikzpicture}
 \begin{feynman}
     \vertex (x);
     \vertex[below= 1cmof x](z);
     \vertex[below =1cm of z](u);
     \vertex[right = 1cm of z](v);
     \vertex[right =1 cm of v](r);
     \vertex[right=  3cm of x](y);
     \vertex[right = 3cm of  u](l);
     \diagram{
     (x)--(v) --[half left](r)-- [half left](v)--(u)
     };
     \diagram{
     (y)--(r)--(l)
     };
 \end{feynman} 
\end{tikzpicture}
\end{gathered} + \text{$(t,u)$-channels} \\
&=-i \l -i \l \d_\l^{(1)}  + \frac{(-i\l)^2}{2} \sum_{p^2=s,t,u} \int \frac{\mathrm{d}^dk}{(2\pi)^d} \frac{i}{k^2 } \frac{i}{(k-p)^2 }\\
&= -i\l -i\d_\l^{(1)} \l - \frac{(-i\l)^2}{2} \sum_{p=s,t,u} L_{1}(p^2)
   \eal
   \ee
with $L_1$ the loop integral. It is $L_1=i B_0$ in the Passarivo-Veltman language.
For $d=4-\epsilon$ it takes the following form in the $\epsilon$-expansion:
\be 
L_1(p^2)= \frac{i}{16\pi^2} \left[\frac{2}{\epsilon} - \ln \left(\frac{-p^2 e^\g}{4\pi}  \right) +2 \right]
\ee
Applying the renormalization condition \eqref{amp ren con} we obtain the 1-loop counterterm
\be \label{vertex ct 1} 
\d_\l^{(1)} =-i\frac{3 \l}{2} L_1(-\m^2)\, .
\ee
Using this result we obtain the expression for the renormalized amputated 4-point function:
\be
{\mathcal{M} = -i\l  + \frac{\l^2}{2} \sum_{p^2=s,t,u} \left[L_1(p^2) - L_1(-\m^2) \right]}\, .
\ee
In the context of $\epsilon$-expansion, for $\epsilon \rightarrow 0$, the leading order result is:
\be
{\mathcal{M} = -i\l - i\frac{\l^2}{2(4\pi)^2}\sum_{p^2=s,t,u}\ln \left(\frac{-p^2}{\m^2} \right)}\, .
\ee
Instead of using the standard definition $\b_\l = \m \frac{\partial \l}{\partial \m}$, we extract it from the Callan-Symanzik equation
\be 
\bal
\left[ \frac{\partial}{\partial \ln \m} + \b_\l \partial_\l + 4\g_\phi \right] \braket{ \phi \phi \phi \phi } =0 
\eal
\ee
A short computation yields the well-known result
\be \label{beta function leading}
\b_\l = \frac{3 \l^2}{ (4\pi)^2} + O(\l^3)\, .
\ee
All this is trivial textbook material, but we presented it because it illustrates the method used in the main text in more complicated cases.

The above is an expression valid in $d=4$. In general dimensions, we just write $\l = \hat{\l} \m^\epsilon$ and repeat the process, which 
adds a classical term that reflects the dimensions of $\l$ in general $d$:
\be \label{hat b}
\hat{\b}_\l = -\epsilon \l +  \frac{3  \l^2}{(4\pi)^2}  + O(\l^3)
\ee
The Wilson-Fisher fixed point is defined as the point on the phase diagram where ${\hat \b}_\l=0$, that is when
\be
\l^* = \frac{\epsilon}{3} (4\pi)^2+ O(\epsilon^2)
\ee
This point is believed to be a conformal point where the QFT description gives its place to a CFT description.

\subsection{2-loop renormalization}

The 2-loop renromalization introduces some of the main loop integrals encountered throughout this paper.

\subsubsection{The propagator}

The first non-vanishing contribution to the propagator is given by the following diagrams:
 \be \label{prop corr}
 \left< \phi(p) \phi(-p) \right> = 
 \begin{tikzpicture}
 \begin{feynman}
 \vertex(x);
 \vertex[right = 2cm of x] (y);
 \diagram{ 
 (x)--(y);
 };
 \end{feynman}
 \end{tikzpicture} +
 \begin{gathered}
 \begin{tikzpicture}
 \begin{feynman}
 \vertex(x);
 \vertex[right = 0.5cm of x] (z);
 \vertex[right =0.5cm of z](h);
 \vertex[right = 0.5cm of h](l);
 \vertex[right =  0.5cm of l](y);
 \diagram{ 
 (x)--(z)--[half left](l)--[half left](z)--(y)
 };

 \end{feynman}
\end{tikzpicture}
  \end{gathered} +
  \begin{tikzpicture}
 \begin{feynman}
 \vertex(x);
 \vertex[right = 1cm of x] (y);
 \vertex[right = 1cm of y] (y1);
 \diagram{ (x)--(y1)};
 \end{feynman}
  \draw[fill=white] (y) circle (3pt); 
 \draw (y) ++(-2pt, -2pt) -- ++(4pt, 4pt); 
  \draw (y) ++(-2pt, 2pt) -- ++(4pt, -4pt); 
   \end{tikzpicture}
 \ee
Incuding the extrenal legs, we have
\be
\bal
\begin{gathered}
 \begin{tikzpicture}
 \begin{feynman}
 \vertex(x);
 \vertex[right = 0.5cm of x] (z);
 \vertex[right =0.5cm of z](h);
 \vertex[right = 0.5cm of h](l);
 \vertex[right =  0.5cm of l](y);
 \diagram{ 
 (x)--[thick](z)--[half left](l)--[half left](z)--[thick](y)
 };
 \end{feynman}
\end{tikzpicture}
  \end{gathered} & = -\frac{i}{p^4} \frac{\l^2}{6} S_1(p^2)\\
\begin{tikzpicture}
 \begin{feynman}
 \vertex(x);
 \vertex[right = 1cm of x] (y);
 \vertex[right = 1cm of y] (y1);
 \diagram{ (x)--[thick](y1)};
 \end{feynman}
 \draw[fill=white] (y) circle (3pt); 
 \draw (y) ++(-2pt, -2pt) -- ++(4pt, 4pt); 
  \draw (y) ++(-2pt, 2pt) -- ++(4pt, -4pt); 
 \end{tikzpicture} &= -\frac{i}{p^2}  \d_\phi
\eal
\ee
where $S_1(p^2)$ is the so called Sunset integral. In the context of $\epsilon$-expansion it is equal to [\eqref{s1 first} -\eqref{s1 exp}] :

\be \label{sun diag}\bal
S_1(p^2)=\frac{p^2}{2(4\pi)^4} \left[ \frac{2}{\epsilon} - \ln \left( \frac{-p^2e^\g}{4\pi} \right) + \frac{13}{4} \right]\, .
\eal\ee
Using the renormalization condition \eqref{prop renorm cond} and solving for the counterterm we get
$\d_\phi = \frac{1}{\m^2} \frac{\l^2}{6}S_1(-\m^2)$ or
\be\label{prop counter}
{\d_\phi = - \frac{\l^2}{12 (4\pi)^4}\left[ \frac{2}{\epsilon} - \ln \left( \frac{\m^2e^\g}{4\pi} \right) + \frac{13}{4} \right]}\, .
\ee
Substituting the results \eqref{sun diag} and \eqref{prop counter} in \eqref{prop corr}  we obtain
\be
{\left< \phi (p) \phi(-p) \right> = \frac{i }{p^2} \left[1 + \frac{\l^2}{12(4\pi)^2} \ln \left(\frac{-p^2}{\m^2}\right) \right] + O(\l^3)}\, .
\ee
As above, we can calculate the anomalous dimension $\g_\phi$ of the primary field $\phi$ from the Callan-Symanzik equation
\be\bal
\left[\frac{\partial}{\partial \ln \m}+\b_\l \partial_\l + 2\g_\phi \right] \left< \phi (p) \phi (-p) \right>  =0 \, .
\eal\ee
This reproduces the famous result
\be
{\g_\phi = \frac{\l^2}{12 (4\pi)^2} + O(\l^3)}
\ee

\subsubsection{The vertex}

We consider the 2-loop contributions of the $s$-channel of the amputated 4-point function. Taking into account the result \eqref{vertex ct 1} of the 1-loop renormalization we get that: 

\be \label{order 3 amp}
\bal
\mathcal{M}^{(2)}(s) &= \begin{gathered}
\begin{tikzpicture}
\begin{feynman}
  \vertex (x);
     \vertex[right= 1cm of x](z);
     \vertex[below = 0.5cm of x ](u);
     \vertex[right = 0.5cm of u](v);
     \vertex[below=  0.5cm of v](y);
     \vertex[below = 0.5cm of  y](l);
     \vertex[below=2cm of x](r);
     \vertex[right = 1 cm of r](p);
     \diagram{
     (x)--(v) --(z)
     };
     \diagram{
     (v)--[half left](y)--[half left](v)
     };
     \diagram{(y)--[half left](l)--[half left](y) };
     \diagram{(r)--(l)--(p)};
\end{feynman}
\end{tikzpicture}
\end{gathered} +
\begin{gathered}
\begin{tikzpicture}
\begin{feynman}
 \vertex (x);
     \vertex[right= 1cm of x](z);
     \vertex[below = 0.5cm of x ](u);
     \vertex[right = 0.2 cm of u](v);
     \vertex[right = 0.5cm of v](q);
     \vertex[below=  1cm of v](y);
     \vertex[right = 0.25cm of y](l);
     \vertex[below=2cm of x](r);
     \vertex[right = 1 cm of r](p);
     \diagram{
     (x)-- (v)--[half left](q)--[half left](v)
     };
     \diagram{
     (z)--(q)
     };
     \diagram{(v)--(l)-- (q)};
      \diagram{(r)--(l)--(p)};
\end{feynman}
\end{tikzpicture}
\end{gathered} +
\begin{gathered}
\begin{tikzpicture}
\begin{feynman}
 \vertex (x);
     \vertex[right= 1cm of x](z);
     \vertex[below = 0.5cm of x ](u);
     \vertex[right = 0.5cm of u](v);
     \vertex[below= 1 cm of v](k);
     \vertex[left = 0.25cm of k](l);
	\vertex[right = 0.25cm of k](s);
	 \vertex[below=2cm of x](r);
     \vertex[right = 1 cm of r](p);
     \diagram{(x)--(v)--(z) };
     \diagram{(l)--(v) --(s) --(p)};
     \diagram{(r)--(l)};
     \diagram{(l)--[half left](s) -- [half left](l) };
\end{feynman}
\end{tikzpicture}
\end{gathered}+
\begin{gathered}
\begin{tikzpicture}
\begin{feynman}
\vertex(x);
\vertex[right = 1 cm of x](z);
 \vertex[right= 1cm of x](z);
     \vertex[below = 0.5cm of x ](u);
     \vertex[right = 0.5cm of u](v);
     \vertex[below= 1cm of v](y);
     \vertex[below=2cm of x](r);
     \vertex[right = 1 cm of r](p);
     \diagram{(x)--(v)--(z)};
     \diagram{(v)--[half left](y)--[half left](v)};
     \diagram{(r)--(y)--(p)};
\end{feynman}
 \filldraw[fill=black] (y) ++(-0.1cm, -0.1cm) rectangle ++(0.2cm, 0.2cm);
  \draw[thick, white] (y) ++(-0.1cm, 0) -- ++(0.2cm, 0);
  \draw[thick, white] (y) ++(0, -0.1cm) -- ++(0, 0.2cm);\end{tikzpicture}
\end{gathered} +
\begin{gathered}
\begin{tikzpicture}
\begin{feynman}
\vertex(x);
\vertex[right = 1 cm of x](z);
 \vertex[right= 1cm of x](z);
     \vertex[below = 0.5cm of x ](u);
     \vertex[right = 0.5cm of u](v);
     \vertex[below= 1cm of v](y);
     \vertex[below=2cm of x](r);
     \vertex[right = 1 cm of r](p);
     \diagram{(x)--(v)--(z)};
     \diagram{(v)--[half left](y)--[half left](v)};
     \diagram{(r)--(y)--(p)};
\end{feynman}
 \filldraw[fill=black] (v) ++(-0.1cm, -0.1cm) rectangle ++(0.2cm, 0.2cm);
  \draw[thick, white] (v) ++(-0.1cm, 0) -- ++(0.2cm, 0);
  \draw[thick, white] (v) ++(0, -0.1cm) -- ++(0, 0.2cm);
\end{tikzpicture}
\end{gathered} +
\begin{gathered}
\begin{tikzpicture}
\begin{feynman}
\vertex(x);
\vertex[below = 1cm of x](u);
\vertex[right = 0.5cm of u](v);
\vertex[right = 1 cm of x](z);
  \vertex[below=2cm of x](r);
     \vertex[right = 1 cm of r](p);
     \diagram{(x)--(p)};
     \diagram{(z)--(r)};
\end{feynman}
 \filldraw[fill=black] (v) ++(-0.1cm, -0.1cm) rectangle ++(0.2cm, 0.2cm);
  \draw[thick, white] (v) ++(-0.1cm, 0) -- ++(0.2cm, 0);
  \draw[thick, white] (v) ++(0, -0.1cm) -- ++(0, 0.2cm);
\end{tikzpicture}
\end{gathered} _{O(\l^3)} +
\begin{gathered}
\begin{tikzpicture}
\begin{feynman}
\vertex(x);
\vertex[right =0.25cm of x](x1);
\vertex[right =0.5cm of x1](x2);
\vertex[right =1cm of x2](x3);
\vertex[right =0.25cm of x3](r);
\vertex[right =0.5cm of x3](x4);
\vertex[above =0.5cm of x4](y);
\vertex[below=0.5cm of x4](y1);
\diagram{(x1)--(x2)--[half left](x3)--[half left](x2) --(x4)};
\diagram{(y)--(r)--(y1)};
\end{feynman}
\end{tikzpicture}
\end{gathered}
+
\begin{gathered}
\begin{tikzpicture}
\begin{feynman}
\vertex(x);
\vertex[right =0.25cm of x](x1);
\vertex[right =0.5cm of x1](x2);
\vertex[right =0.3cm of x2](x3);
\vertex[right =0.25cm of x3](x4);
\vertex[above =0.5cm of x4](y);
\vertex[below=0.5cm of x4](y1);
\diagram{(x1) --(x4)};
\diagram{(y)--(x3)--(y1)};
\end{feynman}
\draw[fill=white] (x2) circle (3pt); 
 \draw (x2) ++(-2pt, -2pt) -- ++(4pt, 4pt); 
  \draw (x2) ++(-2pt, 2pt) -- ++(4pt, -4pt); 
\end{tikzpicture}
\end{gathered}
 \\
&=\frac{i \l^3}{4}\left[L_1(s) \right]^2 - i \l^3 D(s) +2\frac{\d_\l^{(1)} \l^2}{2}L_1(s) - i \d_{\l,s}^{(2)}\l  - i \frac{\l^3}{6}\frac{S_1(q^2)}{q^2} -i \l^3 \d_\phi 
\eal
\ee
with $q$ the momentum carried by the external leg.

We have to consider both the one loop renormalization of the coupling constant and the two loop renormalization of the propagator. Recalling the result \eqref{prop counter} from the renormalization of the propagator we get that: 

\be
\begin{gathered}
\begin{tikzpicture}
\begin{feynman}
\vertex(x);
\vertex[right =0.25cm of x](x1);
\vertex[right =0.5cm of x1](x2);
\vertex[right =1cm of x2](x3);
\vertex[right =0.25cm of x3](r);
\vertex[right =0.5cm of x3](x4);
\vertex[above =0.5cm of x4](y);
\vertex[below=0.5cm of x4](y1);
\diagram{(x1)--(x2)--[half left](x3)--[half left](x2) --(x4)};
\diagram{(y)--(r)--(y1)};
\end{feynman}
\end{tikzpicture}
\end{gathered}
+
\begin{gathered}
\begin{tikzpicture}
\begin{feynman}
\vertex(x);
\vertex[right =0.25cm of x](x1);
\vertex[right =0.5cm of x1](x2);
\vertex[right =0.3cm of x2](x3);
\vertex[right =0.25cm of x3](x4);
\vertex[above =0.5cm of x4](y);
\vertex[below=0.5cm of x4](y1);
\diagram{(x1) --(x4)};
\diagram{(y)--(x3)--(y1)};
\end{feynman}
\draw[fill=white] (x2) circle (3pt); 
 \draw (x2) ++(-2pt, -2pt) -- ++(4pt, 4pt); 
  \draw (x2) ++(-2pt, 2pt) -- ++(4pt, -4pt); 
\end{tikzpicture}
\end{gathered} = -i \frac{\l^3}{12(4\pi)^4} \ln \left(\frac{-4q^2}{3\m^2} \right) 
\ee
The origin of the factor $\frac{4}{3}$ is the  renormalization condition at the $S.P$.
Regarding the Candy counterterms, using \eqref{vertex ct 1} from the  1-loop renormalization of the coupling constant, we have
\be 
\begin{gathered}
\begin{tikzpicture}
\begin{feynman}
\vertex(x);
\vertex[right = 1 cm of x](z);
 \vertex[right= 1cm of x](z);
     \vertex[below = 0.5cm of x ](u);
     \vertex[right = 0.5cm of u](v);
     \vertex[below= 1cm of v](y);
     \vertex[below=2cm of x](r);
     \vertex[right = 1 cm of r](p);
     \diagram{(x)--(v)--(z)};
     \diagram{(v)--[half left](y)--[half left](v)};
     \diagram{(r)--(y)--(p)};
\end{feynman}
 \filldraw[fill=black] (y) ++(-0.1cm, -0.1cm) rectangle ++(0.2cm, 0.2cm);
  \draw[thick, white] (y) ++(-0.1cm, 0) -- ++(0.2cm, 0);
  \draw[thick, white] (y) ++(0, -0.1cm) -- ++(0, 0.2cm);
\end{tikzpicture}
\end{gathered}  =\begin{gathered}
\begin{tikzpicture}
\begin{feynman}
\vertex(x);
\vertex[right = 1 cm of x](z);
 \vertex[right= 1cm of x](z);
     \vertex[below = 0.5cm of x ](u);
     \vertex[right = 0.5cm of u](v);
     \vertex[below= 1cm of v](y);
     \vertex[below=2cm of x](r);
     \vertex[right = 1 cm of r](p);
     \diagram{(x)--(v)--(z)};
     \diagram{(v)--[half left](y)--[half left](v)};
     \diagram{(r)--(y)--(p)};
\end{feynman}
 \filldraw[fill=black] (v) ++(-0.1cm, -0.1cm) rectangle ++(0.2cm, 0.2cm);
  \draw[thick, white] (v) ++(-0.1cm, 0) -- ++(0.2cm, 0);
  \draw[thick, white] (v) ++(0, -0.1cm) -- ++(0, 0.2cm);
  \end{tikzpicture}
\end{gathered}=\frac{\l}{2}\d_\l L_1(p^2) = \frac{-i}{2} \frac{3\l^3}{2} L_1(p^2)L_1(-\m^2)\, .
\ee
These contributions are very important for the elimination of the overlapping divergences that will appear later.
Combining the above relations we conclude that:
\be\label{M2s no exp}
\bal
\mathcal{M}^{(2)}(s)= &\frac{i \l^3}{4} \left\{\left[L_1(s) - L_1(-\m^2) \right]^2 - \left[ L_1(-\m^2)]\right]^2 \right\} \\
&- i\l^3 \left[ L_1(-\m^2) L_1(s) + D(s) \right]\\
  &- i \d_{\l,s}^{(2)}\l  -i \frac{\l^3}{12(4\pi)^4} \ln \left(\frac{-4q^2}{3\m^2} \right)
\eal
\ee

with: 

\be
\bal
- i \l^3 D(s) = \begin{gathered}
\begin{tikzpicture}
\begin{feynman}
 \vertex (x);
     \vertex[right= 1cm of x](z);
     \vertex[below = 0.5cm of x ](u);
     \vertex[right = 0.2 cm of u](v);
     \vertex[right = 0.5cm of v](q);
     \vertex[below=  1cm of v](y);
     \vertex[right = 0.25cm of y](l);
     \vertex[below=2cm of x](r);
     \vertex[right = 1 cm of r](p);
     \diagram{
     (x)-- (v)--[half left](q)--[half left](v)
     };
     \diagram{
     (z)--(q)
     };
     \diagram{(v)--(l)-- (q)};
      \diagram{(r)--(l)--(p)};
\end{feynman}
\end{tikzpicture}
\end{gathered} +
\begin{gathered}
\begin{tikzpicture}
\begin{feynman}
 \vertex (x);
     \vertex[right= 1cm of x](z);
     \vertex[below = 0.5cm of x ](u);
     \vertex[right = 0.5cm of u](v);
     \vertex[below= 1 cm of v](k);
     \vertex[left = 0.25cm of k](l);
	\vertex[right = 0.25cm of k](s);
	 \vertex[below=2cm of x](r);
     \vertex[right = 1 cm of r](p);
     \diagram{(x)--(v)--(z) };
     \diagram{(l)--(v) --(s) --(p)};
     \diagram{(r)--(l)};
     \diagram{(l)--[half left](s) -- [half left](l) };
\end{feynman}
\end{tikzpicture}
\end{gathered} &= \frac{i \l^3}{2} \left[ \int \frac{\mathrm{d}^dk}{(2\pi)^d} \frac{L_1(k+p_2) + L_1(k+p_3) }{ k^2 \left( k+p_s \right)^2} \right]    \\
&= \frac{i \l^3}{2} \left[I_4(p_s,p_2) + I_4(p_s,p_3) \right]\; ,\;p_s^2 = s
\eal
\ee
\be
{D(s) = - \frac{ 1}{2} \left[I_4(p_s,p_2) + I_4(p_s,p_3) \right] \; ,\; p_s^2 = s}
\ee
We will call $D(s)$ the Ice cream integral. The definition of $I_4(p_s,p_2)$ is given in \eqref{I4 exp}. Its value in the $\epsilon$-expansion is
\bea
{I_4(p_s,p_2)}&=&\frac{1}{(4\pi)^4} \left[- \frac{2}{\epsilon ^2}+ \frac{2 \ln \left(\frac{-s e^\g}{4\pi}\right) - 5}{\epsilon } - 
\ln^2 \left(\frac{-s e^\g}{4\pi}\right) + 5 \ln \left(\frac{-s e^\g}{4\pi} \right)- \frac{\pi^2}{4} - \frac{23}{2}\right] \nonumber\\
&+& \frac{G(p_s,p_2)}{(4\pi)^4}\, ,
\eea
where the $G(p_s,p_2)$ is given by \eqref{Gpp2}:
\be \label{Gpsp2}
G(p_s,p_2)=\int_0^1 \mathrm{d}z \mathrm{d}y \frac{z}{1-z} \ln \left( \frac{-2 yz(1-z) p_s \cdot p_2 +yz(1-yz)s+ z(1-z)p_2^2}{y(1-y)s} \right) + O(\epsilon)
\ee
The $\frac{1}{\epsilon}$ divergence of the above expression contains the 'overlapping divergences', corresponding to the term $\frac{\ln(-p^2)}{\epsilon}$.
Responsible for the cancellation of these overlapping  divergences is the $L_1(-\m^2) L_1(s)$ contribution in \eqref{order 3 amp}, since: 
\be \bal
(4\pi)^4 \left. L_1(p^2)L_1(-\m^2) \right.= &-\frac{4}{\epsilon^2} + \frac{2\ln \left( \frac{-p^2e^\g}{4\pi}\right) + 2\ln \left( \frac{\m^2e^\g}{4\pi}\right) -8}{\epsilon}\\
& -2 \left[\frac{1}{2} \ln \left( \frac{-p^2e^\g}{4\pi}\right) + \frac{1}{2} \ln \left( \frac{\m^2e^\g}{4\pi}\right)  \right]^2 \\
&+  8 \left[\frac{1}{2} \ln \left( \frac{-p^2e^\g}{4\pi}\right) + \frac{1}{2} \ln \left( \frac{\m^2e^\g}{4\pi}\right)  \right]+ \frac{\pi^2}{6} -12
\eal
\ee
In the $\epsilon$  expansion,  using \eqref{I4 epsilon} \eqref{double candy p m}, the middle term of  \eqref{order 3 amp} gets the following form:
\be \label{ll d s}
\bal
\left[ L_1(-\m^2) L_1(s) + D(s) \right] (4\pi)^4 &= -\frac{2}{\epsilon^2} + \frac{2\ln \left(\frac{\m^2 e^\g}{4\pi} \right)  -3}{\epsilon} \\
 & \; \; \; + \frac{1}{2} \ln^2 \left( \frac{-s}{\m^2} \right) - \ln \left( \frac{-s}{\m^2} \right)\\
 &\; \; \;- \frac{1}{2}\left[ G(p_s,p_2) + G(p_s,p_3) \right] \\
 &\; \; \; +\text{(momentum independent terms)} 
 \eal
\ee 
where the overlapping  divergences have been cancelled out.  
The remaining divergences are momentum independent and they will be absorbed by the counterterm $\d_\l^{(2)}$.

Using the renormalization condition \eqref{amp ren con}, which is equivalent to the vanishing of \eqref{order 3 amp} at the 
$S.P.(\m^2)$ and solving for the counterterm $\d_{\l,s}^{(2)}$, we obtain
\be
\d_{\l,s}^{(2)} =-\frac{\l^2}{4} \left[ L_1 (-\m^2) \right]^2 - \l^2 \left[ L_1(-\m^2) L_1(-\m^2) + D(-\m^2) \right]
\ee
Thus  the  renormalized $O(\l^3)$ s-channel is:
\bea\label{M2s}
\mathcal{M}^{(2)}(s) &=& \frac{i \l^3}{4} \left[ L_1(s)-L_1(-\m^2) \right]^2 -i  \frac{\l^3}{(4\pi)^4} \left[ \frac{1}{2} \ln^2 \left( \frac{-s}{\m^2} \right) - \ln \left( \frac{-s}{\m^2} \right) - G \right] \nonumber\\
&-&i \frac{\l^3}{12(4\pi)^4} \ln \left(\frac{-4q^2}{3\m^2} \right)\nonumber\\
&=& -\frac{i 3 \l^3}{4(4\pi)^4} \ln^2 \left( \frac{-s}{\m^2}\right) +\frac{i\l^3}{ (4\pi)^4}\ln \left( \frac{-s}{\m^2} \right)  - \frac{i\l^3}{12(4\pi)^4} \ln \left(\frac{-4q^2}{3\m^2} \right) \nonumber\\
&+& \frac{i\l^3}{2(4\pi)^4} \hat{G}(p_s,p_2,p_3) \, .
\eea
where
\be
\hat{G}(p_s,p_2,p_3) =\left[ G(p_s,p_2) +G(p_s,p_3) - 2 G_{S.P.} \right]
\ee
and $2G_{S.P.}$ is  given by substituting $s=-\m^2$, $p_2^2=p_3^2 = -\frac{3}{4}\m^2$ and $p_s \cdot p_2 = - p_s \cdot p_3 = -\frac{1}{2}\m^2$ in \eqref{Gpsp2}:
\be
\bal
2 G_{S.P.} &=\left.G(p_s,p_2)\right|_{S.P.} +\left. G(p_s,p_3) \right|_{S.P.} \\
&= \int_0^1\mathrm{d}y \mathrm{d}z\frac{z}{z-1} \left[ \ln \left(\frac{z \left(\left(4 y^2-4 y+3\right) z-3\right)}{4 (y-1) y} \right) + \ln \left( \frac{z \left(4 y^2 z+4 y (z-2)+3 (z-1)\right)}{4 (y-1) y} \right) \right]
\eal
\ee
So $\hat{G}(p_s,p_2,p_3)$ takes the following form: 
\be
\bal
\hat{G}(p_s,p_2,p_3) = \int_0^1 \mathrm{d}y \mathrm{d}z\frac{z}{z-1}& \left\{\ln \left(4 \frac{yz(1-yz)s+ z(1-z)p_2^2 -2 yz(1-z) p_s \cdot p_2 }{s  \left[\left(4 y^2-4 y+3\right) z^2-3z \right]} \right)  \right.\\
& \; \; \; \left. +  \ln \left(4 \frac{yz(1-yz)s+ z(1-z)p_3^2 -2 yz(1-z) p_s \cdot p_3 }{s  \left[ z \left(4 y^2 z+4 y (z-2)+3 (z-1)\right)\right]} \right) \right\}\, .
\eal
\ee
The contribution proportional to $\hat{G}(p_s,p_2,p_3)$ in \eqref{M2s}  is $\m$-independent  and as  we will see, this term will be neglected as a higher order term in the Callan-Symanzik equation.
Since the other two channels $(t,u)$ will give exactly the same contribution, the total renormalized magnitude of the 4-point function up to $O(\l^3)$ is:
\be\bal
\mathcal{M}= -i \l &-i \frac{\l^2}{2(4\pi)^2} \sum_{p^2 =s,t,u} \ln \left( \frac{-p^2}{\m^2} \right)    -i \frac{3 \l^3}{4(4\pi)^4}\sum_{p^2 =s,t,u}\ln^2 \left(\frac{-p^2}{\m^2} \right)\\
 &+\frac{i\l^3}{ (4\pi)^4}\sum_{p^2 =s,t,u}\ln \left( \frac{-p^2}{\m^2} \right) -i \frac{\l^3}{12(4\pi)^4} \sum_{i =1,2,3,4} \ln \left(\frac{-4q_i^2}{3\m^2}\right)+ \frac{i\l^3}{2(4\pi)^4}\sum_{k=s,t,u} \hat{G}(p_k,p_2,p_3) 
\eal\ee
From the Callan-Symanzik equation we can calculate the $\b$-function up to order $O(\l^3)$. We will use a perturbative expression for the $\b$-function:
\be
\b_\l = \l \sum_{n=1}^{\infty}b_n\frac{\l^n}{(4\pi)^{2n}} = \sum_{n=1}^{\infty} \b_\l^{(n)}
\ee
with $b_1 = 3$, or $\b_\l^{(1)} = \frac{3\l^2}{(4\pi)^2}$.
We present the explicit calculation:
\be\bal
\frac{\partial}{\partial \ln \m} \left< \phi \phi \phi \phi\right>&= i \frac{3\l^2}{(4\pi)^2} +i \frac{3\l^3}{(4\pi)^4} 
\sum_{p^2=s,t,u}\ln \left( \frac{-p^2}{\m^2} \right) -i \frac{6 \l^3}{(4\pi)^4} + i \frac{8 \l^3}{12(4\pi)^4}  + O(\l^4)\\
&= i \b_\l^{(1)} \left[1 + \frac{\l}{4\pi } \sum_{p^2=s,t,u}\ln \left( \frac{-p^2}{\m^2} \right) \right]  - i \frac{16 \l^3}{3(4\pi^4)}+ O(\l^4)
\eal \ee
\be
\b_\l \partial_\l \left< \phi \phi \phi \phi\right> =-i \b_\l^{(1)} \left[1 + \frac{\l}{4\pi } \sum_{p^2=s,t,u}\ln \left( \frac{-p^2}{\m^2} \right) \right] - i \b_\l^{(2)} + O(\l^4)
\ee
\be
4\g_\phi \left< \phi \phi \phi \phi\right>= -i4 \frac{\l^3}{12(4\pi)^4} + O(\l^4)
\ee
Combining these results, the terms which are  multiplied with $i\b_\l^{(1)}$  get cancelled and we get that:
\be \bal
\left[\m \frac{\partial}{\partial \m} + \b_\l \partial_\l +4\g_\phi \right]\left< \phi \phi \phi \phi\right> =0 \\
\Rightarrow -i\b_\l^{(2)}- i \frac{16 \l^3}{3(4\pi^4)} -i4 \frac{\l^3}{12(4\pi)^4} =0\\
\b_\l^{2} = - \frac{17 \l^3}{3 (4\pi)^4} \rightarrow b_2 = - \frac{17}{3}
 \eal\ee
 Therefore the $\beta$- function up to $O(\l^3)$ is given by:
 \be
 {\b_\l = \frac{3 \l^2}{(4\pi)^2} - \frac{17 \l^3}{3 (4\pi)^4} + O(\l^4)}
 \ee

\subsubsection{The Wilson-Fisher fixed point}

As we have seen, the $b$-function for $d=4-\epsilon$ is given by:

\be
\hat{\b}_\l = -\epsilon \l + \b_\l
\ee
So up to two loops
\be
\hat{\b}_\l = -\epsilon \l + \frac{3 \l^2}{(4\pi)^2}  - \frac{17 \l^3}{3 (4\pi)^4} + O(\l^4)
\ee
The Wilson-Fisher fixed point is determined by the vanishing of the $\b$ function. The solution is
\be
\l^* = \frac{\epsilon}{3} (4\pi)^2 + \frac{17 \epsilon^2}{81}(4\pi)^2 + O(\epsilon^3)
\ee

\section{Loop integrals}

In this section we enlist the loop integrals that we use for our calculations. In addition we provide the intermediate steps for the evaluation of these integrals.

\subsection{Feynman parameters }

Feynman parametrization is a basic tool for the computation of loop integrals. The simplest identity is the following:
    \be
     \frac{1}{A B}= \int_0^1 \mathrm{d}x \mathrm{d}y\d(1-x-y) \frac{1}{\left[A x + B y \right]^2}
     \ee
     Some other useful identities are:   
    \begin{align}
    \frac{1}{AB^\n}& =\int_0^1 \mathrm{d}x \mathrm{d}y \d(1-x-y)\frac{\n y^{\n-1}}{\left[ Ax + By\right]^{\n+1}} \label{n feyn par}\\
    \frac{1}{ABC}&= \int_0^1 \mathrm{d}x \mathrm{d}y \mathrm{d}z \d(1-x-y-z) \frac{2}{\left[Ax +By +Cz \right]^3}
    \end{align}
The most general identity, provided that $Re(\n_k) > 0$  for every $1\leq k \leq n$ is given by
 \be \label{general fen par}
    \frac{1}{A_1^{\n_1}\cdots A_n^{\n_n}} = \frac{\G \left( \sum_{k =1}^{n}\n_k \right)}{\G \left(\n_1\right) \cdots \G \left(\n_n\right) } \int_0^1 \mathrm{d}u_1 \cdots \mathrm{d}u_n \frac{\d(1-\sum_{k =1}^{n}u_k )u_1^{\n_1-1} \cdots u_n^{\n_n -1}}{\left[ \sum_{k=1}^n u_k A_k\right]^{\sum_{k =1}^{n}\n_k}}
    \ee

\subsection{Euler's $\Beta$-function}  

A very useful formula for the evaluation of massless loop integrals is the following:
\be \label{euler b}
\Beta (a,b) \equiv \int_0^1 \mathrm{d}x x^{a-1} (1-x)^{b-1} = \frac{\G(a) \G(b)}{\G(a+b)}
\ee 

\subsection{The standard loop integral}

Here is a basic loop integral in Minkowski spacetime:
\be\label{stand int}
\int \frac{\mathrm{d}^d k}{\left( 2\pi \right)^d} \frac{k^{2a}}{\left(k^2 -\Delta \right)^b} = \frac{i}{(4\pi)^{d/2}} \left(-1\right)^{a-b} 
\frac{1}{\Delta^{b-a - \frac{d}{2}}} \frac{\G \left(a +\frac{d}{2} \right) \G \left( b-a -\frac{d}{2} \right)}{\G \left(b \right) \G \left( \frac{d}{2} \right)}
\ee
A more general class of massless scalar loop integrals encountered in loop computations may have the form
\be
L_{\n_0 \n_1 \n_2 \cdots \n_n}(p_1,p_2 , \cdots ,p_n) = \int \frac{\mathrm{d}^d k}{(2\pi)^d} \frac{1}{\left(k\right)^{2\n_0} \left(k-p_1\right)^{2\n_1} \left(k-p_2\right)^{2\n_2} \cdots\left(k-p_n\right)^{2\n_n} }
\ee

\subsection{Massless 1-loop integrals}
 
\subsubsection{The Candy integral}

The simplest integral used extensively here is the $L_{1,1}(p)$. This integral is associated with the Candy diagram:
\be
\begin{gathered}
\begin{tikzpicture}
\begin{feynman}
\vertex (x);
\vertex[below= 0.5cm of x](z);
\vertex[below= 0.5 cm of z](y);
\vertex[right= 0.5cm of z](r);
\vertex[right= 1cm of r](r1);
\vertex[right= 0.5cm of r1](r2);
\vertex[above= 0.5cm of r2](q);
\vertex[below= 0.5cm of r2](p);
\diagram{(x)--(r)--[half left](r1)--[half left](r)--(y)};
\diagram{(q)--(r1)--(p)};
\end{feynman}
\end{tikzpicture}
\end{gathered} = \frac{\l^2}{2} L_1(p^2)
\ee
For short, we will use following  the following notation:
\be
L_1(p^2) \equiv L_{1,1}(p) = \int \frac{\mathrm{d}^dk}{(2\pi)^d} \frac{1}{k^2 \left(k-p\right)^2}\, .
\ee
This integral is equivalent to the well known $B_0$ integral in the Passarino-Veltman language:
\be
L_1(p^2) = i B_0(p^2,0,0)
\ee
For general $d$-dimensions it is given by:
\be \label{L1 int}
L_1(p^2) = i \frac{\G\left( 2-\frac{d}{2}\right) \left[\G \left(\frac{d}{2}-1 \right)\right]^2}{\left( 4\pi \right)^{d/2} \G\left(d-2 \right)} \left(-p^2 \right)^{\frac{d}{2}-2}
\ee
In the $\epsilon$-expansion for $d=4-\epsilon$ it becomes
\be
{L_1\left(p^2\right)= \frac{i}{16\pi^2} \left[\frac{2}{\epsilon} - \ln \left(\frac{-p^2 e^\g}{4\pi}  \right) +2 \right]}
\ee

\subsection{Massless 2-loop  integrals }

\subsubsection{The Sunset integral}

The sunset integral is associated with the 2-loop correction of the propagator, which is given by the following diagram:
\be 
\bal
\begin{gathered}
\begin{tikzpicture}
\begin{feynman}
\vertex (x);
\vertex[right =  0.5 cm of x](x1);
\vertex[right =  0.6 cm of x1](x2);
\vertex[right =  0.5 cm of x2](x3);
\diagram{(x)--(x1)--[half left](x2)--[half left](x1)--(x3)};
\end{feynman}
\end{tikzpicture}
\end{gathered} &= -i \frac{\l^2}{6}\int \frac{\mathrm{d}^d k_1 \mathrm{d}^dk_2}{(2\pi)^{2d}} \frac{1}{k_1^2 k_2^2 \left(k_1 +k_2 -p \right)^2 } \\
&= -i \frac{\l^2}{6} S_1(p^2)
\eal
\ee
The loop integral with respect to $k_2$ can be evaluated with the use of \eqref{L1 int}. Then,
\be \label{s1 first}
\bal 
S_1(p^2) & \equiv\int  \frac{\mathrm{d}^d k_1 \mathrm{d}^dk_2}{(2\pi)^{2d}} \frac{1}{k_1^2 k_2^2 \left(k_1 +k_2 -p \right)^2 } \\
& =i \frac{\G\left( 2-\frac{d}{2}\right) \left[\G \left(\frac{d}{2}-1 \right)\right]^2}{\left( 4\pi \right)^{d/2} \G\left(d-2 \right)}  
\int  \frac{\mathrm{d}^d k}{(2\pi)^{d}} \frac{1}{k^2 \left[ - (k-p)^2 \right]^{2-\frac{d}{2}}} \\
&=i (-1)^{\frac{d}{2}-2} \frac{\G\left( 2-\frac{d}{2}\right) \left[\G \left(\frac{d}{2}-1 \right)\right]^2}{\left( 4\pi \right)^{d/2} 
\G\left(d-2 \right)}  \int  \frac{\mathrm{d}^d k}{(2\pi)^{d}} \frac{1}{k^2 \left[  (k-p)^2 \right]^{2-\frac{d}{2}}} 
\eal
\ee
Next we introduce a Feynman parameter, by applying \eqref{n feyn par}, in order to evaluate the loop integral with respect to $k$:
\be
\bal
\int \frac{\mathrm{d}^d k}{(2\pi)^{d}} \frac{1}{k^2 \left[ (k-p)^2 \right]^{2-\frac{d}{2}}}  &=  \int_0^1\mathrm{d} x \int \frac{\mathrm{d}^d k}{(2\pi)^{d}}  \frac{\left(2-\frac{d}{2} \right)x^{1-d/2}}{\left[ k^2 - 2k\cdot p x  + p^2 x\right]^{3- \frac{d}{2}}} \\
&=\int_0^1\mathrm{d} x \int \frac{\mathrm{d}^d k}{(2\pi)^{d}}  \frac{\left(2-\frac{d}{2} \right)x^{1-d/2}}{\left[ \left(k - p x \right)^2   + p^2 x(1-x) \right]^{3- \frac{d}{2}}} \\
&= \frac{i}{(4\pi)^{\frac{d}{2}}}(-1)^{\frac{d}{2}-3} \frac{ \G(3-d)}{\G\left( 2- \frac{d}{2}\right)} \int_0^1\mathrm{d}x x^{\frac{d}{2}-2}(1-x)^{d-3}(-p^2)^{d-3}
\eal
\ee
In the last step we shifted $ k\rightarrow k+px$ and evaluated the standard loop integral using \eqref{stand int}. 
The integral with respect to the Feynman parameter $x$ is  an Euler $\Beta$ -function, which is defined in \eqref{euler b}. Thus,
\be
\int \frac{\mathrm{d}^d k}{(2\pi)^{d}} \frac{1}{k^2 \left[ (k-p)^2 \right]^{2-\frac{d}{2}}} = \frac{i}{(4\pi)^{d/2}}(-1)^{\frac{d}{2} -3} \frac{\G \left( \frac{d}{2} -1 \right) \G(d-2)}{\G\left(\frac{3d}{2}-3 \right)} \frac{\G(3-d)}{\G \left(2 - \frac{d}{2} \right)} (-p^2) ^{d-3} \, .
\ee
Substituting the above result in \eqref{s1 first} we get
\be \label{s1}
S_1(p^2) =\frac{(-1)^{d-4}}{(4\pi)^{d}} \frac{\G(3-d)\left[\G \left(\frac{d}{2}-1 \right)\right]^3}{\G \left( \frac{3d}{2}-3 \right)}(-p^2)^{d-3}\, .
\ee
In the $\epsilon$-expansion we finally obtain
\be \label{s1 exp}
S_1(p^2) = \frac{p^2}{2(4\pi)^4} \left[ \frac{2}{\epsilon} - \ln \left( \frac{-p^2e^\g}{4\pi} \right) + \frac{13}{4} \right]\, .
\ee

\subsubsection{The Double Candy integral}

The double candy integral is associated with the  following two diagrams:
\be
\begin{gathered}
\begin{tikzpicture}
\begin{feynman}
  \vertex (x);
     \vertex[right= 1cm of x](z);
     \vertex[below = 0.5cm of x ](u);
     \vertex[right = 0.5cm of u](v);
     \vertex[below=  0.5cm of v](y);
     \vertex[below = 0.5cm of  y](l);
     \vertex[below=2cm of x](r);
     \vertex[right = 1 cm of r](p);
     \diagram{
     (x)--(v) --(z)
     };
     \diagram{
     (v)--[half left](y)--[half left](v)
     };
     \diagram{(y)--[half left](l)--[half left](y) };
     \diagram{(r)--(l)--(p)};
\end{feynman}
\end{tikzpicture}
\end{gathered} =   i \frac{\l^3}{4}\left[ L_1(p^2) \right]^2
\ee

\be \label{2loop counter diags}
\begin{gathered}
\begin{tikzpicture}
\begin{feynman}
\vertex(x);
\vertex[right = 1 cm of x](z);
 \vertex[right= 1cm of x](z);
     \vertex[below = 0.5cm of x ](u);
     \vertex[right = 0.5cm of u](v);
     \vertex[below= 1cm of v](y);
     \vertex[below=2cm of x](r);
     \vertex[right = 1 cm of r](p);
     \diagram{(x)--(v)--(z)};
     \diagram{(v)--[half left](y)--[half left](v)};
     \diagram{(r)--(y)--(p)};
\end{feynman}
 \filldraw[fill=black] (y) ++(-0.1cm, -0.1cm) rectangle ++(0.2cm, 0.2cm);
  \draw[thick, white] (y) ++(-0.1cm, 0) -- ++(0.2cm, 0);
  \draw[thick, white] (y) ++(0, -0.1cm) -- ++(0, 0.2cm);
\end{tikzpicture}
\end{gathered}  =\begin{gathered}
\begin{tikzpicture}
\begin{feynman}
\vertex(x);
\vertex[right = 1 cm of x](z);
 \vertex[right= 1cm of x](z);
     \vertex[below = 0.5cm of x ](u);
     \vertex[right = 0.5cm of u](v);
     \vertex[below= 1cm of v](y);
     \vertex[below=2cm of x](r);
     \vertex[right = 1 cm of r](p);
     \diagram{(x)--(v)--(z)};
     \diagram{(v)--[half left](y)--[half left](v)};
     \diagram{(r)--(y)--(p)};
\end{feynman}
 \filldraw[fill=black] (v) ++(-0.1cm, -0.1cm) rectangle ++(0.2cm, 0.2cm);
  \draw[thick, white] (v) ++(-0.1cm, 0) -- ++(0.2cm, 0);
  \draw[thick, white] (v) ++(0, -0.1cm) -- ++(0, 0.2cm);
  \end{tikzpicture}
\end{gathered}=\frac{\l}{2}\d_\l L_1(p^2) = \frac{-i}{2} \frac{3\l^3}{2} L_1(p^2)L_1(-\m^2)
\ee
For the diagrams in \eqref{2loop counter diags} we use the result \eqref{vertex ct 1}.
As we discussed in the previous section, these diagrams are responsible for the cancellation of the non-local divergences.
Using \eqref{L1 int} we get
\be \label{double candy pp} {
\left[ L_1(p^2) \right]^2 =  - \frac{\left[\G\left( 2-\frac{d}{2}\right)\right]^2 \left[\G \left(\frac{d}{2}-1 \right)\right]^4}{\left( 4\pi \right)^{d} 
\left[\G\left(d-2 \right)\right]^2} \left(-p^2 \right)^{d-4}}
\ee
\be \label{double candy p m} {
L_1(p^2)L_1(-\m^2) = - \frac{\left[\G\left( 2-\frac{d}{2}\right)\right]^2 \left[\G \left(\frac{d}{2}-1 \right)\right]^4}{\left( 4\pi \right)^{d} 
\left[\G\left(d-2 \right)\right]^2} \left(-p^2 \right)^{\frac{d}{2}-2}\left(\m^2 \right)^{\frac{d}{2}-2}}
\ee
In the $\epsilon$-expansion these are:
\be 
\left[ L_1(p^2) \right]^2 =\frac{1}{(4\pi)^4} \left[ - \frac{4}{\epsilon^2} + \frac{4 \ln \left(\frac{-p^2e^\g}{4\pi} \right) -8}{\epsilon} - 
2 \ln^2 \left( \frac{-p^2e^\g}{4\pi} \right) + 8 \ln \left( \frac{-p^2e^\g}{4\pi} \right) + \frac{\pi^2}{6} -12\right]
\ee
and
\be \bal
\left. L_1(p^2)L_1(-\m^2) \right.= &-\frac{4}{\epsilon^2} + \frac{2\ln \left( \frac{-p^2e^\g}{4\pi}\right) + 2\ln \left( \frac{\m^2e^\g}{4\pi}\right) -8}{\epsilon}\\
& -2 \left[\frac{1}{2} \ln \left( \frac{-p^2e^\g}{4\pi}\right) + \frac{1}{2} \ln \left( \frac{\m^2e^\g}{4\pi}\right)  \right]^2 \\
&+  8 \left[\frac{1}{2} \ln \left( \frac{-p^2e^\g}{4\pi}\right) + \frac{1}{2} \ln \left( \frac{\m^2e^\g}{4\pi}\right)  \right]+ \frac{\pi^2}{6} -12
\eal
\ee
where the terms $\frac{\ln{(-p^2)}}{\epsilon}$ are the non-local divergences.

\subsubsection {The Ice Cream intergral}

We will now evaluate the loop integral given by the following diagram:
\be
\bal
\begin{gathered}
\begin{tikzpicture}
\begin{feynman}
 \vertex (x){$p_1$};
     \vertex[right= 1cm of x](z){$p_2$};
     \vertex[below = 0.5cm of x ](u);
     \vertex[right = 0.2 cm of u](v);
     \vertex[right = 0.5cm of v](q);
     \vertex[below=  1cm of v](y);
     \vertex[right = 0.25cm of y](l);
     \vertex[below=2cm of x](r){$p_3$};
     \vertex[right = 1 cm of r](p){$p_4$};
     \diagram{
     (x)-- (v)--[half left](q)--[half left](v)
     };
     \diagram{
     (z)--(q)
     };
     \diagram{(v)--(l)-- (q)};
      \diagram{(r)--(l)--(p)};
\end{feynman}
\end{tikzpicture}
\end{gathered} &=\frac{i \l^3}{2} \int \frac{\mathrm{d}^dk_1\mathrm{d}^dk_2}{\left(2\pi\right)^{2d}} \frac{1}{k_1^2 \left( k_1 +p\right)^2 k_2^2 \left( k_1 + p_2 - k_2 \right)^2} \; , \; p =p_3 +p_4 \\
&=\frac{i \l^3}{2}I_4 (p, p_2)
\eal
\ee
We can evaluate the integral with respect to $k_2$ using the result \eqref{L1 int}:
\be \label{ice int}
\bal
I_4 (p, p_2)\equiv &\int \frac{\mathrm{d}^dk_{1,2}}{\left(2\pi\right)^{2d}} \frac{1}{k_1^2 \left( k_1 +p\right)^2 k_2^2 \left( k_1 + p_2 - k_2 \right)^2}  \\
&=\int \frac{\mathrm{d}^d k}{\left(2\pi\right)^d} \frac{L_1(k+p_2)}{k^2 \left( k +p\right)^2}\\
&= \frac{i\G\left( 2-\frac{d}{2}\right) \left[\G \left(\frac{d}{2}-1 \right)\right]^2}{\left( 4\pi \right)^{d/2} \G\left(d-2 \right)}  
\int \frac{\mathrm{d}^d k}{\left(2\pi\right)^d} \frac{1}{k^2 \left( k +p\right)^2  \left[\left( k + p_2 \right)^2\right]^{2- \frac{d}{2}}}\, .
\eal
\ee
Following the same procedure as in the Appendix of Chapter 9 of \cite{Amit:1984ms} we introduce, consecutively, two Feynman parameters.
The first Feynman parametrization  gives
\be
\bal
 \int \frac{\mathrm{d}^d k}{\left(2\pi\right)^d} \frac{1}{k^2 \left( k +p\right)^2 k_2^2 \left[\left( k_1 + p_2 \right)^2\right]^{2- \frac{d}{2}}} =  
 \int_0^1 \mathrm{d}y \int \frac{\mathrm{d}^d k}{\left(2\pi\right)^d} \frac{1}{\left[k^2 +2k \cdot p y + p^2 y\right]^2 \left[\left( k + p_2 \right)^2\right]^{2- \frac{d}{2}}}
\eal
\ee
The second Feynman parametrizatation, after completing the squares in the denominator, gives
\be\bal
\frac{\G \left(4 - \frac{d}{2} \right)}{\G \left(2- \frac{d}{2} \right)}&\int_0^1\int_k  \frac{z(1-z)^{1-\frac{d}{2}}}{\left[  \left( k+ pyz +(1-z)p_2 \right)^2 + yz(1-yz) p^2 + z(1-z)p_2^2 - 2 yz(1-z) p \cdot p_2 \right]^{4- \frac{d}{2}} }
\eal\ee
Now we can evaluate the standard loop integral with respect to $k$ using \eqref{stand int}:
\be
i (-1)^{\frac{d}{2} -4} \frac{\G (4-d)}{\G \left(2 - \frac{d}{2} \right)} \int_0^1 \mathrm{d}y \mathrm{d}z  \frac{z(1-z)^{1-\frac{d}{2}}}{\left[2 yz(1-z)p \cdot p_2 - yz(1-yz)p^2 - z(1-z)p_2^2 \right]^{4-d} }
\ee
For $d=4-\epsilon$ we get the following expression which is in agreement, up to normalization constants, with eqs. (A9-37) of \cite{Amit:1984ms}: 
\be\label{num pref} \bal
i\frac{\G(\epsilon)}{\G\left( \frac{\epsilon}{2} \right)}\int_0^1 \mathrm{d}y \mathrm{d}z  
\frac{z(1-z)^{-1+\frac{\epsilon}{2}}}{\left[-2 yz(1-z)p\cdot p_2 +yz(1-yz)p^2 + z(1-z)p_2^2 \right]^{\epsilon} }=\\
= i\frac{\G(\epsilon)}{\G\left( \frac{\epsilon}{2} \right)}\int_0^1 \mathrm{d}y \mathrm{d}z z(1-z)^{-1+\frac{\epsilon}{2}} f(z,y,p,p_2)
\eal
\ee
In the integral we cannot set $\epsilon = 0$ because of the singularity for $z=1$. Instead, we add and subtract $f(1,y,p,p_2)$: 
\be
\bal
&f(z,y,p,p_2) = f(1,y,p,p_2) + \left[ f(z,y,p,p_2)-f(1,y,p,p_2)\right] \\
&= \left[ y(1-y)p^2 \right]^{-\epsilon} + \left\{\left[-2 yz(1-z) p \cdot p_2 +yz(1-yz)p^2 + z(1-z)p_2^2 \right]^{-\epsilon} - \left[ y(1-y)p^2 \right]^{-\epsilon} \right\} \\
&=  \left[ y(1-y)p^2 \right]^{-\epsilon} - \epsilon \ln \left( \frac{-2 yz(1-z) p \cdot p_2 +yz(1-yz)p^2 + z(1-z)p_2^2}{y(1-y)p^2} \right) + O(\epsilon^2)
\eal
\ee
For $z\rightarrow 1$ the logarithm in the above expression vanishes, so the integral with respect to $z$ is convergent. 
The factor of $\epsilon$ will be cancelled out by the $\frac{1}{\epsilon}$ from expansion of the $\G$-function  in \eqref{ice int}. 
We define the contribution associated with the complicated logarithm as: 
\be\label{G-contr}
G(p,p_2)=- \frac{\G(\epsilon) \left[ \G \left(1-\frac{\epsilon}{2}\right) \right]^2}{(4\pi)^d\G\left( 2-\epsilon \right)} \int_0^1 
\mathrm{d}y \mathrm{d}z (1-z)^{-1+\frac{\epsilon}{2}} z  \left[ f(z,y,p,p_2)-f(1,y,p,p_2)\right]\, .
\ee
Expanding in $\epsilon$ we obtain: 
\be\label{Gpp2}
G(p,p_2)=\int_0^1 \mathrm{d}z \mathrm{d}y \frac{z}{1-z} \ln \left( \frac{-2 yz(1-z) p \cdot p_2 +yz(1-yz)p^2 + z(1-z)p_2^2}{y(1-y)p^2} \right) + O(\epsilon)
\ee
We can see that  the pole for $z\rightarrow 1$ has been eliminated since
\be
\lim_{z\rightarrow 1}\frac{z}{1-z} \ln \left( \frac{-2 yz(1-z) p \cdot p_2 +yz(1-yz)p^2 + z(1-z)p_2^2}{y(1-y)p^2} \right) =  \frac{p^2 y (1-2 y)+2 p \cdotp_2 y-p_2^2}{p^2 (y-1) y}
\ee
If we consider the renormalization conditions at the $S.P.$, we  have that $p\cdot p_2 \sim p^2 \sim p_2^2 \sim -\m^2$
so the argument of the logarithm is a function of only $y$ and $z$.  
This has a strong  impact on  the renormalization procedure, since $G$, does not contribute any $\m$-dependent term in the renormalized expression.
Substituting \eqref{G-contr} in \eqref{ice int}  for $d=4-\epsilon$, we get: 
\be \label{I4 first}
\bal
I_4 (p,p_2) &=- \frac{\G(\epsilon) \left[ \G \left(1-\frac{\epsilon}{2}\right) \right]^2}{(4\pi)^d\G\left( 2-\epsilon \right)} \int_0^1 
\mathrm{d}y \mathrm{d}z (1-z)^{-1+\frac{\epsilon}{2}} z \left \{ f(1,y,p,p_2) + \left[ f(z,y,p,p_2)-f(1,y,p,p_2)\right] \right\} \\
&=- \frac{\G(\epsilon) \left[ \G \left(1-\frac{\epsilon}{2}\right) \right]^2}{(4\pi)^d\G\left( 2-\epsilon \right)} \int_0^1 \mathrm{d}y 
\mathrm{d}z (1-z)^{-1+\frac{\epsilon}{2}} z \left[ y(1-y)p^2 \right]^{-\epsilon} + G (p,p_2)\\
&=- \frac{\G(\epsilon) \left[ \G \left(1-\frac{\epsilon}{2}\right) \right]^2}{(4\pi)^d\G\left( 2-\epsilon \right)} J + G(p,p_2)
\eal
\ee
where : 
\be
\bal
J&= \int_0^1 \mathrm{d}y \mathrm{d}z (1-z)^{-1+\frac{\epsilon}{2}} z \left[ y(1-y)p^2 \right]^{-\epsilon} \\
\eal
\ee
Using Euler's $\Beta$-function \eqref{euler b} we can evaluate the integrals with respect to the Feynman parameters.
Then we arrive at the following result: 
\be
J =\frac{\G \left(\frac{\epsilon}{2} \right) \G (2)}{\G \left( 2 + \frac{\epsilon}{2} \right)}\frac{\left[\G (1-\epsilon )\right]^2}{\G (2-2 \epsilon )}\left(p^2\right)^{-\epsilon } 
\ee
Substituting the above expression in\eqref{I4 first} we get 
\be \label{I4 epsilon}
I_4 = - \frac{\G(\epsilon) \left[ \G \left(1-\frac{\epsilon}{2}\right) \right]^2}{(4\pi)^d\G\left( 2-\epsilon \right)}
\frac{\G \left(\frac{\epsilon}{2} \right) \G (2)}{\G \left( 2 + \frac{\epsilon}{2} \right)}\frac{\left[\G (1-\epsilon )\right]^2}{\G (2-2 \epsilon )}\left(p^2\right)^{-\epsilon } + G(p,p_2)
\ee
We perform the $\epsilon$-expansion and we finally obtain
\bea \label{I4 exp}
{I_4(p,p_2)}&=&\frac{1}{(4\pi)^4} \left[- \frac{2}{\epsilon ^2}+ \frac{2 \ln \left(\frac{p^2 e^\g}{4\pi}\right) - 5}{\epsilon } - \ln^2 \left(\frac{p^2e^\g}{4\pi}\right) +
 5 \ln \left(\frac{p^2e^\g}{4\pi} \right)- \frac{\pi^2}{4} - \frac{23}{2}\right]\nonumber\\ 
&+& G(p,p_2)
\eea

\subsection{Massless 3-loop integrals}

\subsubsection{The Watermelon integral}

This integral is associated with the 3-loop diagram that appears in the two point  function $\braket{\phi^4 \phi^4}$, which is related to 
$\braket{K_3 K_3}$ through the equation of motion.
It is the diagram
\be
 \begin{gathered}
\begin{tikzpicture}
\begin{feynman}
    \vertex (x);
    \vertex[below = 1cm of x] (y);
    \diagram{
        (x) -- [half left] (y) -- [half left] (x) -- [quarter left] (y) -- [quarter left] (x)
    };
\end{feynman}
\filldraw[fill = black] (x) ++(-0.1cm, -0.1cm) rectangle ++(0.2cm, 0.2cm);
\filldraw[fill = black] (y) ++(-0.1cm, -0.1cm) rectangle ++(0.2cm, 0.2cm);

\end{tikzpicture}
\end{gathered} = 4! \int \frac{\mathrm{d}^dk}{(2\pi)^d} \frac{1}{k^2}S_1(p-k)= 4! W(p^2)\, .
\ee
Using  \eqref{s1} the above integral reduces to the following one loop integral:
\be
W(p^2)=\frac{(-1)^{2d-7}}{(4\pi)^{d}} \frac{\G(3-d)\left[\G \left(\frac{d}{2}-1 \right)\right]^3}{\G \left( \frac{3d}{2}-3 \right)}\int \frac{\mathrm{d}^dk}{(2\pi)^d} \frac{1}{k^2 \left[(k-p)^2 \right]^{3-d}}\, ,
\ee
which is easy to evaluate. We introduce a Feynman parameter and the we perform the standard loop integral:
\be
\int \frac{\mathrm{d}^dk}{(2\pi)^d} \frac{1}{k^2 \left[(k-p)^2 \right]^{3-d}} = \int_0^1 \mathrm{d}y \int \frac{\mathrm{d}^dk}{(2\pi)^d}  \frac{(3-d)y^{2-d}}{\left[ k^2 - 2k\cdot p y + p^2 y \right]^{4-d}}
\ee
We then complete the squares in the denominator and we apply the standard shift in the loop momenta $k \to k+py$.  
Then we evaluate the standard loop integral using \eqref{stand int} and we obtain 
\be
\int \frac{\mathrm{d}^dk}{(2\pi)^d} \frac{1}{k^2 \left[(k-p)^2 \right]^{3-d}} = \frac{i(-1)^{d-4}}{(4\pi)^{d/2}} 
\frac{\G \left(4-\frac{3d}{2} \right)}{\G(4-d)} (3-d) \int_0^1 \mathrm{d}y  y^{\frac{d}{2}-2}\left(1-y\right)^{\frac{3d}{2} -4} (-p^2)^{\frac{3d}{2} -4} 
\ee
The integral with respect to the Feynman parameter is an Euler's $B$-function:
\be
\int \frac{\mathrm{d}^dk}{(2\pi)^d} \frac{1}{k^2 \left[(k-p)^2 \right]^{3-d}}  =\frac{i(-1)^{d-4}}{(4\pi)^{d/2}} 
\frac{\G \left(4-\frac{3d}{2} \right)}{\G(3-d)} \frac{\G \left(\frac{d}{2} -1 \right)\G \left( \frac{3d}{2}-3 \right) }{\G (2d-4)}(-p^2)^{\frac{3d}{2}-4}
\ee
and we conclude that:
\be
W(p^2) = \frac{i (-1)^{3d-11}}{(4\pi)^{3d/2}} \frac{\left[\G \left( \frac{d}{2}-1\right) \right]^4 \G\left(4-\frac{3d}{2} \right)}{\G(2d-4)} (-p^2)^{\frac{3d}{2}-4} \, .
\ee
Expanding in $\epsilon$, we obtain:
\be
W(p^2)  = - \frac{p^4}{(4\pi)^6} \left[ \frac{1}{18 \epsilon} - \frac{\ln \left( \frac{-p^2 e^\g}{4\pi} \right) - 71}{216} \right]\, .
\ee

\subsubsection{The Sunset-Tadpole integral}

This integral is associated with the following 3-loop diagram which appears in the $\left< \mathcal{O}_4(p_1)\phi(p_2) \phi(p_3) \right>$ correlator..
\be
\bal
\begin{gathered}
\begin{tikzpicture}
\begin{feynman}
\vertex(x);
\diagram [horizontal=a to b, layered layout] {
  a -- b --b--c
};
\path (b)--++(90:0.5) coordinate (A);
\path  (b)--++(90:1) coordinate (B);
\draw (A) circle(0.5);
\draw (B)circle(0.25);
\path (B)--++(1 :-0.2)coordinate (C);
\end{feynman}
\filldraw[fill=black] (C) ++(-0.13cm, -0.1cm) rectangle ++(0.15cm, 0.15cm);
 \end{tikzpicture}
\end{gathered} =\begin{gathered}
\begin{tikzpicture}
\begin{feynman}
\vertex(x);
\diagram [horizontal=a to b, layered layout] {
  a -- b --b--c
};
\path (b)--++(90:0.5) coordinate (A);
\path  (b)--++(90:1) coordinate (B);
\draw (A) circle(0.5);
\draw (B)circle(0.25);
\path (B)--++(1 :0.2)coordinate (C);
\end{feynman}
\filldraw[fill=black] (C) ++(-0.04cm, -0.1cm) rectangle ++(0.15cm, 0.15cm);
 \end{tikzpicture}
\end{gathered}&= -i\l^2\frac{i}{p_2^2}\frac{i}{p_3^2} \int \frac{\mathrm{d}^dk_{1,2}}{(2\pi)^{2d}} \frac{\mathrm{d}^dl}{(2\pi)^d} \frac{1}{k_1^2 k_2^2 \left( k_1+ k_2 -l\right)^2 l^2 \left(l-p_1\right)^2} \\
&=-i\l^2\frac{i}{p_2^2}\frac{i}{p_3^2} \int \frac{\mathrm{d}^dl}{(2\pi)^d} \frac{S_1(l)}{l^2 \left(l-p_1 \right)^2}\\
&=-i\l^2\frac{i}{p_2^2}\frac{i}{p_3^2} ST(p_1^2)
\eal
\ee
where
\be
ST(p_1^2)=\int \frac{\mathrm{d}^dl}{(2\pi)^d} \frac{S_1(l)}{l^2 \left(l-p_1 \right)^2}.
\ee
Substituting \eqref{s1} for the Sunset this becomes
\be\label{ST def}
ST(p_1^2)=\frac{(-1)^{d-4}}{(4\pi)^{d}} \frac{\G(3-d)\left[\G \left(\frac{d}{2}-1 \right)\right]^3}{\G \left( \frac{3d}{2}-3 \right)}
\int \frac{\mathrm{d}^dl}{(2\pi)^d} \frac{1}{l^2 \left(l-p_1\right)^2 \left(-l^2\right)^{3-d}}\, .
\ee
We introduce a Feynman parameter. After the Feynman parametrization we get for the integral
\be
\bal
\int \frac{\mathrm{d}^dl}{(2\pi)^d} \frac{1}{l^2 \left(l-p_1\right)^2 \left(-l^2\right)^{3-d}} &=(-1)^{d-3} 
\int_0^1 \mathrm{d}y\int \frac{\mathrm{d}^dl}{(2\pi)^d} \frac{(1-y)^{3-d}(4-d)}{\left[l^2 -2l\cdot p_1 y +p_1^2y \right]^{5-d} }\\
&= \frac{i (-1)^{2d-8}}{(4\pi)^{d/2}}  \frac{\G\left(5-\frac{3d}{2} \right)}{\G(4-d)}\int_0^1 \mathrm{d}y  
\frac{(1-y)^{3-d}}{\left[y(1-y) \right]^{5-\frac{3d}{2}} \left( -p_1^2 \right)^{\frac{3d}{2}-5}}\, ,
\eal
\ee
where we have completed the squares in the denominator and shifted the loop momenta as $l \rightarrow l-p_1y$. 
The integral with respect to the Feynman parameter is an Euler $\Beta$-function. Substituting in \eqref{ST def} we conclude that
\be\label{st general d} \bal
ST(p_1^2) &= \frac{i(-1)^{3d-12}} {(4\pi)^{3d/2}} \frac{\left[\G \left( \frac{d}{2}-1 \right) \right]^{4} \G(3-d) \G \left(5 - \frac{3d}{2} \right) 
\G\left( \frac{3d}{2} -4\right)}{\G \left( \frac{3d}{2}-3 \right) \G(4-d) \G(2d-5)}\left(-p_1^2 \right)^{\frac{3d}{2}-5} 
\eal
\ee
In the $\epsilon$-expansion this is
\be \label{st e exp}
ST({p_1}^2) = i\frac{p_1^2}{(4\pi)^6} \left[ -\frac{1}{6\epsilon} + \frac{1}{4} \ln \left(\frac{-p_1^2e^\g}{4\pi} \right) - \frac{25}{24} \right] \, .
\ee
This integral is also part of the following two point function diagram of $\braket{\phi^2 \phi^4}$:
\be
\begin{tikzpicture}
\begin{feynman}
\vertex(x);
\vertex[right = 1cm of x](z);
\vertex[above =0.8cm of z](x1);
\vertex[below =0.8cm of z](x2);
\vertex[right = 1 cm of z](y);
\diagram{(x)--(x1)--[quarter left](x)--[quarter left](x1)-- (y)--[quarter left](x)};
\end{feynman}
\filldraw[fill=black] (x) ++(-0.1cm, -0.1cm) rectangle ++(0.2cm, 0.2cm);
\filldraw[fill =black](y) circle(3pt );
\end{tikzpicture}
\ee

\subsubsection{The Tent integrals}

The Tent integrals arise as double limits of the following diagrams:
\be
\bal
\begin{gathered}
\begin{tikzpicture}
\begin{feynman}
\vertex(x);
\vertex[right = 1cm of x](x1);
\vertex[right = 0.5cm of x1](y1);
\vertex[below = 1.5cm of y1](y2);
\vertex[below = 0.5cm of x](r1);
\vertex[right = 0.5cm of r1](r);
\vertex[below = 0.5cm of r](z);
\vertex[below=0.5 of z](z1);
\vertex[right = 0.5cm of z1](x3);
\vertex[left = 0.5cm of z1](x4);
\diagram{(x)--(r)--(x1)};
\diagram{(r)--[half left](z)--[half left](r)};
\diagram{(x3)--(z)--(x4)};
\diagram{(y1)--(y2)};
\end{feynman}
\end{tikzpicture}
\end{gathered}\longrightarrow
\begin{gathered}
\begin{tikzpicture}
\begin{feynman}
\vertex (x);
\vertex[right= 0.5cm of x](x1);
\vertex[right= 1cm of x1](x2);
\vertex[right= 1cm of x2](x3);
\vertex[right= 0.5cm of x3](x4);
\vertex[above= 1cm of x2](x5);
\diagram{(x)--(x1)--[quarter left](x5)--[quarter left](x1)--(x3)--[quarter left](x5)--[quarter left](x3)--(x4)};
\end{feynman}
\filldraw[fill=black] (x1) ++(-0.1cm, -0.1cm) rectangle ++(0.2cm, 0.2cm);
\end{tikzpicture}
\end{gathered}
\eal
\ee
\be\label{top tent limit}
\bal
\begin{gathered}
\begin{tikzpicture}
\begin{feynman}
\vertex(x);
\vertex [right = 0.5cm of x](x1);
\vertex [right = 1 cm of x1](x2);
\vertex [right = 0.5 cm of x2](x3);
\vertex [above = 0.5cm of x1](x4);
\vertex [below= 0.5cm of x1](x5);
\vertex [above = 0.5cm of x2](x6);
\vertex [below= 0.5cm of x2](x7);
\diagram{(x)--(x3)};
\diagram{(x4)--(x5)};
\diagram{(x6)--(x7)};
\end{feynman}
\end{tikzpicture} 
\end{gathered}\longrightarrow \begin{gathered}
\begin{tikzpicture}
\begin{feynman}
\vertex (x);
\vertex[right= 0.5cm of x](x1);
\vertex[right= 1cm of x1](x2);
\vertex[right= 1cm of x2](x3);
\vertex[right= 0.5cm of x3](x4);
\vertex[above= 1cm of x2](x5);
\diagram{(x)--(x1)--[quarter left](x5)--[quarter left](x1)--(x3)--[quarter left](x5)--[quarter left](x3)--(x4)};
\end{feynman}
\filldraw[fill=black] (x5) ++(-0.1cm, -0.1cm) rectangle ++(0.2cm, 0.2cm);
\end{tikzpicture}
\end{gathered}
\eal
\ee
They arise from the 3-loop diagram in the $\braket{ \mathcal{O}_4(p_1)\phi(p_2) \phi(p_3) }$ correlator.
There are two types of Tents.

{\bf Tent with an insertion on the base }

This Tent diagram can be expressed as:
\be
\bal
\begin{gathered}
\begin{tikzpicture}
\begin{feynman}
\vertex (x);
\vertex[right= 0.5cm of x](x1);
\vertex[right= 1cm of x1](x2);
\vertex[right= 1cm of x2](x3);
\vertex[right= 0.5cm of x3](x4);
\vertex[above= 1cm of x2](x5);
\diagram{(x)--(x1)--[quarter left](x5)--[quarter left](x1)--(x3)--[quarter left](x5)--[quarter left](x3)--(x4)};
\end{feynman}
\filldraw[fill=black] (x1) ++(-0.1cm, -0.1cm) rectangle ++(0.2cm, 0.2cm);
\end{tikzpicture}
\end{gathered} &=\frac{3}{2}i^9 \l^2 \int \frac{\mathrm{d}^dk_{1,2,3 }\mathrm{d}^d l}{(2\pi)^{3d}}
\frac{\d(p_1 -k_1-k_2+p_3+k_3) \d(k_1+k_2-k_3 +p_2)}{p_2^2 p_3^2 k_1^2 k_2^2 k_3^2 l^2 (l-k_1-k_2)^2} \\
&=  -\frac{3}{2}i \l^2 \frac{i}{p_2^2}\frac{i}{p_3^2}\tilde{\d}(p_1 +p_2 +p_3) \int \frac{\mathrm{d}^dk_{1,2,}}{(2\pi)^{2d}} 
\frac{L_1(k_1+k_2)}{k_1^2 k_2^2 (k_1+k_2+p_2)^2} \; , \; k_1 \rightarrow k_1 -k_2  \\
&=-\frac{3}{2}i\l^2 \frac{i}{p_2^2}\frac{i}{p_2^2} \int  \frac{\mathrm{d}^dk_{1,2,}}{(2\pi)^{2d}}
\frac{L_1(k_1)}{(k_1-k_2)^2 k_2^2 (k_1+p_2)^2} \tilde{\d}(p_1 +p_2 +p_3) \\
&= -\frac{3}{2}i\l^2 \frac{i}{p_2^2}\frac{i}{p_2^2} \int  \frac{\mathrm{d}^dk_{1}}{(2\pi)^{d}}\frac{\left[L_1(k_1)\right]^2}{ (k_1+p_2)^2} \tilde{\d}(p_1 +p_2 +p_3)\\
&= -\frac{3}{2}i\l^2 \frac{i}{p_2^2}\frac{i}{p_2^2}  TB(p_2^2)\tilde{\d}(p_1 +p_2 +p_3)
\eal
\ee
with :
\be\bal
TB(p_2^2) &= \int  \frac{\mathrm{d}^dk}{(2\pi)^{d}}\frac{\left[L_1(k)\right]^2}{ (k+p_2)^2} \\
&= - \frac{\left[\G\left( 2-\frac{d}{2}\right)\right]^2 \left[\G \left(\frac{d}{2}-1 \right)\right]^4}{\left( 4\pi \right)^{d} 
\left[\G\left(d-2 \right)\right]^2}  \int \frac{\mathrm{d}^d k}{(2\pi)^d}   \frac{1}{\left(-k^2\right)^{4-d}(k+p_2)^2}  \\
&= - \frac{\left[\G\left( 2-\frac{d}{2}\right)\right]^2 \left[\G \left(\frac{d}{2}-1 \right)\right]^4}{\left( 4\pi \right)^{d} 
\left[\G\left(d-2 \right)\right]^2} \int_0^1 \mathrm{d}y \int \frac{\mathrm{d}^d k}{(2\pi)^d}  \frac{(-1)^{d-4}(4-d)(1-y)^{3-d}}{\left[k^2 + 2k\cdot p_2 y  + p_2^2y \right]^{5-d}}
\eal\ee
where we have used \eqref{double candy pp}. We shift $k \rightarrow k-p_2y$ and we get:
\be
\bal
TB(p_2^2) &= - \frac{\left[\G\left( 2-\frac{d}{2}\right)\right]^2 \left[\G \left(\frac{d}{2}-1 \right)\right]^4}{\left( 4\pi \right)^{d} \left[\G\left(d-2 \right)\right]^2} \int_0^1 \mathrm{d}y \int \frac{\mathrm{d}^d k}{(2\pi)^d}  \frac{(-1)^{d-4}(4-d)(1-y)^{3-d}}{\left[k^2    + p_2^2y(1-y) \right]^{5-d}}
\eal
\ee
This is a standard loop integral that can be evaluated using \eqref{stand int}:
\be
TB(p_2^2)=(-1)^{2d-10}i \frac{\left[\G\left( 2-\frac{d}{2}\right)\right]^2 \left[\G \left(\frac{d}{2}-1 \right)\right]^4}{\left( 4\pi \right)^{3d/2} 
\left[\G\left(d-2 \right)\right]^2} \frac{\G\left( 5-\frac{3d}{2} \right)}{\G (4-d)} \int_0^1  \mathrm{d}y 
\frac{(1-y)^{3-d}}{\left[y(1-y) \right]^{5-\frac{3d}{2}}} \left(- p_2^2 \right)^{\frac{3d}{2}-5}
\ee
The integral with respect to the Feynman parameter can be evaluated using Euler's Beta function:
\be\label{TB general d}
{TB(p^2)=i(-1)^{2d-10} \frac{\left[\G\left( 2-\frac{d}{2}\right)\right]^2 \left[\G \left(\frac{d}{2}-1 \right)\right]^5}
{\left( 4\pi \right)^{3d/2} \left[\G\left(d-2 \right)\right]^2} \frac{\G\left( 5-\frac{3d}{2} \right)}{\G (4-d)} \frac{\G \left(\frac{3d}{2}-4 \right)}{\G (2d-5)} \left(- p^2 \right)^{\frac{3d}{2}-5}}
\ee
In the $\epsilon$ expansion it takes the form: 
\be
\bal
TB(p^2)&=\frac{ip^2}{(4\pi)^6} \left[ \frac{4}{3 \epsilon^2} -\frac{2\ln\left( \frac{-p^2e^\g}{4\pi} \right) -\frac{20}{3}}{\epsilon}  + \frac{3}{2} \ln^2 \left( \frac{-p^2e^\g}{4\pi}\right) -10 \ln \left(\frac{-p^2e^\g}{4\pi}\right)  -\frac{\pi^2}{12} +\frac{64}{3}\right]
\eal
\ee

{\bf Tent with an insertion on the top}

It is important to note that, from topological point of view, this integrals has to be crossing symmetric.  
So we have to  carefully write down the loop integral of this diagram. Recalling that we have to consider the limit \eqref{top tent limit} in  order to get this diagram, we have that:

\bea 
&& \begin{gathered}
\begin{tikzpicture}
\begin{feynman}
\vertex (x);
\vertex[right= 0.5cm of x](x1);
\vertex[right= 1cm of x1](x2);
\vertex[right= 1cm of x2](x3);
\vertex[right= 0.5cm of x3](x4);
\vertex[above= 1cm of x2](x5);
\diagram{(x)--(x1)--[quarter left](x5)--[quarter left](x1)--(x3)--[quarter left](x5)--[quarter left](x3)--(x4)};
\end{feynman}
\filldraw[fill=black] (x5) ++(-0.1cm, -0.1cm) rectangle ++(0.2cm, 0.2cm);
\end{tikzpicture}
\end{gathered}\nonumber\\
&=& -i 12 \l^2 \frac{i} {p_2^2} \frac{i}{p_3^2} \int \frac{\mathrm{d}^dk_{1,2,3,4}}{(2\pi)^{3d}}
\frac{\d(p_1-k_1 -k_2 +k_3 +k_4) \d(k_1+k_2-k_3-k_4 +p_2+p_3)}{k_1^2 k_2^2 k_3^2 k_4^2 (k_1+k_2+p_2)^2}\nonumber\\
&=& -i6\l^2 \frac{i}{p_2^2}\frac{i}{p_3^2}  \int \frac{\mathrm{d}^dk_{1,2}}{(2\pi)^{2d}} \left[ \frac{L_1 (k_1 +k_2 -p_1)}{k_1^2 k_2^2 (k_1 +k_2 +p_2)^2} +\frac{L_1 (k_1 +k_2 -p_1)}{k_1^2 k_2^2 (k_1 +k_2 +p_3)^2}\right] \tilde{\d}(p_1 +p_2+p_3)\nonumber\\
\eea
We can shift the loop momenta in the above expression as $k_1 \rightarrow k_1-k_2$ and get the following form:

\be\bal
\begin{gathered}
\begin{tikzpicture}
\begin{feynman}
\vertex (x);
\vertex[right= 0.5cm of x](x1);
\vertex[right= 1cm of x1](x2);
\vertex[right= 1cm of x2](x3);
\vertex[right= 0.5cm of x3](x4);
\vertex[above= 1cm of x2](x5);
\diagram{(x)--(x1)--[quarter left](x5)--[quarter left](x1)--(x3)--[quarter left](x5)--[quarter left](x3)--(x4)};
\end{feynman}
\filldraw[fill=black] (x5) ++(-0.1cm, -0.1cm) rectangle ++(0.2cm, 0.2cm);
\end{tikzpicture}
\end{gathered} &= -i6\l^2 \frac{i}{p_2^2}\frac{i}{p_3^2}   \int \frac{\mathrm{d}^dk}{(2\pi)^d}L_1(k)L_1(k-p_1) \left[\frac{1}{(k+p_2)^2} + \frac{1}{(k+p_3)^2} \right] \tilde{\d}(p_1 + p_2 +p_3)\\
&= -i6\l^2 \frac{i}{p_2^2}\frac{i}{p_3^2}   \int \frac{\mathrm{d}^dk}{(2\pi)^d}\left[ \frac{L_1(k^2) L_1((k-p_1)^2)}{\left(k+p_2 \right)^2} + \left(p_2 \leftrightarrow p_3\right) \right] \tilde{\d}(p_1 + p_2 +p_3)\\
&=-i6\l^2 \frac{i}{p_2^2}\frac{i}{p_3^2}   \int \frac{\mathrm{d}^dk}{(2\pi)^d}\left[ \frac{L_1((k+p_3)^2) L_1((k-p_2)^2)}{k^2} + \left(p_2 \leftrightarrow p_3\right) \right] \tilde{\d}(p_1 + p_2 +p_3)\\
&= -i6\l^2 \frac{i}{p_2^2}\frac{i}{p_3^2} \left[T(p_2,p_3)+\left(p_2 \leftrightarrow p_3\right) \right] \tilde{\d}(p_1+p_2+p_3)
\eal\ee
with:
\be
\bal
T(p_2,p_3) &= \int \frac{\mathrm{d}^dk}{(2\pi)^{d}} \frac{L_1\left((k+p_3)^2\right) L_1\left((k-p_2)^2\right)}{k^2}  \\
&=  - \frac{\left[\G\left( 2-\frac{d}{2}\right)\right]^2 \left[\G \left(\frac{d}{2}-1 \right)\right]^4}{\left( 4\pi \right)^{d} \left[\G\left(d-2 \right)\right]^2}\int \frac{\mathrm{d}^dk}{(2\pi)^{d}} \frac{(-1)^{d-4}}{k^2 \left[(k+p_3^2) \right]^{2-\frac{d}{2}} \left[ (k-p_2)^2 \right]^{2-\frac{d}{2}}}\, ,
\eal 
\ee
where we have used \eqref{double candy p m}. This loop integral has a similar form with the loop integral introduced in \cite{Skenderis 3K, Skenderis CFT}, 
which is associated with the 3-point function of a CFT in momentum space. 
We will continue the evaluation by introducing consecutively two Feynman parameters:
\be\bal
T(p_2,p_3) =- \frac{ \left[\G \left(\frac{d}{2}-1 \right)\right]^4\G(5-d)}{\left( 4\pi \right)^{d} \left[\G\left(d-2 \right)\right]^2} &
\int_0^1 \mathrm{d}y \mathrm{d}z \int \frac{\mathrm{d}^dk}{(2\pi)^d} \frac{y^{1-\frac{d}{2}} (1-z)^{1-\frac{d}{2}} z^{2-\frac{d}{2}}}{\left[(1-z)(k-p_2)^2 + z (k^2 +2kp_3y +p_3^2y) \right]^{5-d}}
\eal\ee 
The integral with respect to $k$ is a standard one loop integral. Performing it, we obtain
\be
T(p_2,p_3) =\frac{i(-1)^{2d-10}}{(4\pi)^{3d/2}} \frac{\left[\G \left( \frac{d}{2}-1 \right)\right]^4 \G \left( 5-\frac{3d}{2} \right)}
{ \left[\G (d-2)\right]^2} \int_0^1 \mathrm{d}y \mathrm{d}z y^{1-d/2}(1-z)^{1-d/2} f(p_2,p_3,y,z)
 \ee
where
\be
f(p_2,p_3,y,z)=\frac{z^{2-d/2}}{\left[ -p_3^2 yz(1-yz)-p_2^2z(1-z) -2p_2\cdot p_3yz(1-z) \right]^{5-\frac{3d}{2}}}
\ee
For $d\rightarrow 4$ the above integral has poles at $y=0$ and $z=1$. We isolate the poles:
\be \label{T general d div plus finite}
T(p_2,p_3) = i \frac{(-1)^{2d-10}}{(4\pi)^{3d/2}}\frac{ \left[\G \left(\frac{d}{2}-1 \right)\right]^4\G\left(5-\frac{3d}{2}\right)}{ \left[\G\left(d-2 \right)\right]^2} \left( I_d(p_2,p_3) +I_f(p_2,p_3) \right)
\ee
with
\be
I_d(p_2,p_3)=  \int_0^1 \mathrm{d}y \mathrm{d}z y^{1-\frac{d}{2} } (1-z)^{1-\frac{d}{2}} \left[f(p_2,p_3,0,z)+ f(p_2,p_3,y,1) \right]
\ee
and
\be
I_f(p_2,p_3)= \int_0^1 \mathrm{d}y \mathrm{d}z y^{1-\frac{d}{2} } (1-z)^{1-\frac{d}{2}} \left[f(p_2,p_3,y,z)-f(p_2,p_3,0,z)- f(p_2,p_3,y,1) \right]
\ee
We first evaluate the $I_d$ integral .
\be
I_d(p_2,p_3) = \int_0^1 \mathrm{d}y \mathrm{d}z\; y^{1-\frac{d}{2}} (1-z)^{1-\frac{d}{2}} \left\{\frac{z^{2-d/2}}{\left[z(1-z) \right]^{5-\frac{3d}{2}}} 
(-p_2^2)^{\frac{3d}{2} -5} + \frac{1}{\left[y(1-y)\right]^{5-\frac{3d}{2}}} (-p_3^2)^{\frac{3d}{2}-5} \right\}
\ee
The integrals with respect to the Feynman parameters can be evaluated with the use of Euler's $\Beta$-function.:
\be \label{Id}
I_d(p_2,p_3) =2\frac{\G(d-3)\G(d-2)}{(4-d)\G(2d-5)} (-p_2^2)^{\frac{3d}{2}-5} + 2\frac{\G(d-3)\G \left(\frac{3d}{2}-4\right)}{(4-d)\G \left(\frac{5d}{2}-7 \right)} \left(-p_3^2 \right)^{\frac{3d}{2}-5}
\ee
In the $\epsilon$-expansion $T$ becomes
\be\bal
T(p_2,p_3) = \frac{i}{(4\pi)^6} &\left\{ \frac{2}{3\epsilon^2} \left(p_2^2 +p_3^2 \right)  -p_2^2 
\frac{\ln \left(\frac{-p_2^2 e^\g}{4\pi} \right)}{\epsilon} -p_3^2 \frac{\ln \left(\frac{-p_3^2 e^\g}{4\pi} \right)}{\epsilon}  + \frac{\frac{11}{3}p_2^2 +\frac{23}{6} p_3^2}{\epsilon} \right.  \\
& \;+p_2^2\left[ \frac{18}{24} \left(\ln \left(\frac{-p_2^2 e^\g}{4\pi} \right) - \frac{11}{3}\right)^2 + \frac{11}{4} - \frac{\pi^2}{24} \right]\\
&\left. + p_3^2 \left[ \frac{54}{72} \left(\ln \left(\frac{-p_3^2 e^\g}{4\pi} \right) - \frac{23}{6}\right)^2 +\frac{157}{48} - \frac{7\pi^2}{72} \right]  \right\} + I_{f}(p_2,p_3)
\eal \ee
The $I_f$ integral is simpler than $I_d$, since it is free from poles:
\be
I_f(p_2,p_3) = \int_0^1 \mathrm{d}y \mathrm{d}z \; \left[p_3^2 (1-y-yz) -2 p_2\cdot p_3 yz\right] = \frac{p_3^2}{4} -p_2 \cdot p_3
\ee
This is finite and as we discuss in the renormalization of $\braket{\mathcal{O}_4 \phi \phi }$, 
does not contribute in the renormalized expression, since it will become an $O(\epsilon)$ term.
We also add the crossing symmetric term and we finally arrive at
\be
\bal
T(p_2 ,p_3) + (p_2 \leftrightarrow p_3) = \frac{ip_2^2}{(4\pi)^6} &\left[  \frac{4}{\epsilon^2}- \frac{2\ln \left(\frac{-p_2^2 e^\g}{4\pi} \right) - \frac{55}{6} }{\epsilon}\right. \\
& \; \left.  + \frac{3}{4}\ln^2 \left(\frac{-p_2^2 e^\g}{4\pi} \right) - \frac{19}{4}\ln \left(\frac{-p_2^2 e^\g}{4\pi} \right) + \frac{635}{48} - \frac{\pi^2}{9}
\right] \\
&+ I_f(p_2,p_3) + (p_2 \leftrightarrow p_3)
\eal
\ee

\subsection{Massless 4-loop integrals}

These are integrals that arise in the computation of $\braket{K_3 K_2}$ and $\braket{K_3 K_3}$.

\subsubsection{Loop integral with a Tent insertion}
This integral is associated with the following diagram:
\be
 \begin{gathered}
\begin{tikzpicture}
\begin{feynman}
    \vertex (x);
    \vertex[below = 0.5cm of x](z);
    \vertex[below = 0.5cm of z] (y);
    \diagram{
        (x) -- [quarter left] (z) -- [quarter left] (y) -- [quarter left] (z) -- [quarter left] (x)--[half left](y)--[half left](x)
    };
\end{feynman}

\filldraw[fill = black] (x) ++(-0.1cm, -0.1cm) rectangle ++(0.2cm, 0.2cm);
\filldraw[fill = black] (y) ++(-0.1cm, -0.1cm) rectangle ++(0.2cm, 0.2cm);

\end{tikzpicture}
\end{gathered} 
\ee
The corresponding loop integral is given below :
\be
Q(p^2)=\int \frac{\mathrm{d}^d k}{(2\pi)^d} \frac{1}{(k+p)^2} TB(k^2)
\ee
Using the result for $TB(p^2)$ from \eqref{TB general d} we have
\be
i(-1)^{\frac{7d}{2}-15} \frac{\left[\G\left( 2-\frac{d}{2}\right)\right]^2 \left[\G \left(\frac{d}{2}-1 \right)\right]^5}{\left( 4\pi \right)^{3d/2} 
\left[\G\left(d-2 \right)\right]^2} \frac{\G\left( 5-\frac{3d}{2} \right)}{\G (4-d)} \frac{\G \left(\frac{3d}{2}-4 \right)}{\G (2d-5)} 
\int \frac{\mathrm{d}^d k}{(2\pi)^d} \frac{1}{\left(k^2 \right)^{5-\frac{3d}{2}}}\frac{1}{\left(k+p \right)^2}  
\ee
The remaining one-loop integral is easily obtained after one Feynman parametrization:
\be
\int \frac{\mathrm{d}^d k}{(2\pi)^d} \frac{1}{\left(k^2 \right)^{5-\frac{3d}{2}}}\frac{1}{\left(k+p \right)^2}  = 
\frac{i(-1)^{\frac{3d}{2} -6} }{(4\pi)^{d/2}}\frac{\G(6-2d) \G(2d-5) \G \left( \frac{d}{2} -1 \right)}{\G \left(5 - \frac{3d}{2} \right)  \G \left( \frac{5d}{2} -6 \right) }(-p^2)^{2d-6}
\ee
Then,
\be
Q(p^2) =\frac{(-1)^{5d-20}}{(4\pi)^{2d}} 
\frac{\left[ \G \left(2- \frac{d}{2} \right) \right]^2 \left[\G \left(\frac{d}{2}-1 \right) \right]^6  \G \left(\frac{3d}{2}-4 \right) \G(6-2d)}
{\left[\G(d-2)\right]^2 \G (4-d) \G \left( \frac{5d}{2} -6 \right)}  (-p^2)^{2d-6}
\ee
In the $\epsilon$-expansion the result is: 
\be
\bal
Q(p^2)  =\frac{p^4}{(4\pi)^8} &\left[ \frac{1}{6 \epsilon^2} - \frac{24 \ln \left(\frac{-p^2e^\g}{4\pi} \right)  -97}{72 \epsilon}  \right. \\
&\left.+  \ln^2 \left(\frac{-p^2e^\g}{4\pi} \right)  - \frac{97}{36} \ln \left(\frac{-p^2e^\g}{4\pi} \right) + \frac{5659}{864} - \frac{\pi^2}{72}    \right]\, .
\eal
\ee

\subsubsection{Loop integral with a Sunset-Tadpole insertion}

This integral is associated with the following diagram:
\be
\begin{tikzpicture}
\begin{feynman}
\vertex(x);
\vertex[right = 1cm of x](z);
\vertex[above =0.8cm of z](x1);
\vertex[below =0.8cm of z](x2);
\vertex[right = 1 cm of z](y1);
\vertex[right =0.8 cm of y1](y);
\diagram{(x)--(x1)--[quarter left](x)--[quarter left](x1)-- (y1)--[quarter left](x)};
\diagram{(y1)--[half left](y)--[half left](y1)};
\end{feynman}
\filldraw[fill=black] (x) ++(-0.1cm, -0.1cm) rectangle ++(0.2cm, 0.2cm);
\filldraw[fill =black](y) circle(3pt );
\end{tikzpicture}
\ee
The corresponding loop integral is the following:
\be
SC(p^2)=\int \frac{\mathrm{d}^dk}{(2\pi)^d} \frac{ST(k^2)}{k^2 \left(k-p \right)^2} \, .
\ee
Using \eqref{st general d} we obtain the following expression 
\be
SC(p^2) =\frac{i(-1)^{\frac{9d}{2}-17}} {(4\pi)^{3d/2}} 
\frac{\left[\G \left( \frac{d}{2}-1 \right) \right]^{4} \G(3-d) \G \left(5 - \frac{3d}{2} \right) \G\left( \frac{3d}{2} -4\right)}
{\G \left( \frac{3d}{2}-3 \right) \G(4-d) \G(2d-5)}\int \frac{\mathrm{d}^dk}{(2\pi)^d} \frac{1}{\left(k^2\right)^{6-\frac{3d}{2}} (k-p)^2}
\ee
The one loop integral is straightforward:
\be
SC(p^2) = \frac{\left( -1 \right)^{6d-23}}{(4\pi)^{2d}} 
\frac{\Gamma \left(\frac{d}{2}-1\right)^5 \Gamma (3-d) \Gamma \left(5-\frac{3 d}{2}\right) \Gamma \left(\frac{3 d}{2}-4\right) \Gamma (2 d-6) \Gamma (7-2 d)}{\Gamma \left(\frac{3 d}{2}-3\right) \Gamma (4-d) \Gamma (2 d-5) \Gamma \left(6-\frac{3 d}{2}\right) \Gamma \left(\frac{5 d}{2}-7\right)}\left(-p^2\right)^{2d-7}\, .
\ee
Expanding in $\epsilon$ we obtain:
\be
SC(p^2)=-\frac{p^2}{(4\pi)^8} 
\left[ \frac{1}{16\epsilon} - \frac{\ln \left(\frac{-p^2e^\g}{4\pi} \right) - 5}{8} \right]
\ee

\subsubsection{Loop integral with a Tent on the top}
This integral is associated with the following diagram:
\be
\begin{tikzpicture}
\begin{feynman}
\vertex(x);
\vertex[right = 1cm of x](z);
\vertex[above =0.8cm of z](x1);
\vertex[below =0.8cm of z](x2);
\vertex[right = 1 cm of z](y);
\diagram{(x)--(x1)--[quarter left](y)--[quarter left](x1)--(x2)--[quarter left](y)--[quarter left](x2) -- (x)};
\end{feynman}
\filldraw[fill=black] (y) ++(-0.1cm, -0.1cm) rectangle ++(0.2cm, 0.2cm);
\filldraw[fill =black](x) circle(3pt );
\end{tikzpicture}
\ee
Τhe corresponding loop integral is
\be
LT(p^2)=\int \frac{\mathrm{d}^dk }{(2\pi)^{d}} \frac{1}{k^2 (k+p)^2}\int \frac{\mathrm{d}^dl}{(2\pi)^d} \frac{L_1\left(( l+k+p)^2 \right) L_1\left(( l+k)^2 \right)}{l^2}\, .
\ee
Using the formula for $L_1(p^2)$  it becomes
\be\label{LT integral}
\bal
LT(p^2)=(-1)^{d-3}\frac{\left[\G\left( 2-\frac{d}{2}\right)\right]^2 \left[\G \left(\frac{d}{2}-1 \right)\right]^4}{\left( 4\pi \right)^{d} 
\left[\G\left(d-2 \right)\right]^2} 
&\int \frac{\mathrm{d}^dk }{(2\pi)^{d}} \frac{1}{k^2 (k+p)^2}\\
&\int \frac{\mathrm{d}^dl}{(2\pi)^d} 
\frac{1}{l^2 \left[ (l+k)^2\right]^{2- \frac{d}{2}} \left[ (l+k+p)^2\right]^{2- \frac{d}{2}}}
\eal
\ee
A similar loop integral has been evaluated in Appendix A of \cite{Rychkov_Trace}. 
However to see its precise relation to \eqref{LT integral} needs a few steps that we now outline. 
To begin, the integral eq.(A.1) in \cite{Rychkov_Trace} is evaluated in Euclidean position space.
In addition the theory under consideration is a non-local $\phi^4$-theory with its 
propagator of the form:
\be
\braket{\phi(x_E) \phi(0)} = \frac{1}{{|x_E|}^{2\a}} \, , \hskip .5cm \a = \frac{d-\epsilon}{4}
\ee
For the purposes of our analysis we will consider the normalization
\be
\braket{\phi(x_E) \phi(0)} = \frac{c_\phi}{{|x_E|}^{2\a}} 
\ee
with
\be
c_\phi = \pi^{-d/2} 2^{-d+2\a} \frac{ \G (\a)}{\G \left( \frac{d -2 \a}{2} \right)} 
\ee
so that the two point function in momentum space assumes the form
\be
\braket{\phi(p_E) \phi(-p_E)} = |p_E|^{2\a-d}\, .
\ee
Start from the loop integral in \cite{Rychkov_Trace} (with our new normalization):
\be \label{Itot position}
I_{tot}(x_E) =(c_\phi)^7\int \mathrm{d}^dy_E \mathrm{d}^dz_E \frac{1}{|x_E-y_E|^{2\a} 
|x_E-z_E|^{2\a} |y_E -z_E|^{2\a} |y_E|^{2\b} |z_E|^{2\b} } \,  , \b=2\a
\ee
Using the identity
\be\label{FT general d}
\int \mathrm{d}^dx_E\,  e^{i p_E \cdot x_E} \frac{1}{{x_E}^{2\D}} = 
\frac{\pi^{d/2}2^{d-2\D} \G \left( \frac{d-2\D}{2} \right)  }{\G (\D)} {|{p}_E|}^{2\D-d}\, ,
\ee 
this becomes
\be
I_{tot}(x_E) =(c_\phi)^7 \mathcal{A} \int \frac{\mathrm{d}^d k_E \mathrm{d}^d 
q_E \mathrm{d}^d l_E }{(2\pi)^{3d}} \,e^{i (k_E+q_E) \cdot x_E} 
k_E^{2\a-d} q_E^{2\a-d} (l_E- q_E)^{2\a-d} (k_E + q_E -l_E)^{2\b-d} l_E^{2\b -d} \; ,
\ee
where
\be
\mathcal{A} = \pi^{5d/2}\left[ 2^{d-2\a}\right]^{3} \left[ 2^{d-2\b}\right]^{2}
\frac{\left[ \G \left(\frac{d-2\a}{2} \right)\right]^{3}\left[ \G \left(\frac{d-2\b}{2} \right)\right]^{2}}
{\left[\G(\a) \right]^3 \left[ \G (\b) \right]^2}\, .
\ee
Now it is easy to pass to momentum space:
\be\bal
\tilde{I}_{tot}(p_E)& = \int\mathrm{d}^d x_E e^{-i p_E \cdot x_E} I_{tot}(x_E) \\
&= (c_\phi)^7 \mathcal{A}
\int \frac{\mathrm{d}^d k_E  \mathrm{d}^d l_E }{(2\pi)^{2d}} 
\frac{1}{ \left[k_E^2 \right]^{ \frac{d+\epsilon}{4}} \left[(k_E-p_E)^2 \right]^{ \frac{d+\epsilon}{4}} 
\left[(k_E+l_E -p_E)^2 \right]^{ \frac{d+\epsilon}{4}}  \left[l^2 \right]^{ \frac{\epsilon}{2}} \left[(l_E-p_E)^2 \right]^{ \frac{\epsilon}{2}}   }
\eal\ee
where we have substituted the values of $\a$ and $\b$.
Next we apply the three consecutive shifts on the loop momenta 
\be
\bal
k_E &\to -k_E \\
l_E &\to l_E+k_E +p_E \\
l_E &\to -l_E
\eal
\ee
to obtain:
\be
\tilde{I}_{tot}(p_E) = (c_\phi)^7 \mathcal{A}\int \frac{\mathrm{d}^d k_E  \mathrm{d}^d l_E }{(2\pi)^{2d}} 
\frac{1}{ \left[k_E^2 \right]^{ \frac{d+\epsilon}{4}} \left[(k_E+p_E)^2 \right]^{ \frac{d+\epsilon}{4}} 
\left[l_E^2 \right]^{ \frac{d+\epsilon}{4}}  \left[(k_E+l_E +p_E)^2 \right]^{ \frac{\epsilon}{2}} \left[(l_E+k_E)^2 \right]^{ \frac{\epsilon}{2}}   }
\ee
Finally, for $d=4-\epsilon$ we obtain:
\be\bal
\left. \tilde{I}_{tot}(p_E) \right|_{4-\epsilon} =\, &\frac{ \left[\G \left(1-\frac{\epsilon }{2}\right)\right]^4 \left[\G\left(\frac{\epsilon }{2}\right)\right]^2}{\left[\Gamma (2-\epsilon )\right]^2}\\
& \times  \int \frac{\mathrm{d}^{4-\epsilon} k_E  \mathrm{d}^{4-\epsilon} l_E }{(2\pi)^{2(4-\epsilon)}} 
\frac{1}{ k_E^2  \left( k_E+p_E \right)^2  l^2   \left[(k_E+l_E +p_E)^2 \right]^{ \frac{\epsilon}{2}} \left[\left(l_E+k_E \right)^2 \right]^{ \frac{\epsilon}{2}}   }
\eal \ee
This expression is identical to the Euclidean version of \eqref{LT integral} for $d=4-\epsilon$.
This means that we can use the result of the integral $I_{tot}(x_E)$ of \cite{Rychkov_Trace} for $d=4-\epsilon$.
Their result is
\be
I_{tot}(x_E) =(c_\phi)^7 4\pi^d \frac{ \G \left( \frac{-d}{4} \right)}{\G \left( \frac{3d}{4}\right) } 
\frac{1}{|x_E|^{2 \left(\frac{3d}{4} - \frac{7 \epsilon}{4}  \right )}}  + O(\epsilon)
\ee
which can be moved to momentum space using \eqref{FT general d}, to obtain
\be
\tilde{I}_{tot}(p_E) = \frac{4 \G \left(-\frac{d}{4}\right)\left[ \G \left(\frac{d-\epsilon }{4}\right)\right]^7 \G \left(\frac{7 \epsilon }{4}-\frac{d}{4}\right)}
{\left( 4\pi \right) ^{2 d}\G \left(\frac{3 d}{4}\right) \G \left(\frac{3 d}{4}-\frac{7 \epsilon }{4}\right) \left[\G \left(\frac{d+\epsilon }{4}\right)\right]^7} 
 \left(p_E^2\right)^{\frac{1}{4} (d-7 \epsilon )} + O(\epsilon)
\ee
Now we set $d=4-\epsilon$ and get
\be
\bal
\left. \tilde{I}_{tot}(p_E) \right|_{4-\epsilon} &=\frac{4   \left[\G \left(1-\frac{\epsilon }{2}\right)\right]^7 
\G \left(\frac{\epsilon -4}{4}\right) \G (2 \epsilon -1)}
{\left(4\pi \right)^{8-2 \epsilon }\G \left(3-\frac{5 \epsilon }{2}\right) \G \left(3-\frac{3 \epsilon }{4}\right)}
\left(p_E^2\right)^{1-2 \epsilon } + O(\epsilon) \\
& =  \frac{p_E^2}{(4\pi)^8} \left[  \frac{2}{\epsilon^2} - \frac{4 \ln(p_E^2)}{\epsilon}  + \cdots \right]
\eal
\ee
Finally, after a Wick rotation we obtain the $LT(p^2)$ integral in Minkowski space that we are after:
\be
LT(p^2)= -\frac{p^2}{(4\pi)^8} \left[  \frac{2}{\epsilon^2} - \frac{4 \ln(-p^2)}{\epsilon}  + \cdots \right]\, .
\ee

\subsubsection{Loop integral with a Double Candy insertion}

This integral is associated with the following diagram from the 2-point function $\braket{\phi^2 \phi^4}$:
\be
\begin{tikzpicture}
\begin{feynman}
\vertex(x);
\vertex[right = 1cm of x](z);
\vertex[right =1cm of z](z1);
\vertex[right = 1 cm of z1](y);
\diagram{(x)--(z)--[quarter left](z1)--[quarter left](z)--[quarter left](y)--[quarter left](z1) --[quarter left] (y)--[quarter left](x)};
\end{feynman}
\filldraw[fill=black] (y) ++(-0.1cm, -0.1cm) rectangle ++(0.2cm, 0.2cm);
\filldraw[fill =black](x) circle(3pt );
\end{tikzpicture}
\ee
The corresponding loop integral is the following:
\be\bal
QC(p^2)=\int \frac{\mathrm{d}^d k \mathrm{d}^d l}{(2\pi)^{2d}} \frac{\left[L_1\left(l^2\right) \right]^2}{k^2 (k-p)^2 (k-l)^2} 
\eal
\ee
Using \eqref{double candy pp}  the above integral gets the following form:
\be\label{QC 1st}
QC(p^2)=- \frac{\left[\G\left( 2-\frac{d}{2}\right)\right]^2 \left[\G \left(\frac{d}{2}-1 \right)\right]^4}{\left( 4\pi \right)^{d} 
\left[\G\left(d-2 \right)\right]^2}\int \frac{\mathrm{d}^d k }{(2\pi)^{2}} \frac{\left(-1 \right)^{d-4}}{k^2 (k-p)^2 }  \int \frac{\mathrm{d}^d l}{(2\pi)^{d}} \frac{1}{(k-l)^2 (l^2)^{4-d}}
\ee
First we will evaluate the one loop integral with respect to $\l$. After a Feynman parametriazation we obtain the standard one loop integral and we finally conclude to the following result:
\be
\int \frac{\mathrm{d}^d l}{(2\pi)^{d}}\frac{1}{(k-l)^2 (l^2)^{4-d}}= \frac{i}{(4\pi)^{d/2}}(-1)^{d-5}
\frac{ \G \left(5-\frac{3 d}{2}\right) \G \left(\frac{d}{2}-1\right) \G \left(\frac{3 d}{2}-4\right)}{\G (4-d) \G (2 d-5)}
\left(-k^2\right)^{\frac{3 d}{2}-5}
\ee 
Substituting in  \eqref{QC 1st} we obtain:
\be
QC(p^2)=-i(-1)^{\frac{7d}{2}-14} \frac{\left[\G\left( 2-\frac{d}{2}\right)\right]^2 \left[\G \left(\frac{d}{2}-1 \right)\right]^5}{\left( 4\pi \right)^{3d/2} \left[\G (d-2) \right]^2 } 
\frac{ \G \left(5-\frac{3 d}{2}\right)  \G \left(\frac{3 d}{2}-4\right)}{\G (4-d) \G (2 d-5)}
\int \frac{\mathrm{d}^dk}{(2\pi)^d} \frac{1}{(k-p)^2 (k^2)^{6- \frac{3d}{2}}}
\ee
We follow the standard procedure for the one loop integral and we obtain: 
\be
\int \frac{\mathrm{d}^dk}{(2\pi)^d} \frac{1}{(k-p)^2 (k^2)^{6- \frac{3d}{2}}}=
\frac{i(-1)^{\frac{3d}{2}-7 }}{(4\pi)^{d/2}}
\frac{\Gamma (7-2 d)   \Gamma \left(\frac{d}{2}-1\right) \Gamma (2 d-6)}{\Gamma
   \left(6-\frac{3 d}{2}\right) \Gamma \left(\frac{5 d}{2}-7\right)}
(-p^2)^{2d-7}
\ee
So we conclude that:
\be
QC(p^2)= \frac{(-1)^{5d- 21}}{(4\pi)^{2d}} \frac{\left[\G \left(2-\frac{d}{2} \right) \right]^2 \left[\G \left(\frac{d}{2}-1 \right) \right]^6\G 
\left(\frac{3d}{2}-4 \right) \G (7-2d) \G (2d-6)     } { \left[\G(d-2) \right]^2 \G(4-d) \G (2d-5) \G 
\left( \frac{5d}{2}-7 \right)  \left(5 - \frac{3d}{2} \right)  }\left(-p^2 \right)^{2d-7}
\ee
Expanding in $\epsilon$ we obtain:
\be
QC(p^2) = \frac{p^2}{(4\pi)^8} \left[\frac{1}{2 \epsilon^2} - \frac{8 \ln \left(\frac{-p^2e^\g}{4\pi} \right) -35}{8 \epsilon} + \cdots  \right]
\ee

\end{appendix}
\newpage

\end{document}